\documentclass[aps,pre,twocolumn,amsmath,amssymb,floatfix,citeautoscript]{revtex4-1} 
\usepackage{amssymb}
\usepackage{amsmath}
\usepackage{graphicx}
\usepackage{bm}
\usepackage{dcolumn}

\newcommand{\beq}{\begin{equation}}
\newcommand{\enq}{\end{equation}}
\newcommand{\ber}{\begin{eqnarray}}
\newcommand{\enr}{\end{eqnarray}}

\begin{document}

\title{Transitions of tethered chain molecules under tension}
\author{Jutta Luettmer-Strathmann}
\email[]{jutta@uakron.edu}
\affiliation{
Department of Physics and Department of Chemistry, The University of
Akron, Akron, Ohio 44325-4001} 
\author{Kurt Binder}
\affiliation{
Institut f{\"{u}}r Physik, Johannes-Gutenberg-Universit{\"{a}}t,
Staudinger Weg 7, D-55099 Mainz, Germany}
\date{\today}
\begin{abstract}
An applied tension force changes the equilibrium conformations of a 
polymer chain tethered to a planar substrate and thus affects the
adsorption transition as well as  the coil-globule and crystallization
transitions.
Conversely, solvent quality and surface attraction are reflected in 
equilibrium force-extension curves that can be measured in experiments. 
To investigate these effects theoretically, we study tethered chains
under tension with Wang-Landau simulations of a bond-fluctuation
lattice model.
Applying our model to pulling experiments on biological
molecules we obtain a good description of experimental data in the 
intermediate force range, where universal features dominate and finite
size effects are small.  
For tethered chains in poor solvent, we observe the predicted
two-phase coexistence at 
transitions from the globule to stretched conformations
and also discover direct transitions from crystalline to stretched
conformations. 
A phase portrait for finite chains constructed by evaluating the
density of states for 
a broad range of solvent conditions and tensions shows how 
increasing tension leads to a  disappearance of the globular phase.
For chains in good solvents tethered to hard and
attractive surfaces  we find the predicted scaling with the chain
length in the low-force regime and show that our results are well described
by an analytical, independent-bond approximation for the
bond-fluctuation model for the highest tensions.
Finally, for a hard or slightly attractive surface the stretching of a
tethered chain is a conformational change that does not correspond to
a phase transition. 
However, when the surface attraction is sufficient to adsorb
a chain it will undergo a desorption transition at a critical value of
the applied force.
Our results for force-induced desorption show the transition to be
discontinuous with partially desorbed conformations in the coexistence
region. 

\end{abstract}

\maketitle

\section{Introduction}
Experiments that involve the stretching of single chain molecules have 
become an important tool in biological physics \cite{ri06,ku10}. 
In non-equilibrium experiments the
chain may be extended at a constant rate to determine the
rate-dependent ``rupture'' force, i.e.\ the force where 
abrupt changes in conformation take place \cite{ev97,ra07}, 
or chain molecules may be extended at
constant force to observe the step-wise unfolding of parts of the
chain.\cite{do09}
To interpret rupture experiments, a knowledge of the equilibrium
elastic properties of chains under tension is required \cite{ra07}.
In equilibrium experiments, on the other hand, the chain is allowed to
explore all 
possible conformations consistent with the applied tensile force. 
Depending on the force range, the measured extensions reflect 
properties of specific molecules or universal features common to
many types of chain molecules. Equilibrium force-extension
data therefore provide insight into interactions of particular
molecules and chain segments and also serve as tests 
of more general models and theoretical predictions \cite{ri06,ku10}.

The conformations of chain molecules near surfaces are affected by
solvent conditions and interactions of chain segments with the
surface. Depending on the solvent conditions, chains may be in coil,
globule, or ordered, crystalline conformations, where, in each case,
the chains may be adsorbed or desorbed, depending on the surface
interactions. Even for simple chain models, 
the competition between segment-segment and segment
surface interactions leads to complex phase diagrams that are still
being investigated \cite{vr96,ra02b,kr05,ba06,lu08,bi08,mo11}. 
For biomolecules, applications in nano science and biomaterials have
inspired extensive computational research into the conditions under
which proteins adsorb to surfaces and the resulting conformational
changes of the molecules (see, for example,
Refs.~[\onlinecite{kn08,he09,sw12,ra12}]).  
Investigating the effects of tension on chain molecules near adsorbing
surfaces and in poor solvent conditions may help us understand the
effects of multiple interactions on configurational properties of the
molecules. This is a challenging problem since 
three independent thermodynamic variables, related to the strength of
the effective monomer-monomer attraction, the monomer-surface
attraction, and the force acting on the free end of the chain,
govern the states of the chains even for the simplest models.  

The mechanical response of chain molecules to an applied tension force
in equilibrium conditions has been investigated with 
experimental
\cite{ri06,bu94,sm96c,ha02c,de02d,co03,da07,gu07b,sa09,li10,di11}, 
theoretical
\cite{pi76,ha91c,ha91b,ma95c,ha97f,bo99,li03,bh09,ku10,kl11,sk12}, and 
simulation \cite{wi94c,wi96c,gr02,fr02b,ce04,mo07b,bh09b,to10,hs12}
techniques. 
Studies in good and moderate solvent conditions have explored
scaling relations at intermediate extensions
\cite{pi76,sa09,di11,hs12} as well as the high tension regime 
\cite{hs12,to10}. 
Except for the smallest and largest forces, the
mechanical response of a chain depends strongly on its stiffness
\cite{hs12}.  Dittmore {\it et al.}\cite{di11} observed the
complex scaling behavior predicted for semiflexible chains in 
recent measurements on poly(ethylene glycol) (PEG), while Saleh 
{\it et al.}\cite{sa09} investigated the effect of solvent condition
on scaling relations of flexible chain molecules; we discuss flexible
chains in this work.  
In poor solvent conditions, 
the transition from globular to stretched chain conformations has been
the focus of attention
\cite{ha91c,ha91b,wi96c,fr02b,gr02,gu07b,mo07b,li10} and led to
confirmation of the predicted first-order nature of the transition by
simulations and experiment. 
For adsorbing surfaces, an applied tension force perpendicular to the
surface leads to the desorption of the chain at a critical value of
the force\cite{ha02c,ce04,bh09,bh09b,kl11,sk12}.   
The nature of the
force-induced desorption transition has recently been the subject of
some controversy \cite{bh09,sk12} since it shows characteristics of
discontinuous as well as continuous transitions.

The study of chain molecules in the absence of tension has 
benefited greatly from simulations with Wang-Landau type algorithms 
\cite{wa01,wa01b,pa07,wu11,lu08,ta09,de08,ra12,sw12}. These algorithms give
access to the density of states and are well suited to investigate
phase transitions in finite-size systems and to study chain
conformations that are difficult to reach with traditional, Metropolis
Monte Carlo methods. 
In this work, we apply Wang-Landau simulation techniques to a lattice
model for a single chain, with one end
tethered to a planar surface and the other end subject to a constant
applied force in the direction perpendicular to the surface.
We construct two kinds of densities of state: 
The first is over a three-dimensional state space spanned by chain
extensions and energy contributions from interactions of chain
segments with each other and the surface. 
This 3-d density of states allows us to identify interesting 
state points by evaluating properties of
tethered chains for continuous ranges of solvent, surface, and force
parameters. 
For conditions of interest, 
we perform Wang-Landau simulations over one-dimensional state spaces
of chain extensions at fixed surface and solvent conditions.
These 1-d densities of state allow us to reach extreme extensions
and investigate chain stretching in great detail. 

This article is organized as follows: Following this overview 
we briefly review scaling predictions for
flexible 
chain molecules under tension. 
In section \ref{model} we describe the model and thermodynamic
relations employed in this work. Details about the simulation method
are presented in the Appendix.
In section~\ref{athermal} we discuss force-extension relations for
chains in athermal solvent in the presence and absence of a  
hard surface. The effect of solvent quality on chain extension is the
subject of \ref{solvent} while section \ref{adsorption} focuses on 
the effects of attractive surface interactions.
A summary and conclusions are presented in 
section \ref{conclusion}. 

\subsection{Scaling predictions for flexible chain molecules under
  tension}\label{review_tension} 

The response of a chain molecule to a stretching force applied to its
ends is known exactly for many simple polymer models, 
where long range correlations due to excluded volume are ignored 
(see, for example, Refs.~[\onlinecite{bu94,ma95c,bo99,gr94c,hs12}], and
references therein). 
For small forces, the extension varies linearly 
with the applied force and satisfies Hooke's law
\begin{equation}\label{Hooke} 
R_z = k^{-1}f,
\end{equation}
where $f$ is the applied force, $k$ is the spring
constant, and $R_z$ is the extension, that is the component of the
end-to-end vector along the force direction, 
which we take to be the $z$-axis in a Cartesian coordinate system.
Noting that the temperature and the tension force define a length
scale, the so-called tensile screening length $\xi = k_B T/f$,
Pincus\cite{pi76,sa09} derived a general scaling description 
of the extension of a chain under tension, 
$R_z \sim R_0\Psi(R_0/\xi)$. 
Here $\Psi$ is a scaling function whose form depends on the
relative size of $R_0$ and $\xi$. 
For low tension, $\xi > R_0$, Hooke's law is recovered when
$\Psi(x)\sim x$  so that 
\beq\label{lin_scale}
R_z/R_0 \sim R_0 f/k_BT .
\enq 
Since $R_0$ scales with the chain length
$N$ as $R_0\sim N^\nu$, where $\nu\simeq 3/5$ is the good-solvent
exponent in three dimensions \cite{gr94c}, Eq.~(\ref{lin_scale})
implies that the spring constant in good solvent conditions  
decreases with increasing chain length as $N^{-2\nu}$. 

For larger tensions, $\xi < R_0$, there is an intermediate regime
where the extension $R_z$ is larger than $R_0$ but still considerably
smaller than the contour length $L$. In this force regime, 
the extension scales with the contour length $L$, which yields 
$\Psi(x)\sim x^{(1-\nu)/\nu}$ and 
\beq\label{Rgood}
R_z/L \sim (bf/k_BT)^{2/3} ,
\enq
where $b$ is the length of a chain segment.\cite{pi76,sa09} 
The derivation of Eq.~(\ref{Rgood}) assumes only that the chain is in
a good solvent, where excluded volume interactions between the
segments play a role. It is therefore expected to be universal,
independent of the particular model or molecule studied. 
The power law in Eq.~(\ref{Rgood}) has been confirmed experimentally
\cite{sa09} and (for fully flexible chains) is valid until the extension
becomes comparable to the 
contour length. Its range of applicability may be estimated from
Pincus' blob picture \cite{pi76,sa09}, where the polymer is represented by
an ideal chain of blobs of size $\xi$, with the polymer segments
inside each blob subject to excluded volume interactions. As the
tension increases, the 
blob size decreases until it contains a single Kuhn segment and 
details of the chain model become important \cite{to10,hs12}. 
For both intermediate and high tensions, the
normalized extension $R_z/L$ for given tension force is 
independent of the chain length.  

In poor solvent, a polymer chain collapses into a globule in the
absence of tension. 
Halperin and Zhulina \cite{ha91c,ha91b} 
investigated the elastic properties of individual
polymer globules. Considering the increase in surface free energy by
stretching a globule and using the blob picture of Pincus \cite{pi76}
they find three force regions with different 
scaling laws corresponding to three different types of conformations. 
For small tension, the globule is deformed and the force law is
linear, $f\sim R_z$; 
for intermediate tension, the globule unravels and there is
coexistence between the part of the chain that is still globular and
the part that is already extended, in this case the force is
independent of the size $f\sim R_z^0$. Finally, for the largest tension,
the whole chain is extended and linear scaling is predicted for the
tension $f\sim R_z$. 
A number of simulation studies have been performed on stretched
polymers in poor solvent (see, for example,
Refs.~[\onlinecite{ci95,wi96c,gr02,fr02b,br09}] and references
therein). 
The simulations confirm the general picture laid out by Halperin and
Zhulina \cite{ha91b,ha91c} and show evidence of the first-order nature
of the coil-globule transition under high tension and coexistence of
stretched and globular regions along the same chain. 
Recent force-extension measurements by Walker and coworkers
\cite{gu07b,li10} on poly(styrene) in water and other poor solvents
showed the predicted force plateau and thus 
provided experimental confirmation of the discontinuous nature of the 
transition from the globule to stretched conformations. 

The presence of a hard tethering surface is felt most strongly for
small forces and extensions. 
For a tethering point at $z=0$, the $z$-coordinate of the last chain
segment is the 
extension and always non-negative. Its value at zero force is the
perpendicular size of the chain, $z_0 = R_\perp$, which  
serves as the scaling length in the low force regime, 
\beq\label{scalet}
z/z_0 = \Psi(z_0f/k_BT) .
\enq 
In the limit $x\rightarrow 0$, the scaling function $\Psi(x)$ reduces
to $\Psi(x)-1 \sim x$ and Hooke's law for surface-tethered
chains becomes $z-z_0\sim z_0^2 f/k_BT$. As for free chains under
tension, the spring constant is expected to scale with chain length as
$z_0^{-2}\sim N^{-2\nu}$.  

An adsorbing surface changes the force-extension relations. For
adsorbed chains, the perpendicular size $R_\perp$ is 
independent of the chain length and decreases with increasing surface
attraction  until the chain is completely adsorbed (strong coupling
limit).\cite{ei82,de08}  Since $R_\perp$ is independent of $N$,
we expect the low force extension of adsorbed chains to be the same
for all $N$. As the force increases, it eventually reaches the
critical value for force-induced desorption. Once the untethered
segments are removed from the surface, the adsorbing surface does not
affect the scaling behavior any more.

\section{Model and methods}\label{model}

The bond fluctuation (BF) model \cite{ca88b,bi95,la00b} is a
coarse-grained lattice model, where every segment of the model chain 
represents several repeat units of a polymer molecule. 
In this model, beads of a chain occupy 
sites on a simple cubic lattice. The bond lengths are allowed to vary
between $b = 2a$ and $b = \sqrt{10}a$, 
where $a$ is the lattice constant, which we set to unity, $a=1$. 
The advantage of this model, compared to fixed bond-length models like
the self-avoiding walk on a simple cubic lattice, is that the large
number of possible angles between successive bonds gives a
description of polymer chains that is closer to more realistic
off-lattice models, while still providing the computational advantages
of a lattice model. 
A tethered polymer is represented by a chain whose first bead is
fixed just above an impenetrable surface. In the Cartesian coordinate
system employed in this work, the surface spans the $x-y$ plane at
height $z=-1$ and the tethered monomer is at $z=0$. 
All monomers at $z=0$ are considered to be in contact with the surface and
contribute an amount $\epsilon_s$ to the energy.
To compare surface tethered chains with free chains under tensions, 
we have also performed 
simulations for a model where the surface is absent and the first bead
of the chain is fixed to the origin. 

The interactions between monomers have repulsive and attractive
parts. 
Hard core repulsion prevents distances $r_{ij}^2 < 4$ between any two 
monomers $i$ and $j$. 
Attractive interactions are implemented by counting as one bead
contact a pair of monomers with distances in the range 
$4 \leq r_{ij}^2 < r_c^2$, where $r_c^2 = 7$. 
The total energy of the system is given by
\begin{equation}\label{energy}
E(n_s,n_b) = n_s\epsilon_s + n_b\epsilon_b ,
\end{equation}
where $n_s$ and $n_b$ are the number of monomer-surface and
monomer-monomer contacts, respectively. 
We have also tested a larger range of the monomer-monomer attraction
and found that that this causes only minor quantitative differences, the
qualitative behavior remains the same.

\begin{figure}[!htbp]
\begin{center}
\includegraphics[width=3.4in]{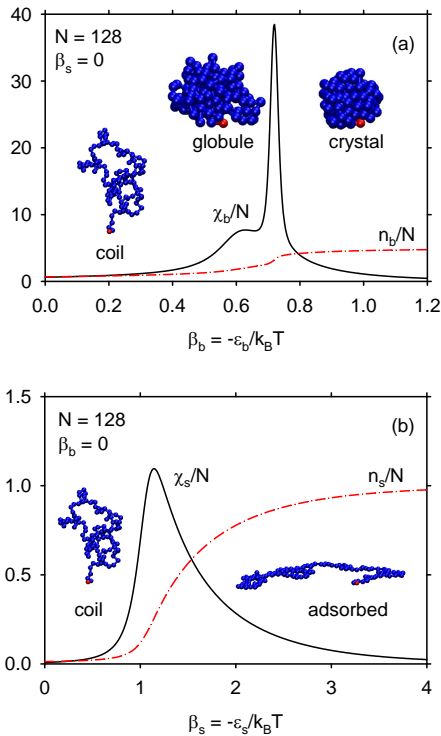}
\caption{
(a) Coil-globule and crystallization transition for chains tethered to
  a hard surface, $\beta_s = 0$. The dash-dotted line 
  shows the normalized number of bead contacts, $n_b/N$, 
  as a function of solvent quality, $\beta_b=-\epsilon_b/k_BT$, for
  chain length $N=128$; the solid line represents the normalized
  bead-contact fluctuations $\chi_b/N$.
  The maxima in $\chi_b$ provide estimates for the coil-globule
  transition, 
  $\beta_c \simeq 0.63$, and the freezing transition $\beta_X \simeq 0.72$. 
  The insets represent simulation snapshots that illustrate expanded
  coil, globule, and crystalline chain conformations. 
 (b)  Adsorption transition in athermal solvent, $\beta_b = 0$.  The
  dash-dotted line shows the normalized number of surface contacts,
  $n_s/N$, as a function of  $\beta_s = -\epsilon_s/k_BT$  for
  $N=128$; the solid line represents the
  normalized surface contact fluctuations $\chi_s/N$.  
  The maximum in $\chi_s$ provides an estimate for the adsorption
  transition, which occurs at $\beta_{sa} \simeq 1.14$ for this chain
  length. 
  The simulation snapshots illustrate expanded coil and adsorbed chain
  conformations.   
 \label{N128_bb0_bs0}}
\end{center}
\end{figure}

When a force $f$ perpendicular to the surface is applied to the last bead
of a tethered chain of length $N$, the chain extension is the 
$z$ coordinate of the last bead and the maximum extension is 
$z_\mathrm{max} = 3(N-1)$.  
A state of the system is described by the triplet $(n_s,n_b,z)$ and 
the density of states $g(n_s,n_b,z)$ is defined as the number of
chain conformations for given $(n_s,n_b,z)$.  The canonical partition
function is given by  
\beq\label{Zpartf}
Z = \sum_{n_s,n_b,z}g(n_s,n_b,z)
\mathrm{e}^{\beta_s n_s+\beta_b n_b+\beta_f z} ,
\enq
where 
\beq\label{betacan}
\beta_s = -\epsilon_s/{k_B T}, \mbox{\hspace{1em}} 
\beta_b = -\epsilon_b/{k_B T}, \mbox{ and } 
\beta_f = {fa}/{k_B T},
\enq
are the thermodynamic fields conjugate to $n_s$, $n_b$, and $z$,
respectively. We refer to $\beta_f$ as the tension field and 
to $\beta_s$ and $\beta_b$ as surface and bead
contact fields, respectively. 
The contact fields describe surface and solvent effects; 
as $\beta_s$ increases, the surface becomes more adsorbing, 
the number of surface contacts increases, and the chain goes through
an adsorption transition, which is accompanied by large fluctuations 
in the number of surface contacts.   
As $\beta_b$ increases, the bead-bead interactions become more
attractive, corresponding to increasingly poor solvent conditions,  
the number of bead contacts increases, and the chain first
goes through a coil-globule and then through a freezing transition.
Fig.~\ref{N128_bb0_bs0} shows the average contact numbers and their
fluctuations along with representative chain conformations as a chain
of length $N=128$ undergoes adsorption and chain collapse. 

The probability to find a
chain subject to contact and tension fields in the state 
$(n_s,n_b,z)$ is obtained from 
\beq\label{probab2}
P(n_s,n_b,z;\beta_s,\beta_b,\beta_f) = \frac{1}{Z}g(n_s,n_b,z)
  \mathrm{e}^{\beta_s n_s+\beta_b n_b+\beta_f z} ,
\enq
which implies that the density of states needs to be determined only up
to a constant prefactor. 
From the probability distribution we calculate canonical expectation
values such as the average height of the last bead 
\beq\label{zavgfields}
\langle z\rangle =  
\sum_{n_s,n_b,z} z P(n_s,n_b,z;\beta_s,\beta_b,\beta_f), 
\enq
the average number of surface contacts, $\langle n_s\rangle$ 
the average number of bead-bead contacts, $\langle n_b\rangle$, 
as well as fluctuations of these quantities 
\ber
\chi_s &=& \langle n_s^2 \rangle - \langle n_s
\rangle^2,\label{chis}\\
\chi_b &=& \langle n_b^2 \rangle - \langle n_b \rangle^2, \label{chib}\\
\chi_z &=& \langle z^2 \rangle - \langle z \rangle^2, \label{chiz} \\
\chi_{zb} &=& \langle zn_b \rangle - \langle z \rangle 
\langle n_b \rangle, \label{chizb} \\
\chi_t &=& \chi_z + \chi_b + 2\chi_{zb}. \label{chit}
\enr
In Appendix \ref{g3} we describe the simulation methods we employed to 
construct the density of state $g(n_s,n_b,z)$ over the
three-dimensional state space spanned by contact numbers and
extensions. 

At fixed contact fields $\beta_s$ and $\beta_b$, the chain extensions,
$z$, form a one-dimensional state space with density of states
$g(z;\beta_s,\beta_b)$, which, after normalization, represents 
the probability distribution for the extension.
The canonical probability distribution 
\beq\label{prob1d}
p(z;\beta_s,\beta_b,\beta_f) =
\frac{1}{Q}g(z;\beta_s,\beta_b)\mathrm{e}^{\beta_f z}, 
\enq
where $Q = \sum_z g(z;\beta_s,\beta_b)\mathrm{e}^{\beta_f z}$ 
is the partition function, may be evaluated to find the average
extension 
\beq\label{zavgWL}
\langle z\rangle = \sum_{z=0}^{z_\mathrm{max}}z
p(z;\beta_s,\beta_b,\beta_f) = \left(\frac{\partial\ln
  Q}{\partial\beta_f}\right)_{\beta_s,\beta_b},  
\enq
and the fluctuations 
\beq\label{chizWL}
\chi_z = \langle z^2\rangle - \langle z\rangle^2 = 
\left(\frac{\partial\langle z\rangle}%
{\partial \beta_f}\right)_{\beta_s,\beta_b}. 
\enq
In this statistical ensemble, the tension field $\beta_f$
is controlled and the extension $z$ is allowed to fluctuate. 
This is the approach we take for most of the results presented
here. 
However, when investigating the nature of a transition, it is
useful to consider a micro-canonical type of evaluation, where the
extension $z$ is controlled and the field $\beta_f$ fluctuates. 
In this formalism, the average tension field $\bar{\beta}_f$ at a
given height $z$ is calculated from the first derivative of the
density of states  
\beq\label{betafmicro}
\bar{\beta}_f(z,\beta_s,\beta_b) = 
-\left(\frac{\partial \ln(g)}{\partial z}\right)_{\beta_s,\beta_b}, 
\enq
and the inverse of the fluctuations from the second derivative
\beq\label{chizmicro}
\chi_z^{-1} =
\left(\frac{\partial \bar{\beta}_f}{\partial z}\right)_{\beta_s,\beta_b} = 
-\left(\frac{\partial^2 \ln(g)}{\partial z^2}\right)_{\beta_s,\beta_b}. 
\enq
In Appendix \ref{g1} we describe our Wang-Landau simulations for the 
1-d densities of states $g(z;\beta_b,\beta_s)$.

When a chain is highly stretched, or when interactions
between non-bonded beads are effectively screened, individual 
bonds respond independently to the applied force.
Wittkop {\it et al.}\cite{wi94c} enumerated the possible orientations
and lengths 
of the bonds in the BF model to determine a high-tension approximation
for the extension of an untethered, athermal chain. 
To include energetic effects, we note that
the beads constituting a bond make a
bead-bead contact ($n_b=1$) when the bond-length is smaller than
$r_c = \sqrt{7}$. 
The number of bond vectors for each pair of $(z_b,n_b)$ values defines  
the single-bond density of states  $g_1(z_b,n_b)$ and is presented in
Table \ref{table_g1}. 
For given bead contact and tension fields, the average extension of a
single bond is given by 
\beq\label{indep}
\langle z_b \rangle = \frac{1}{q_1}\sum_{z_b,n_b} z_b g_1(z_b,n_b) 
\mathrm{e}^{\beta_f z_b} \mathrm{e}^{\beta_b n_b} ,
\enq
where $q_1 = \sum_{z_b,n_b}g_1(z_b,n_b)\exp{(\beta_f z_b + \beta_b n_b)}$
is the the single bond partition function. 
In the independent-bond (IB) approximation, $\langle z_b \rangle$ is
equal to the normalized extension $z/z_\mathrm{max}$, where 
$z_\mathrm{max}=3(N-1)$ is the maximum extension of the chain, 
and may be compared with simulations results.

\begin{table}[!htbp] 
\caption{Single bond density of states; $g_1(z_b,n_b)$ is the number of
  single-bond configurations with $z$-coordinate $z_b$ and $n_b$
  contacts between the beads constituting the bond. The total number
  of bond configurations is 108.}
\label{table_g1}
\vspace*{0.5cm}
\begin{ruledtabular}
\begin{tabular}{l|r|r|r|r|r|r|r|r|}
 $z_b$ & 0& 0 
& $\pm 1$& $\pm 1$ & $\pm 2$& $\pm 2$
& $\pm 3$& $\pm 3$
\\ \hline 
 $n_b$ & 0& 1& 0& 1& 0& 1& 0& 1  \\ \hline 
 $g_1(z_b,n_b)$ & 12& 12& 8& 12& 8& 9& 5& 0 \\
\end{tabular}
\end{ruledtabular}
\end{table}

\section{Results}\label{results}

When presenting our results, we measure all
interaction energies in units of a positive energy $\epsilon$,
temperatures in units of $\epsilon/k_B$, and forces in units of
$\epsilon/a$, where $k_B$ is Boltzmann's constant and $a$ is the
lattice constant. We thus have 
$\beta_f=f/T$, $\beta_b = -\epsilon_b/T$, and $\beta_s = -\epsilon_s/T$, 
where $\epsilon_b<0$ and $\epsilon_s<0$ for
attractive bead-bead and bead-surface interactions, respectively.  
Unless otherwise stated, we evaluate our density-of-states results in
the canonical ensemble, where the fields 
$\beta_f$, $\beta_b$, and $\beta_s$ are constant and the extension 
and the contact numbers fluctuate. To ease notation, we omit the
angular brackets and write $z$ for $\langle z\rangle$, etc.
Similarly, when we perform a microcanonical evaluation of the density
of states $g(z,\beta_s,\beta_b)$, where the extension $z$ is
controlled and the tension field fluctuates, we omit the overbar and
write $\beta_f$ for $\bar{\beta}_f$. We state in the text and in the
figure captions when microcanonical evaluations have been performed.

\subsection{Force-extension relations for hard surface and athermal
  solvent conditions}\label{athermal} 

\begin{figure}[!htbp]
\includegraphics[width=3.4in]{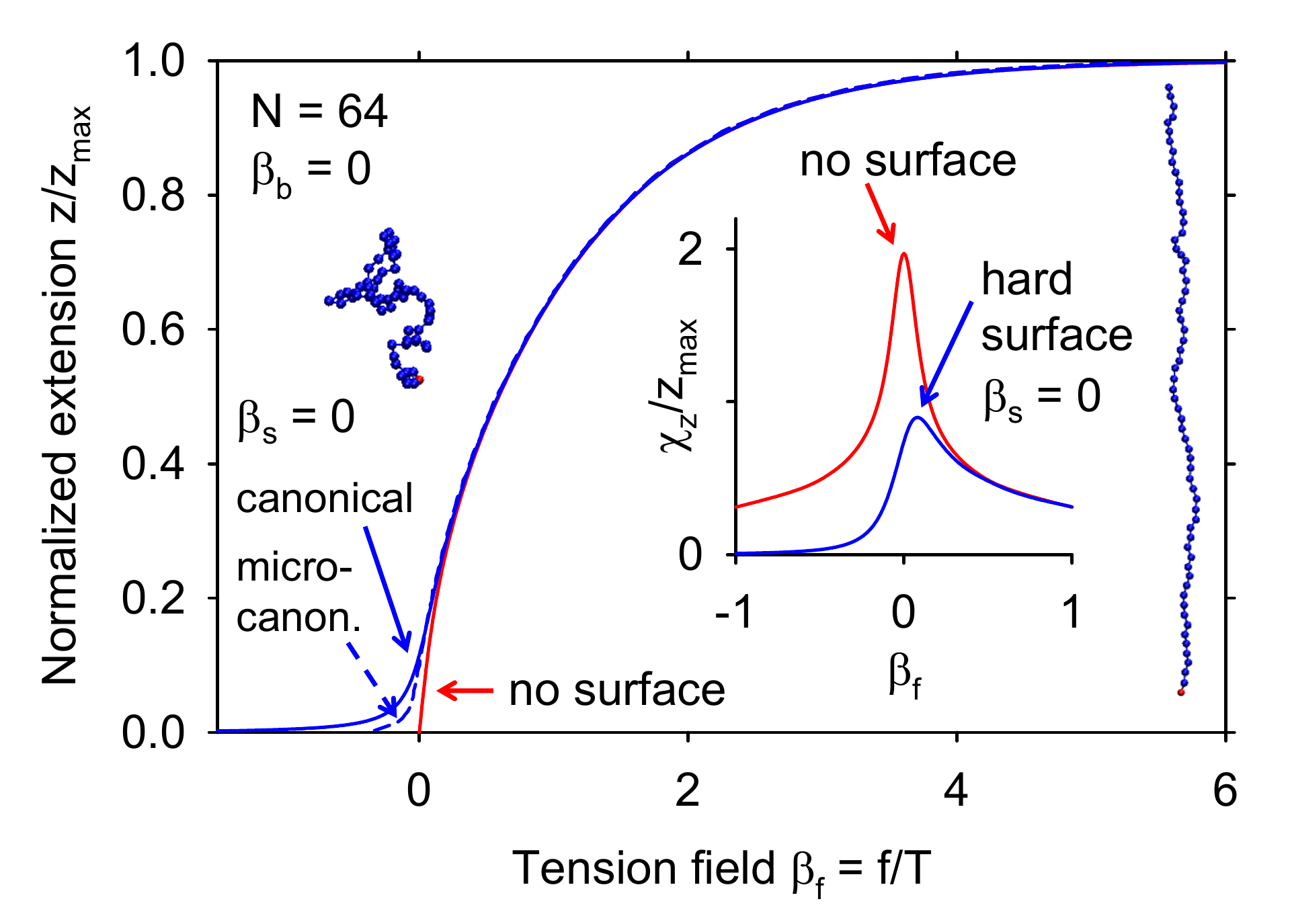}
\caption{Force-extension relations for chains of length
  $N=64$ tethered to a hard surface, $\beta_s=0$, or to a single point
  (no surface); both systems are athermal, $\beta_b=0$. 
  The solid lines represent results for the normalized  
  extension, $z/z_\mathrm{max}$, from the canonical 
  evaluation, Eq.~(\ref{zavgWL}), of the densities of states. 
  The dashed line represents the microcanonical result for the
  force-extension relation of the surface-tethered chain evaluated
  with Eq.~(\ref{betafmicro}). 
  The inset shows the normalized extension fluctuations,
  $\chi_z/z_\mathrm{max}$, for chains in the presence and absence of
  the hard surface.  
  Simulation snapshots of chain conformations at low and high tension
  force are 
  shown in the left and right part of the figure, respectively.
\label{microcan}}
\end{figure}

To investigate the mechanical response of a chain to an applied
tension force we determine the normalized extension,
$z/z_\mathrm{max}$, where $z$ is the average height of the last bead,
as a function of the tension field $\beta_f = f/T$.
In Fig.~\ref{microcan} we present force-extension curves for chains of
length $N=64$ with hard-core bead-bead interactions, $\beta_b = 0$,
that are tethered either to a hard surface  $\beta_s = 0$, or to a
single point (no surface).  The inset shows the normalized
fluctuations, $\chi_z/z_\mathrm{max}$, calculated with the aid of
Eq.~(\ref{chizWL}). 
In the absence of a surface,
the force-extension curve is antisymmetric with
respect to the origin, the magnitude of its slope decreases
monotonically with increasing force, and the curve has an inflection
point at the origin. In the presence 
of a hard surface, the extensions are always positive; the 
average extension is finite when $f=0$, decreases when the chain is
pressed against the surface by a negative applied force, and increases
when the chain is pulled by a positive force. The force extension
curve for the hard surface has an inflection point at a positive
applied force value and becomes indistinguishable from the curve for a  
point-tethered chain soon after.

\begin{figure}[!htbp]
\includegraphics[width=3.4in]{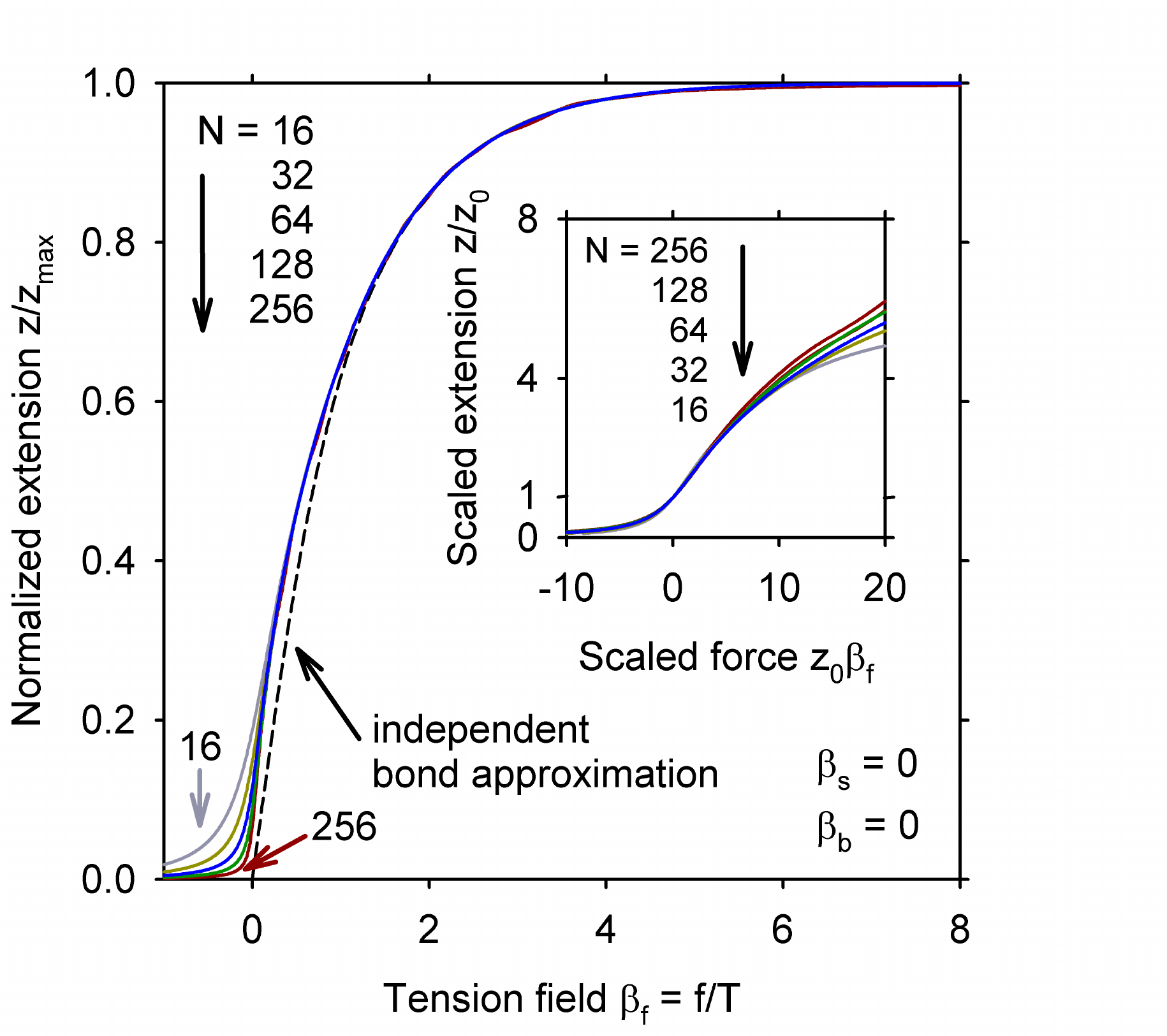}
\caption{Force-extension relation for contact fields $\beta_s =
  \beta_b = 0$ and chain lengths $N=16$, 32, 64, 128, and 256. 
The solid lines represent results for the normalized extension 
$z/z_\mathrm{max}$ from Eq.~(\ref{zavgWL}) as a
  function of the tension field $\beta_f$.
The dashed line is the independent bond approximation 
$\langle z_b\rangle$ of Eq.~(\ref{indep}). 
The inset shows the scaled extension $z/z_0$ as a function of the scaled
  tension force $z_0\beta_f$; the data are seen to collapse onto a
  single curve for small forces. 
\label{allNforce}}
\end{figure}

The fluctuations represent the slope of the force-extension
curves, $\chi_z = \partial \langle z\rangle/\partial \beta_f$, and are 
the inverse of the (differential) spring constant; the smaller $\chi_z$
the more force is required to increase the extension of the chain.
A linear regime, corresponding to Hookean springs, requires a constant
slope, which is approximately true for a narrow range of near-zero 
forces (see inset of Fig.~\ref{allNforce}). 
In the absence of an applied force, the chain conformations of a free
chain have, on average, spherical symmetry while those of a chain
tethered to a hard surface are elongated in the direction
perpendicular to the surface. This elongation is akin to a
pre-stretching of the chain and contributes to the higher spring
constant for small forces. Once the chains are sufficiently extended,
the effect of the hard surface disappears and the chains show the same
response to a further increase in the applied force.
Representative chain configurations of unstretched and nearly
fully stretched surface tethered chains are included in
Fig.~\ref{microcan}. 

The force-extension curves represented by solid lines in 
Fig.~\ref{microcan} have been calculated from a canonical evaluation
of the density of states with the aid of Eq.~(\ref{zavgWL}). 
The dashed line shows the microcanonical result for the
force-extension relation of the surface-tethered chain calculated
with Eq.~(\ref{betafmicro}) by taking a numerical derivative
(centered-difference approximation without smoothing) of the density
of states.  
The differences between the two evaluation methods
are most significant at the lowest forces and extensions. 
This is not surprising since the 
canonical and micro-canonical approaches are equivalent in the 
thermodynamic ($N\rightarrow\infty$) limit and the effects of chain
length on the force-extension curves are most significant in the low
tension regime, as Fig.~\ref{allNforce} shows.  

The chain-length dependence of force-extension relations is expected
to be different in different force regimes. 
Figure \ref{allNforce} shows force-extension relations for contact
fields $\beta_s = \beta_b = 0$ and different chain lengths. 
For high and intermediate tension fields, the normalized extension
$z/z_\mathrm{max}$ is seen to be independent of chain length, 
in agreement with theoretical predictions that the extension is
proportional to the contour length for sufficiently large applied
forces. 
For the highest extensions, the force extension relation is expected
to be dominated by the single-bond effects since there are
very few interactions between non-bonded monomers.
In Fig.~\ref{allNforce} we include normalized 
extensions calculated from the independent-bond (IB) approximation in
Eq.~(\ref{indep}) with $\beta_b = 0$.  
The agreement between simulation and IB results 
is excellent at the highest tensions, validating our simulation
method and showing that interactions between non-bonded beads are not
significant for tensions larger than $\beta_f \simeq 2$.

\begin{figure}[!htbp]
\includegraphics[width=3.0in]{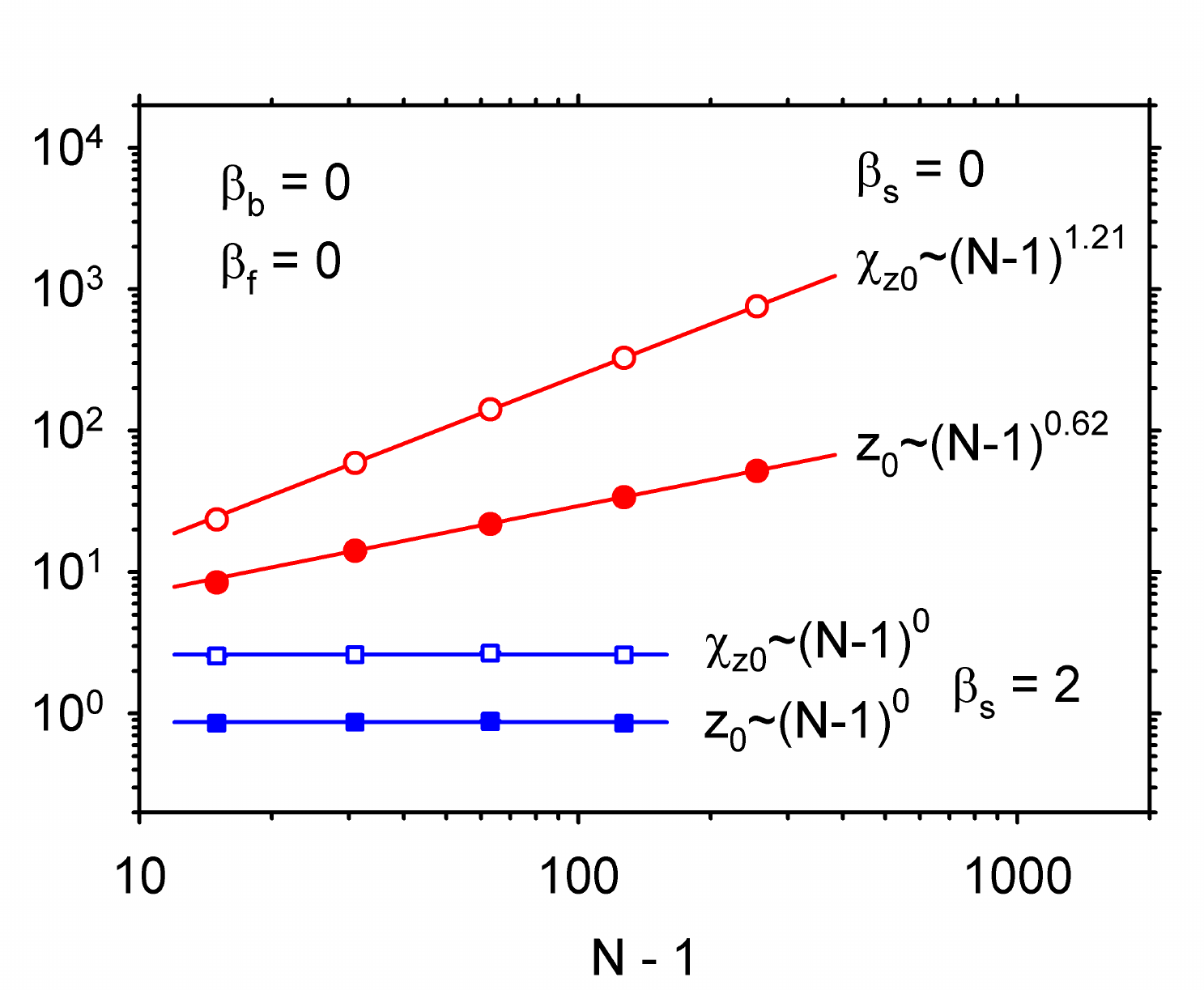}
\caption{Scaling with chain length of the zero-force 
 extension, $z_0$, and fluctuations, $\chi_{z0}$, for hard surface,
 $\beta_s=0$ (circles), and adsorbing surface, $\beta_s=2$ (squares),
 for athermal solvent conditions, $\beta_b = 0$. 
 The filled and open symbols represent simulation 
 results of $\chi_{z0}$ and $z_0$, respectively, for chain lengths
 $N=16$, 32, 64, 128, and 256 ($\beta_s=0$). 
 The solid lines represent power laws with the indicated exponents,
 for $\beta_s=0$ they were obtained from fits to the data for chain
 lengths $N>16$.  
\label{z0chiz0}}
\end{figure}

As the tension force decreases, the relative extensions
$z/z_\mathrm{max}$ for different chain lengths start to deviate from
each other. Due to the presence of the hard surface, the average
height at zero force has a finite value, $z_0$, which is expected to
scale as $z_0\sim N^\nu$, where
$\nu\simeq 0.6$ is the scaling exponent for good solvent conditions
\cite{gr94c}. From Hooke's law for tethered chains, 
$z-z_0 \sim z_0^2\beta_f$, 
we expect the slope of the force-extension curve at zero force
to scale as $N^{2\nu}$.  According to
Eq.~(\ref{chizWL}), this implies that the height fluctuations scale as 
$\chi_{z0} \sim N^{2\nu}$ at $\beta_f = 0$. 
In Fig.~\ref{z0chiz0} we present results for the height $z_0$ and
fluctuations $\chi_{z0}$ as a function of chain length, $(N-1)$. The 
double-logarithmic plot shows good agreement with the scaling
predictions given our limited chain lengths. A power-law fit of the
$z_0$ data for chain lengths  
larger than 16 yields $\nu = 0.62$, while the corresponding fit of
$\chi_{z0}$ yields $2\nu = 1.21$. 

As described in Eq.~(\ref{scalet}), zero-force extension, $z_0$, is the
scaling length for low forces.  
The inset of Fig.~\ref{allNforce} shows force extension data in scaled
representation, $z/z_0$ as a function of $z_0\beta_f$. 
While the linear regime where Hooke's law holds is very small, we find
the scaled force-extension graphs to collapse onto a single curve for
a sizable range of positive and negative tensions that includes the
inflection points of the force-extension curves. 
Since the fluctuation maxima occur in the low-force scaling region,
the peak fluctuations, $\chi_z^*$, are expected to scale in the same
way as the zero-force fluctuations, i.e.\ $\chi_{z}^*\sim z_0^2\sim
N^{2\nu}$. 
This is confirmed by the chain-length independence of the 
$\chi_z^*/z_0^2$ results for $\beta_b = 0$
presented in Fig.~S1 of the supplementary
material\cite{supp}. 
The location of the inflection point scales with the inverse of the 
zero-force extension, $\beta_f^* \sim z_0^{-1} \sim N^{-\nu}$.
Since $\beta_f^* \rightarrow 0$ as $N\rightarrow \infty$, 
the effect of the tethering surface disappears in the infinite
chain limit as already suggested by the result of Fig.~\ref{allNforce}.

\subsection{Effect of solvent quality}\label{solvent}

To investigate the effect of solvent quality, we start in the
intermediate force regime and apply our model to force-extension
experiments on biological molecules (Sec.~\ref{intermediate}. We then
focus on force-induced 
transitions from compact chain conformations (Sec.~\ref{poor}) before
constructing a phase portrait for finite chains in the
$\beta_f$--$\beta_b$ plane (Sec.~\ref{phase}). 

\subsubsection{Intermediate force regime -- Application to
  biomolecules}\label{intermediate} 

Force-extension measurements have become an important tool in
investigating conformational properties of polymers. Since many
experiments are carried out on chains tethered to non-adsorbing
surfaces we set the bead-surface interaction parameter to zero,
$\epsilon_s=0$. 
Since we discuss flexible chains in this work, our
results apply to molecules such as single-stranded DNA (ss-DNA) but not
to the very stiff double-stranded DNA (ds-DNA) or the 
PEG chains investigated by Dittmore {\it et al.}\cite{di11}. 

Saleh {\it et al.}\cite{sa09} investigated the scaling behavior of
single chains under tension by measuring 
force-extension curves of denatured single-stranded DNA
molecules in solvents of different salt concentrations, spanning the
range from good to poor solvent conditions. 
For intermediate tensions, Saleh {\it et al.}\cite{sa09} observed 
power law behavior for the extension as a function of tension, 
$R_z\sim f^\gamma$, with exponents $\gamma$ near the predicted value 
$\gamma=2/3$ for good to moderate solvent conditions (see
Eq.~(\ref{Rgood})).  
For very good solvents, the experiments yielded somewhat smaller
exponents, $\gamma\simeq 0.6$, while the transition to poor solvent
conditions resulted in a large increase of the exponents. 

\begin{figure}[!htbp]
\includegraphics[width=3.4in]{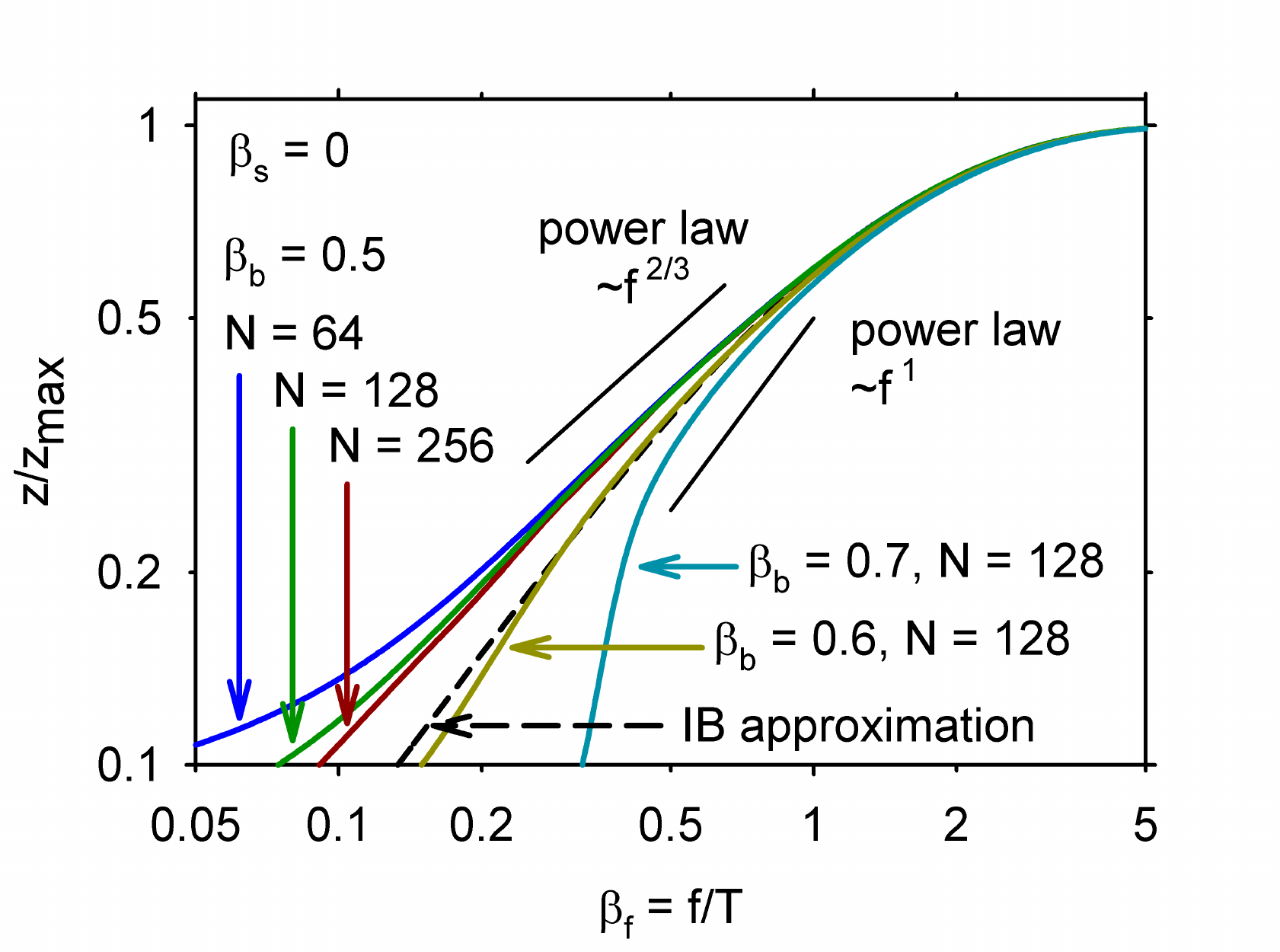}
\caption{Normalized extension $z/z_\mathrm{max}$ as a function of 
  tension field $\beta_f$ for a hard surface ($\epsilon_s = 0$) and
  three different solvent conditions.
 The solid lines represent simulation results for chains of
  length $N=64$, 128, and 256 for $\beta_b = 0.5$, 
  and $N = 128$ for $\beta_b = 0.6$, and $\beta_b = 0.7$ as indicated
  in the figure.      
  The dashed line represents the independent bond (IB) approximation, 
  Eq.~(\ref{indep}), for $\beta_b = 0.6$, which is barely
  distinguishable on this scale from IB results for $\beta_b = 0.5$
  and $\beta_b = 0.7$.  
  Line segments of slope $2/3$ and 1 illustrate predicted power
  laws.
\label{logforce}}
\end{figure}

\begin{figure}[!htbp]
\includegraphics[width=3.4in]{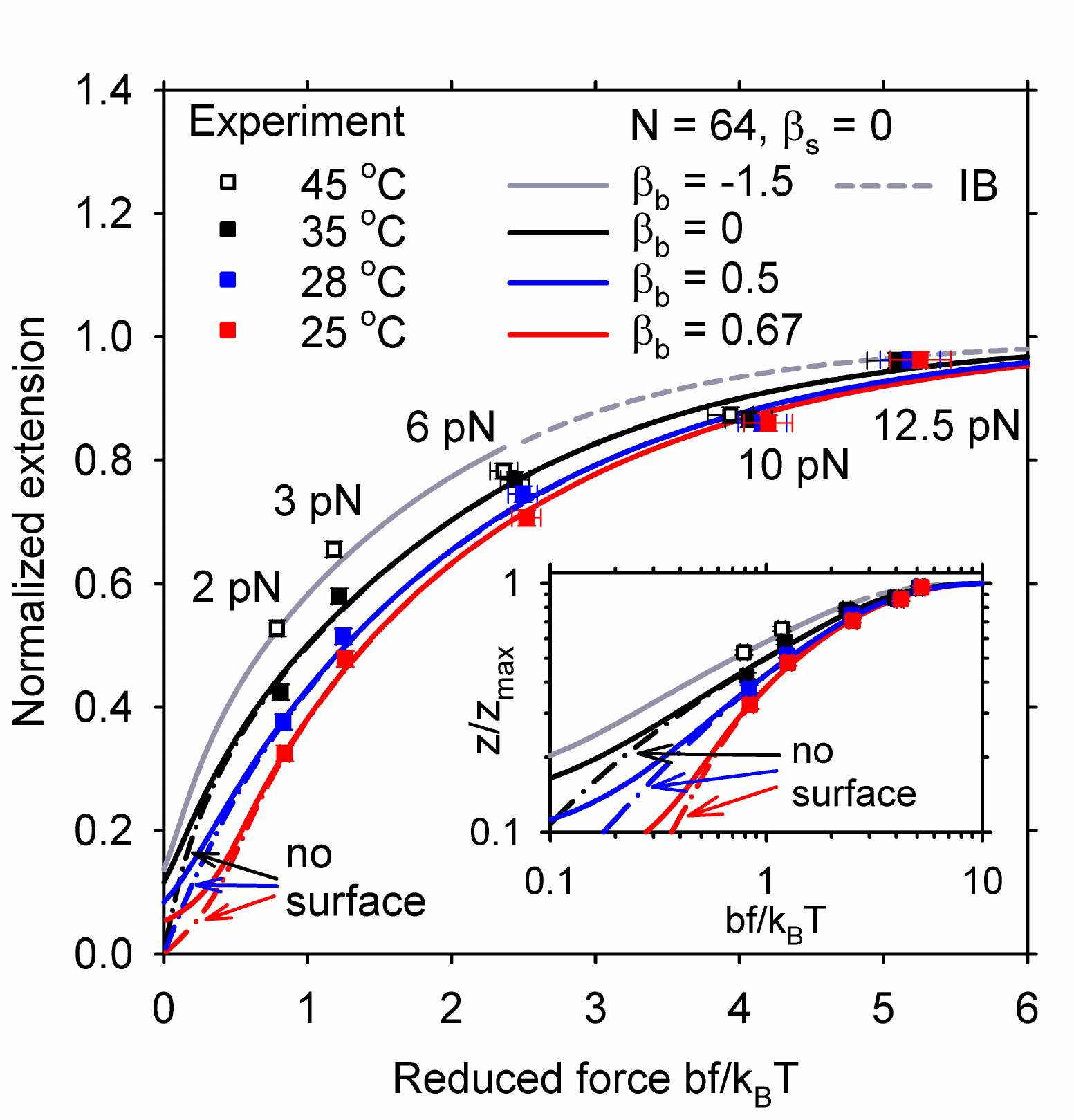}
\caption{Normalized extension as a function of reduced force
  $bf/k_BT$ in linear scale (main figure) and log-scale (inset). The
  symbols represent experimental data on ssDNA by Danilowicz {\it et
  al.}\protect\cite{da07} at 
  temperatures 25~$^0$C, 28~$^0$C, 35~$^0$C, and 45~$^0$C and forces
  between 2 and 12.5 pN. 
  The solid lines represent our results for $N=64$ at contact fields
  $\beta_b = -1.5$, 0, 0.5, and 0.67. 
  The force extension curves for 
 $\beta_b = 0$ and 0.5, and 0.67 are calculated from dedicated (1-d)
  simulations at these contact fields; for $\beta_b = -1.5$, the 
  force-extension curve evaluated from the 3-d density of states 
  is supplemented at high tension with results from the
  independent bond (IB) approximation.
  The segment lengths are $b=1.73$~nm and $b=1.7$
  lattice constants for experimental and simulation data,
  respectively. 
  The dash-dotted lines represent results for point-tethered chains
  (no surface). Note that graph for $\beta_b=0.67$ has an inflection
  point near $b\beta_f = 0.54$. 
\label{Danil_comp}}
\end{figure}

To explore the elastic response of chains in moderate and poor
solvents, we determined force-extension curves for three values 
of the bead-contact field $\beta_b$ near the coil-globule transition
($\beta_c=0.63$) of a chain of length $N=128$.  
In Fig.~\ref{logforce} we present these results in a log-log plot to
facilitate comparison with Fig.~1(a) of  Ref. [\onlinecite{sa09}]. 
The inset of Fig.~\ref{Danil_comp} also includes good-solvent data in 
log-log representation.
In qualitative agreement with experimental data \cite{de02d,sa09} 
we find that the effect of solvent quality increases with
decreasing tension force. For high tensions the independent bond (IB) 
approximation of Eq.~(\ref{indep}), indicated by the dashed line, 
describes the simulation data well. As the tension decreases,
interactions between different chain segments become important and the
IB approximation begins to fail.  
For $\beta_b = 0.5$, excluded volume interactions expand the chain and 
the IB approximation underestimates the extension.
Ideal (IB) behavior is observed over
the largest force range for $\beta_b = 0.6$, where the chain is 
closest to the coil-globule transition 
and excluded volume interactions are expected to be
screened by attractive interactions. The IB curve varies nearly
linearly with force for the lower tension range shown in
this graph.  
For poor solvent condition, $\beta_b$ = 0.7, the actual extension is
smaller than in the IB approximation since the strongly attractive
interactions between chain segments favor compact configurations. 
The steep part of the force-extension curve for $\beta_b=0.7$ 
belongs to a force-induced transition from the globule to the extended
chain conformations (see Sec.~\ref{poor}). 
This is followed at higher tension by the stretching of an extended
chain in poor solvent.  
In this regime, scaling arguments predict a linear dependence of the
extension on the force \cite{ha91c}. A comparison with
a line segment of unit slope in Fig.~\ref{logforce} shows
approximately linear behavior of our results in a narrow force range
following the transition. 

To illustrate the power law prediction of Eq.~(\ref{Rgood}) at
intermediate tensions, we have 
included a straight line segment of slope 2/3 in Fig.~\ref{logforce}.
An inspection of our results in Fig.~\ref{logforce} shows that the 
calculated curves straighten for intermediate tensions but never 
quite lose their curvature. 
While much larger chain lengths are required to observe power law
behavior \cite{hs12}, our results for different solvent conditions in
Fig.~\ref{logforce} and the inset of Fig.~\ref{Danil_comp} exhibit 
important properties of the intermediate force regime. 
A comparison of results for chain lengths 
$N=64$, 128, and 256 at $\beta_b = 0.5$ in Fig.~\ref{logforce} shows
that the chain length dependence of the normalized extension is most
pronounced at low tensions and disappears as the tension
increases. This agrees with our observations on chains in athermal
($\beta_b=0$) solvents and is related to the low-force scaling
behavior of the extension. 
To illustrate the effect of the hard tethering surface, we include
results for point-tethered chains (no surface) in the inset of 
Fig.~\ref{Danil_comp}. They show that the effect of the tethering
surface decreases rapidly with increasing tension even for relatively
short chains. 
Hence, in the intermediate force regime, neither the chain length nor
the presence of the tethering surface affect the relative extension of
the chain. 
In this regime, the solvent quality determines the mechanical response
of the chain with a decrease in solvent quality leading to a steeper
increase of the extension. 

The universal nature of force-extension relations in the intermediate
force regime encourages us to compare our simple, implicit solvent model
with experimental data on a complex biomolecule. 
Danilowicz and coworkers \cite{da07} investigated the effect of
temperature on the force-extension relation of single-stranded DNA
($\lambda$-phage ssDNA) in phosphate buffer saline solution by
performing magnetic tweezer experiments at several temperatures between 
25~$^0$C and 50~$^0$C. 
At the lower temperatures, base pairing leads to the formation of
hairpins, which reduce the size of the coil. As the temperature
increases, the number of hairpins decreases and the coil swells until
a temperature of about 40$^0$C, where no more hairpins are formed and
the chain dimensions are comparable to chemically denatured
single-stranded DNA (dssDNA). 
To compare simulation results for our model with experimental data 
we select four isotherms, 25~$^0$C, 28~$^0$C, 35~$^0$C,
and 45~$^0$C, from the data presented in Fig.~3 of
Danilowicz {\it et  al.}\cite{da07}. We reduce the measured extension
$x$ by the contour length $L$, $\hat{z} = x/L$,  and calculate the
dimensionless force $bf/k_B T$, where $b$ is a Kuhn 
segment length and $k_B$ is Boltzmann's constant. 
Values of $L = 23.5$~$\mu$m and $b = 1.73$~nm are obtained 
from a description of experimental data at 25~$^0$C in terms of an
extensible freely-jointed chain model \cite{sm96c}, which also yields 
$S = 300$~pN for the stretch modulus \cite{da07}. 
From our simulation data for chain length $N=64$, we calculate
force-extension curves for several values of the bead-contact field
$\beta_b$ corresponding to a range of solvent conditions. For this
chain length, the collapse transition occurs at about
$\beta_c=0.69$ and we choose $\beta_b=0.67$ as our highest
contact field (poorest solvent condition). 
To convert the tension field $\beta_f$ to the reduced force
$b\beta_f$, we consider the chain dimensions at $\beta_b=0.67$ and 
estimate an effective segment length of $b=1.7$ from
$b=R_e^2/(N-1)b_l$, where $b_l$ is the average bond length. 
As before, we reduce the extension $z$ by the maximum extension 
$z_\mathrm{max}=3(N-1)$.
In Fig.~\ref{Danil_comp} we present experimental and simulation 
force-extension data at four temperatures and $\beta_b$ values, which
were chosen to approximate the experimental data at 2 and 3 pN.
The inset shows the data in log-log presentation for easier comparison
with Fig.~\ref{logforce}. 
Three of the curves in Fig.~\ref{Danil_comp} represent
results from Wang-Landau simulations at fixed fields $\beta_s=0$ and 
$\beta_b = 0.67$, $\beta_b = 0.5$, and $\beta_b=0$, respectively. 
The result for $\beta_b=-1.5$, was calculated from the 3-d density of
states $g(n_s,n_b,z)$, which is reliable in a limited range of
tensions. At the highest tensions we therefore supplement the data
with results from the independent bond approximation.

Fig.~\ref{Danil_comp} shows that our model is able to describe 
experimental data for moderately high forces. 
The behavior of the biomolecule at the highest forces is not captured 
by our model. In this regime, the bonds of ssDNA molecules become
extensible\cite{sm96c}, while the bond-fluctuation model has a limited
bond length.  
The lowest extensions in Fig.~\ref{Danil_comp} correspond to the
lowest temperature, 25~$^0$C and highest $\beta_b$ value, $\beta_b=0.67$.
The relatively steep decline at low force shows the proximity to the
collapsed state and indicates that many bead-bead contacts are formed,
approximating the formation of hairpins in the ssDNA. 
Since $\beta_b=-\epsilon_b/k_BT$ and $\epsilon_b<0$ for net attractive
interactions, a decrease in $\beta_b$ corresponds to an increase in
temperature. The next isotherm in the physical system is at 
28~$^0$C, only about 1\% above the first in absolute temperature. 
The corresponding isotherm in the model, on the other hand, is at
$\beta_b=0.5$, about 25\% from the first. Part of the reason for the
larger change in the model temperature is the size of the
chain. The shorter a chain, the larger the transition region near the
collapse transition, which implies that the biological molecule
requires a much smaller change in temperature for an equivalent change
in solvent conditions. Taking this finite-size effect into account is
not sufficient to reach the highest isotherms; 
for the 35~$^0$C degree isotherm, for example, athermal contact
conditions ($\beta_b=0$) in the model barely match the experimental
data, while net repulsive bead-bead interactions $\beta_b<0$ are
required to reach even larger extensions. 
The biological molecule is a polyelectrolyte in a
complex aqueous solvent mixture \cite{co03}
where changes in temperature affect,
for example, ion concentration and hydrogen bonding rates, so that the
net segment-segment interactions vary with temperature. At the highest
temperature, 45~$^0$C, where the molecule is denatured, the
solvent-segment interactions are much more attractive than the 
segment-segment interactions leading to a swelling of the coil.
In an implicit solvent model, such as the bond-fluctuation model
employed in this work, solvent-solvent, bead-bead, and
bead-solvent interactions are described by a single, net interaction
parameter $\epsilon_b$, which has to be adjusted if a complex solvent
system is to be described in a temperature range where the solvent
quality changes rapidly.

\subsubsection{Force-induced transitions from globular and crystalline
  states}\label{poor}

For a collapsed polymer chain under tension, 
Halperin and Zhulina\cite{ha91c}  
predicted that the force induces a 
discontinuous transition between a deformed globule at low
tension and a stretched chain at high tension. 
At the transition force, which we call $f^*$, coexistence between
states leads to a plateau in the force as a function of extension (or
a vertical jump in the extension as a function of force). 
A number of simulation studies (see
e.g. Refs.~[\onlinecite{wi96c,gr02,fr02b}]) have confirmed the
first-order nature of the transition and identified chain conformations
with globular and string-like sections on the same molecule in the
transition region.
More recently, Walker and coworkers \cite{gu07b,li10} performed
single-chain pulling experiments on poly(styrene) (PS) in water, a poor
solvent, and a range of solvents of different quality. The measured  
force-extension curves of PS in water \cite{gu07b} show clearly a 
plateau in the force and confirmed experimentally the theoretical
prediction of a first-order transition. A careful study of the
transition force $f^*$ for PS in 
a range of solvent mixtures shows a linear dependence of $f^*$ on the
interfacial energy between the polymer and the solvent.\cite{li10}

\begin{figure}[!htbp]
\includegraphics[width=3.4in]{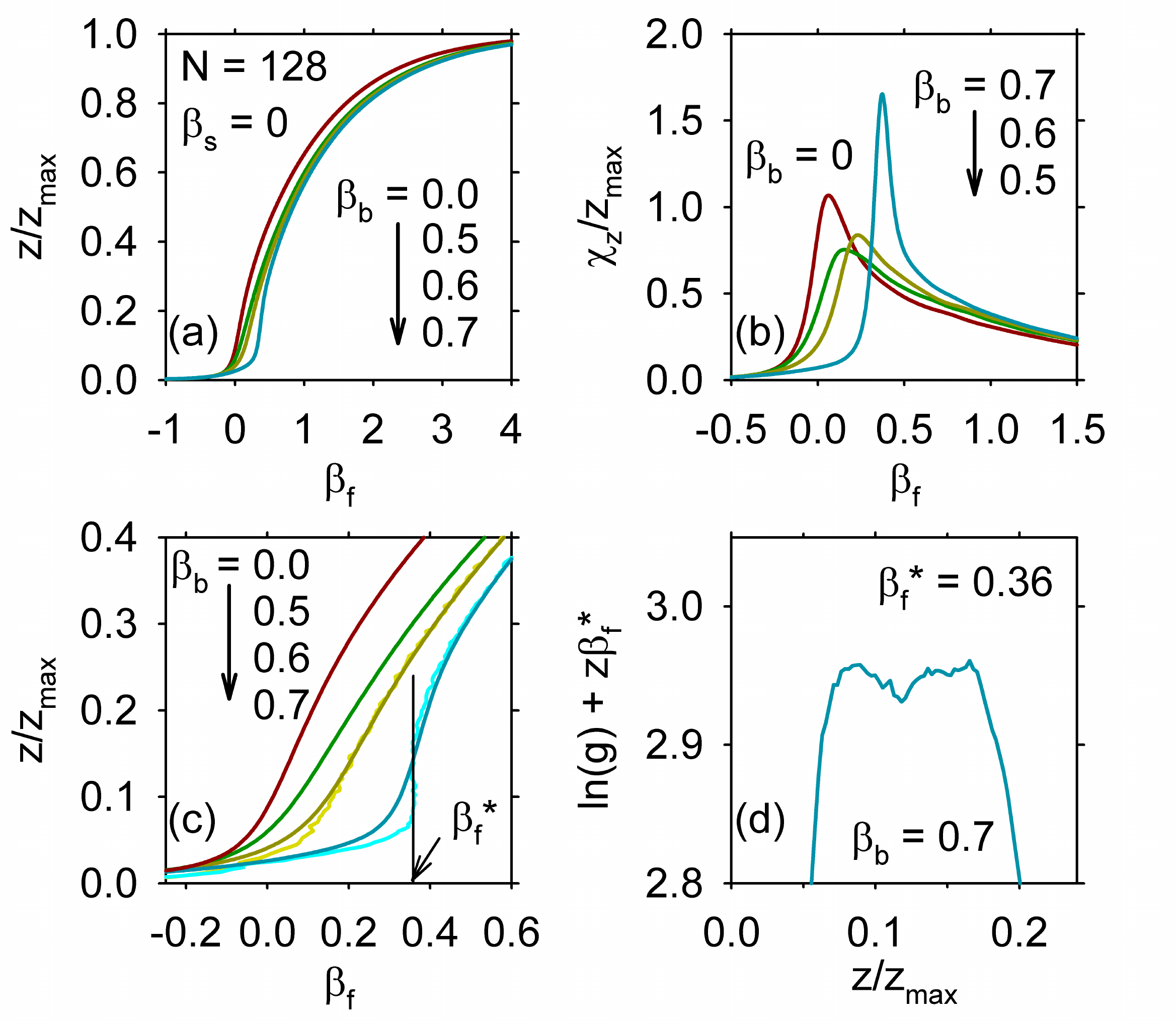}
\caption{Chain stretching in good and poor solvent. 
  (a) Force-extension curves, in linear scale, for chains of length
  $N=128$ tethered to a hard surface $\beta_s=0$. The bead
  contact fields represent athermal, $\beta_b=0$,
  moderate, $\beta_b=0.5$, near $\theta$,
  $\beta_b=0.6$, and poor solvent, $\beta_b=0.7$, conditions
  respectively. 
  (b) The normalized fluctuations of the extension $\chi_z/z_\mathrm{max}$, are
  equal to the slopes of the force extension curves in (a). 
  (c) Transition region of the force-extension curves in (a). 
  The bright (somewhat noisy) lines represent 
  results from a microcanonical evaluation of the densities of state 
  for $\beta_b = 0.6$ and 0.7. The vertical line segment indicates the 
  tension field at the transition, $\beta_f^*$, and highlights the
  vertical part of the microcanonical extension curve. (d)
  Evidence for a bimodal probability distribution for the 
  tension field $\beta_f^*=0.36$ at $\beta_b=0.7$; the solid line
  shows the reweighted density of states, $\ln(g)+z\beta_f^*$ as a
  function of normalized extension, $z/z_\mathrm{max}$. 
\label{N128goodpoor}}
\end{figure}

In Fig.~\ref{N128goodpoor} we present stretching results for chains 
of length $N=128$ in four solvent conditions ranging from athermal to
poor. Panels (a) and (c) show force-extension curves in linear scale
for a large and a restricted range of extensions, respectively.
Panel (b) shows graphs of the normalized fluctuations 
$\chi_z/z_\mathrm{max}$ which, according to Eq.~(\ref{chizWL}) 
represent the slopes of the force-extension curves in (a).
The fluctuation maxima indicate the inflection points in the
force-extension curves. With increasing $\beta_b$, the location of the
maximum, which we call $\beta_f^*$, moves to higher tensions while the 
height of the peak, $\chi_z^*/z_\mathrm{max}$, first decreases and then
increases as the solvent quality becomes poorer. 

For the poor-solvent case, $\beta_b=0.7$, our results agree 
with the predictions of Halperin and Zhulina\cite{ha91c}: 
The force-extension curve in panel (c) shows an extended linear region at
low tension corresponding to the deformation of
the globule, followed by a steep increase as the chain transitions
from globule to extended conformations. Beyond the transition, 
the extended chain is stretched further and differences between chains
in different solvents gradually diminish. 
For $\beta_b=0.7$, the extension fluctuations in panel (b) show a tall
and narrow peak at $\beta_f^*\simeq 0.37$, which is consistent with a
discontinuous transition of a finite chain.
To investigate the transition further, we include results from a
microcanonical evaluation of $g(z;\beta_s,\beta_b)$ 
for $\beta_b = 0.6$ and $\beta_b=0.7$ in panel (c) (bright lines).   
For finite chain lengths, canonical and microcanonical results are not
identical. 
In the canonical evaluation of $g(z;\beta_s,\beta_b)$ with
Eq.~(\ref{zavgWL}),  
the summation over all states (weighted by the appropriate Boltzmann
factors) leads to a smoothing of calculated extensions, which may
obscure localized features, such as the vertical rise in extension 
expected  near a discontinuous transition. 
In the microcanonical evaluation with Eq.~(\ref{betafmicro}),  
the numerical derivative of $g$ that yields $\beta_f(z)$ involves only
the two neighboring values of $g$ at $z \pm 1$; the results are
therefore highly local but also somewhat noisy. 
For $\beta_b=0.6$, neither
evaluation method suggests a discontinuous transition. For
$\beta_b=0.7$, however, the microcanonical force extension curve shows
a nearly vertical rise at $\beta_{f}^* = 0.36$, suggesting a
discontinuous transition. 
For a finite-size system, a first-order
transition is accompanied by changes in curvature of the density of
states, which we describe in more detail below. For the microcanonical
force extension curve, this leads to an ``S''-shaped rather than 
vertical line in the coexistence region, which is barely visible
through the noise in the force-extension curve presented in panel (c).  
The coexistence of states at the transition field is reflected in the
probability distribution for the extensions, which
according to Eq.~(\ref{prob1d}), is given by $\ln(p) = \ln(g) +
z\beta_{f}^* - Q$, where $Q$ is the normalization constant. 
The results for $\ln(g)+z\beta_{f}^*$ presented in panel (d) show the 
bimodal nature of the probability distribution a the transition field
$\beta_{f}^*$. 

\begin{figure}[!htbp]
\includegraphics[width=3.4in]{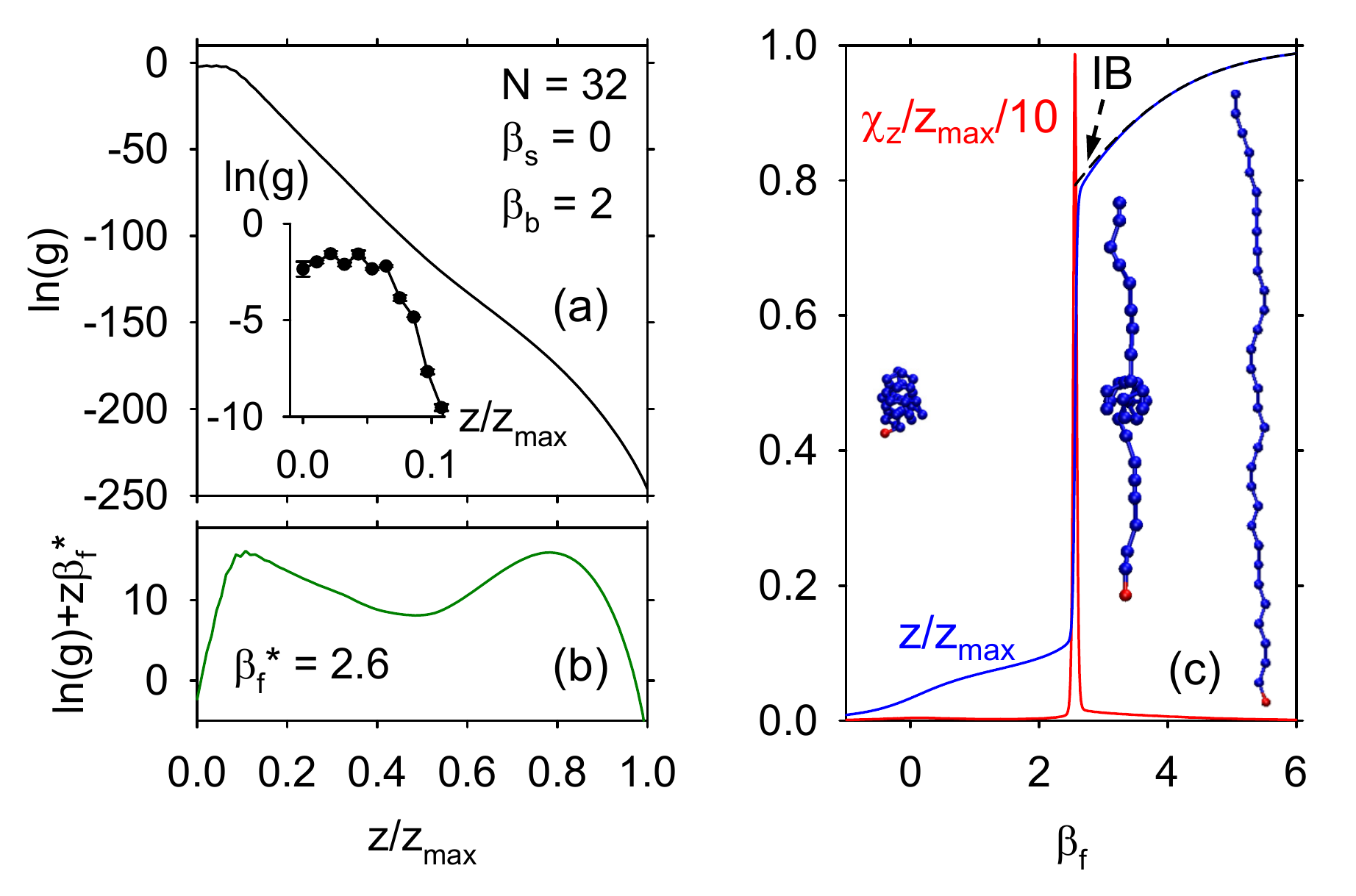}
\caption{Chain stretching from the crystalline state. 
  (a) Logarithmic density of states $\ln(g(z))$ as a function of the
  normalized extension $z/z_\mathrm{max}$ (solid line) for a chain of
  length $N=32$ with crystalline order, $\beta_b = 2.0$. The inset
  shows the low extension region, where the crystalline order leads to
  discrete jumps in the density of states; the symbols connected by
  lines represent the $\ln(g(z))$ values. (b) Logarithmic density
  of states reweighted with the transition tension $\beta_f^* =
  2.6$. (c) Normalized extension $z/z_\mathrm{max}$ and 
  fluctuations $\chi_z$ as a function of the
  tension field $\beta_f$. The blue solid line represents
  $z/z_\mathrm{max}$ values obtained from a canonical evaluation of
  the density of states; the dashed line at high tension shows the
  independent bond (IB) approximation for $\beta_b = 2$. The red solid
  line shows fluctuation values that have been divided by
  $10z_\mathrm{max}$ to fit into the figure; the fluctuation have a tall
  and very narrow peak at the transition field $\beta_f^*$.
  The simulation snapshots show chain conformations corresponding
  to the crystalline state before the transition (left), a transition
  state with coexisting crystallite and stretched chain segments
  (center), and a stretched chain (right).  
\label{N32crystal}
}
\end{figure}

The first order nature of the transition becomes more pronounced as
the solvent condition worsens and the chains crystallize. 
We would like to stress that the crystal structures found in our model
reflect the symmetry of the underlying simple cubic lattice and have 
nothing to do with the crystal structure of biological and synthetic
polymers\cite{lu08}. Similarly, the transition to the crystalline state in our 
model involves the spatial rearrangement of flexible chain segments
and is quite different from the crystallization transition observed in 
synthetic polymers, where sections of the chain stiffen and 
fold back on themselves to form crystallites.
In Fig.~\ref{N32crystal} we show results for a chain of length $N=32$
that is pulled out of the crystalline phase. The density of states
for $\beta_b=2$ presented in panel (a) and its inset has two
characteristic features. First, the values of $\ln(g)$
show discrete jumps up and down at low extensions due to the
crystalline order of the chain. Second, the curvature of the
$\ln(g)$ graph changes from concave to convex and back again as the
extension increases. As is well known, 
such a ``convex intruder'' indicates the presence
of a discontinuous transition in a finite-size system. Upon reweighting
with the transition field $\beta_{f}^* = 2.6$ we obtain
the bimodal probability distribution shown in panel (b). Note that the
maximum in the probability distribution at low extension shows the
discrete jumps characteristic for crystalline states. 
Panel (c) shows extensions $z/z_\mathrm{max}$ and fluctuations
$\chi_z$ as a function of the tension field $\beta_f$. The
fluctuations are very small in the crystalline state and have a 
tall and narrow peak at the transition field. 
At the lower extensions, the response to the applied force
is almost entirely due to reorientation of the crystal without
significant loss of bead-bead contacts (of all the available
crystalline chain conformations, those that are oriented with their
long axis perpendicular to the surface become increasingly probable
with increasing tension force). Since the bead-bead interactions are
highly attractive, high tensions field, larger than about
$\beta_f = 2.0$, are required to induce conformational
changes (defects) that lead to a break-up of the compact crystal at
the transition field $\beta_{f}^* = 2.6$. 
The transition states are 
chain conformations with stretched chain segments attached to compact
crystallites; an example is shown in Fig.~\ref{N32crystal}. 
The force-extension curve beyond the stretching transition is well
described by the independent bond approximation, shown as a dashed
line at high tension, indicating that essentially all contacts 
between non-bonded beads are lost in the stretching transition.

\subsubsection{Phase portrait for finite chains in solvent}\label{phase}

\begin{figure}[!htbp]
\includegraphics[width=3.4in]{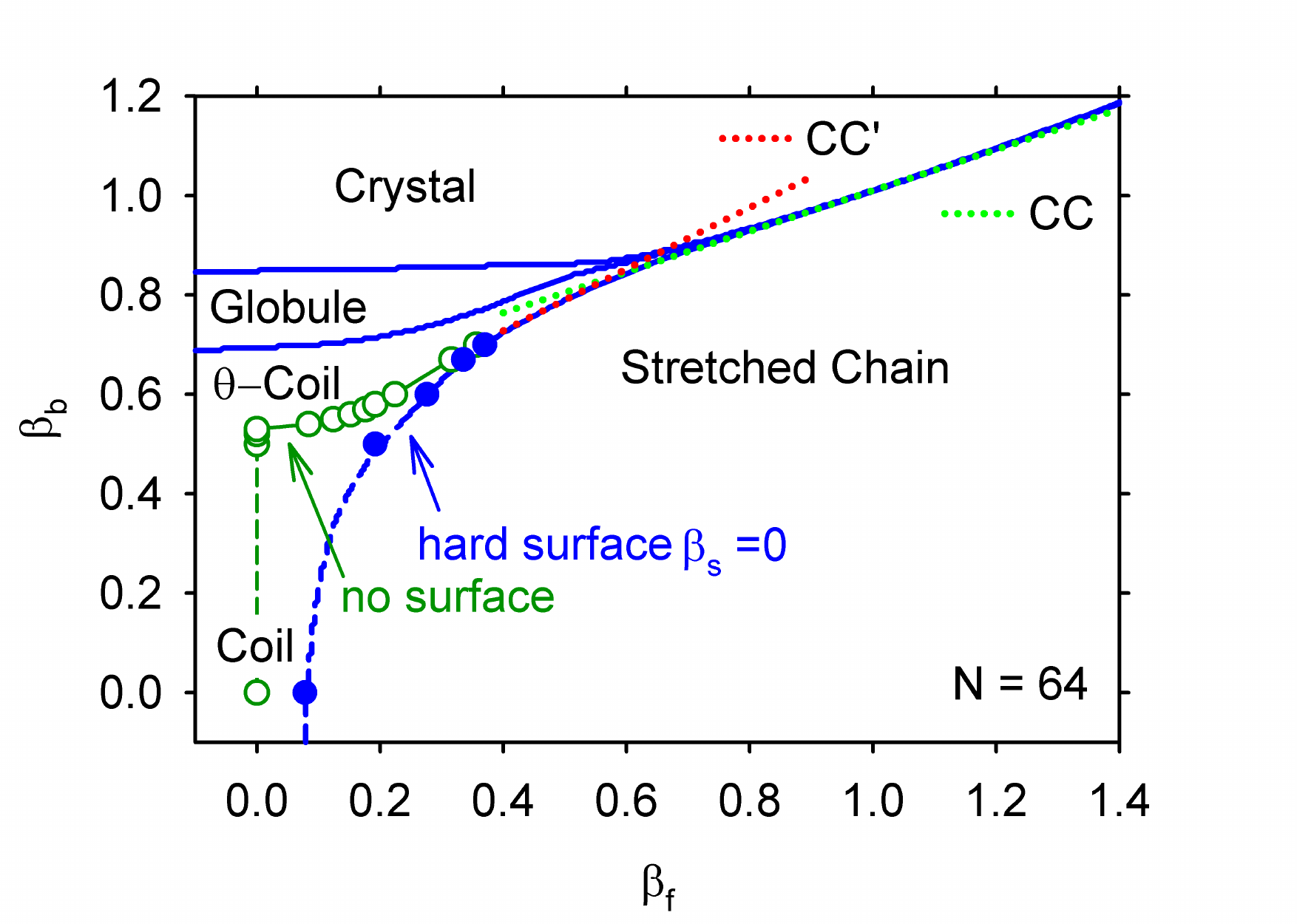}
\caption{Phase portrait for tethered chains of length $N=64$ in the
  space of tension field, $\beta_f$, and bead-contact field,
  $\beta_b$. 
  Crystal and globule are ordered phases at high bead-contact fields,
  i.e. in poor solvent. In the disordered phase we indicate regions of
  stretched, $\theta$-coil, and expanded coil conformations. 
  The solid and dashed blue lines show the location of
  fluctuation maxima determined from an evaluation of the 3-d
  density of states $g(n_s,n_b,z)$ for chains tethered to a hard 
  surface, $\beta_s=0$. The dashed line show the location of
  inflection points of the force extension curves at low $\beta_b$,
  which are not associated with a phase transition. 
  The filled symbols represent the location of inflection points
  ($\chi_z$ maxima) obtained from 1-d Wang-Landau simulations at fixed 
  $\beta_b$ and $\beta_s=0$. The open symbols connected by green lines
  show inflection point locations from simulations in the absence
  of a hard surface. 
  The dotted lines labeled CC and CC' are straight lines with slopes
  of 0.41 and 0.62, respectively, estimated from generalized 
  Clausius-Clapeyron relations. 
\label{N64_bs0phase}}
\end{figure}

To provide a comprehensive description of 
the effect of solvent quality on chain stretching and 
the effects of tension on chain collapse
we construct a phase portrait in the $\beta_f$--$\beta_b$ plane for
chains of length $N=64$ tethered to a hard surface, $\beta_s=0$. 
In Fig.~\ref{N64_bs0phase} we show locations of
fluctuation maxima separating regions of different conformations of
the chain. 
In the absence of tension, the good solvent region at small $\beta_b$
is separated from the poor solvent region at high $\beta_b$ 
by a region where the solvent is near $\theta$ conditions. 
The $\theta$ point for free chains of this model, estimated by
extrapolating coil-globule transition temperatures to infinite chain
lengths, is about $\beta_\theta=0.46$ \cite{ra05b}. 
The finite chains considered here enter a $\theta$ region
around $\beta_b=0.5$; we find, for example, that 
the chain dimensions of surface tethered chains 
scale approximately as $N^{1/2}$ at $\beta_b=0.5$. 
The $\theta$-region ends with the collapse transition, which occurs at 
$\beta_c = 0.69$ for $N=64$. 

\noindent
{\bf Good and $\theta$ solvent, $\mathbf{\beta_b \lesssim 0.69}$}

The dashed blue line and filled symbols in Fig.~\ref{N64_bs0phase} indicate
$\chi_z$ maxima for chains tethered to a hard surface 
obtained from an evaluation of the density of states 
$g(n_s,n_b,z)$ and from 1-d Wang-Landau simulations, respectively. The
dashed green line and open symbols show the corresponding results in
the absence of the hard surface. As discussed in Sec.~\ref{athermal},
the inflection points occur at zero force for free (no surface) chains
and at positive forces for surface-tethered chains. They are shown as dashed
lines in the diagram since they do not represent phase transitions,
even in the infinite chain limit.

It is interesting to compare free and surface tethered chains
under tension as the solvent quality decreases.  
For small increases in $\beta_b$, the inflection
points of the free chains remain at $\beta_f=0$, while those 
for tethered chains move to slightly higher tension fields
$\beta_f$. 
With increasing $\beta_b$, the peak heights $\chi_{z}^*$ for
tethered chains decrease until they reach a minimum 
when the chains enter the $\theta$-solvent region 
(this is seen in Fig.~\ref{N128goodpoor} for $N=128$);  
for chain length $N=64$, the minimum occurs at $\beta_b=0.53$. 
At $\beta_b=0.53$, the fluctuations $\chi_z(\beta_f)$ of
free chains show a very broad maximum at $\beta_f=0$, which
splits into a minimum at $\beta_f=0$ and two symmetric maxima at
finite $\beta_f$ values as $\beta_b$ increases further. The inflection
point corresponding to the 
positive tension maximum is just visible in the point-tethered
force-extension curve for $\beta_b=0.67$ included in
Fig.~\ref{Danil_comp}. In Fig.~\ref{N64_bs0phase} we show the positive
branch of peak locations for free chains.  
As the solvent conditions change from $\theta$ solvent to poor
solvent, the peak locations of the free chains approach those of
the tethered chains. 

The scaling with chain length 
of the scaled fluctuation maxima, $\chi_z^*/z_0^2$, where $z_0$
is the zero-force extension, 
depends on the solvent conditions and is discussed in 
the supplemental materials\cite{supp}. The results presented in
Fig.~S1 suggest that chain stretching acquires the character
of a phase transition as the solvent quality decreases. 
The transition from $\theta$--region coils to stretched chains appears
to be continuous. After a small initial deformation of the coil at low 
forces, the extension and number of bead contacts change rapidly
through the transition region, before the extended chain is stretched
further at higher forces.
Since the $\theta$--region decreases in size with increasing chain
length this kind of transition is not expected to exist in the limit
of infinite chain length. 

\noindent
{\bf Poor solvent -- coil-globule and freezing
transition under tension}

\begin{figure}[!htbp]
\includegraphics[width=3.4in]{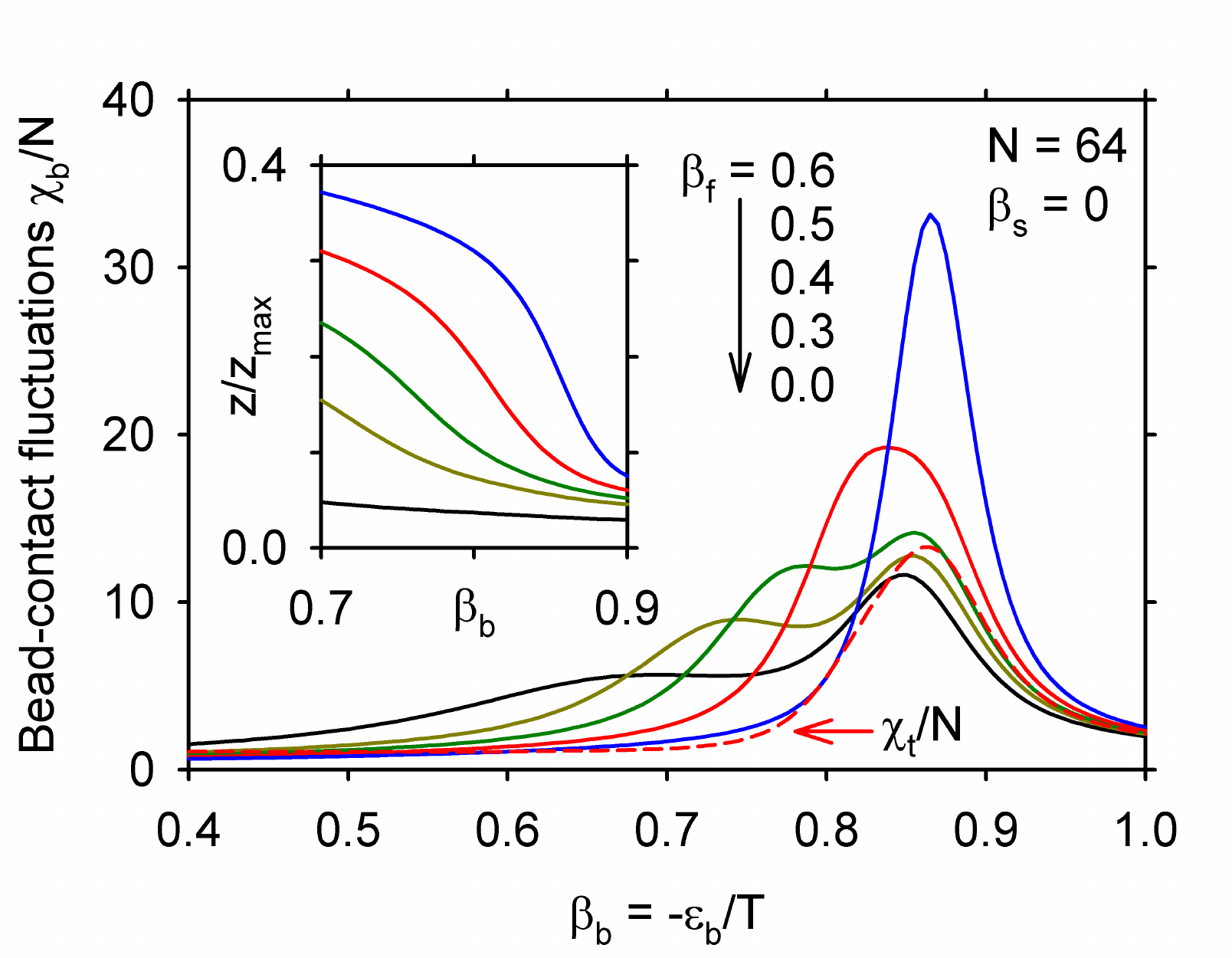}
\caption{Chain collapse under tension. The solid lines represent  
normalized fluctuations in the number of bead-contacts, $\chi_b/N$,
as a function of bead-contact field, $\beta_b$, for
applied tension fields $\beta_f = 0.6$, 0.5, 0.4, 0.3, and 0.0. 
The dashed red line represents the combined fluctuations, $\chi_t$ defined
in Eq.~(\ref{chit}), for $\beta_f=0.5$. 
The inset shows the corresponding normalized extension
$z/z_\mathrm{max}$ in the range of $\beta_b$ values, where the tension
has the largest effect on the collapse transition.
\label{N64ftavg}}
\end{figure}

In poor solvent conditions, we find two types
of compact states: the globule, a high-density, amorphous state and 
the crystal, characterized by order of the chain segments\cite{lu08}. 
To construct the lines in the phase portrait separating the 
regions of coil, globule, and crystal conformations 
we consider chains under fixed tension field as the solvent
quality changes.
In Fig.~\ref{N64ftavg} we present 
results for the fluctuations in the number of bead contacts $\chi_b$
as a function of $\beta_b$ at constant tension field $\beta_f = f/T$
for chains of length $N=64$. 
The inset of Fig.~\ref{N64ftavg} shows normalized extensions values,
$z/z_\mathrm{max}$, in the transition region. 
In the absence of tension (black solid line),
the fluctuations show two peaks, one at $\beta_c = 0.69$ indicating
the coil-globule transition and another at $\beta_X=0.85$
corresponding to the crystallization transition. When a tension field
of $\beta_f = 0.3$ is applied, the coil-globule peak moves
to a higher $\beta_b$ value and sharpens, while
the tension has little effect on the crystallization transition. 
The graphs in the inset show a strong decrease of the extension 
during the coil-globule transition while only small changes in height
are associated with the crystallization transition.
As the tension increases further, the coil-globule transition peak
continues to move closer to the crystallization peak until the peaks 
overlap completely; from then on a single peak continues to
grow in height and move slowly to higher $\beta_b$ values. 
At the highest tension, the transition is between a highly stretched
strand and a crystalline phase, as we saw in the example for
$N=32$. For intermediate tensions, around $\beta_f = 0.5$, 
collapse and crystallization are distinct even though the contact
fluctuations show one broad maximum. 
A good estimate for the crystallization transition may be obtained
from the combined fluctuations $\chi_t$, defined in Eq.~(\ref{chit}),
and shown as a dashed line for $\beta_f=0.5$ in Fig.~\ref{N64ftavg}. 
The quantity $\chi_t$ represents fluctuations of the sum $(z+n_b)$.
Near the coil-globule transition, the extension decreases
rapidly as the number of bead contact grows (see inset of
Fig.~\ref{N64ftavg}). This makes $\chi_t$ smaller than $\chi_b$ and
more sensitive to fluctuations due to local rearrangements of chain
segments, which characterize the crystallization transition. 
In the phase portrait, Fig.~\ref{N64_bs0phase},  we show as solid blue
lines locations of $\chi_b$ maxima separating coil and globule regions
and $\chi_t$ maxima separating globule and crystal regions.
Fig.~\ref{N64_bs0phase} shows a narrow $\theta$--coil region between
the globule and stretched chain regions, which is a finite-size
effect. The lines bordering 
this region represent maxima of different fluctuations: bead-contact
fluctuations for the boundary of the globule region and extension
fluctuations for 
the stretched chain region. For finite systems, it is not uncommon
that different quantities show transitions at somewhat different
fields and that the differences decrease with system
size. \cite{fe91c}  
We find for our model, too, that the size of the intervening
region decreases markedly from $N=32$ over $N=64$ to $N=128$.

\noindent
{\bf Poor solvent -- transitions 
  to stretched chain conformations}

In the infinite chain limit, transitions from crystalline 
to amorphous states are always discontinuous.  
For the chains of length $N=64$ considered in Fig.~\ref{N64_bs0phase},
we observe two-phase coexistence, characteristic of first-order
transitions, 
between crystal and globule and between crystal and stretched chains. 

The line separating crystal and stretched chains
increases nearly linearly with increasing $\beta_f$ in the tension
range considered here. For first-order transitions, the slope of the
transition line in a diagram of field variables may be estimated from
a Clausius-Clapeyron (CC) type equation.
For chains tethered to a hard surface it takes the form
$d\beta_b/d\beta_f = -\Delta z/\Delta n_b$, where $\Delta z$ and
$\Delta n_b$ are the differences in extension and number of bead 
contacts of the coexisting states, respectively. 
Simulation results for crystalline and stretched states for
$N=64$, $\beta_f=1.0$, $\beta_b\simeq 1.0$ yield a slope of
0.41 for the coexistence line at $\beta_f = 1.0$. A straight line
segment of slope 0.41 is superimposed as a dotted green
line on the coexistence curve in Fig.~\ref{N64_bs0phase} and seen to
describe the transition line well up to almost the highest tension
fields in the figure. 
For higher tensions, the slope of the coexistence curve
increases gradually. In the limit as both $\beta_b\rightarrow\infty$
and $\beta_f\rightarrow\infty$, the coexisting states are a stretched
chain of independent bonds and a crystal that has the maximum number
of contacts for the given chain length.
For $N=64$, $n_{b,\mathrm{max}}/N = 4.5$, predicting a limiting value
of 0.57 for the slope. For $N\rightarrow\infty$, 
the crystal structure yields $n_{b,\mathrm{max}}/N = 6$ and 
a limiting slope of 0.4.
In general, the CC analysis predicts a decreasing slope of the
crystal-stretched chain coexistence curve with increasing $N$, which we
observe for the chain lengths considered in this work.
For $N=32$, where we also have 1-d Wang-Landau results
for crystalline chains, we find good agreement between the
high-field CC prediction and the slope of the coexistence curve. 

The line of first-order transitions separating crystalline and
globular states in Fig.~\ref{N64_bs0phase} is almost independent of
the applied tension field. In 
contrast, the coil-globule transition occurs at increasing $\beta_b$
values as $\beta_f$ increases, thus reducing the $\beta_b$ range of
the globule until the globular phase disappears at high tension.
For long chains, the force-induced transition is discontinuous with
two-state coexistence, as we have seen, for example, in
Fig.~\ref{N128goodpoor} for $N=128$ at $\beta_b=0.7$. 
For shorter chains and at $\beta_b$--values closer to the coil-globule
transition, the fluctuations in the system are too large for phase
coexistence to occur and the stretching transition becomes
continuous. In this 
case, we may estimate the slope of the transition line from a modified
Clausius-Clapeyron equation, which reads $d\beta_b/d\beta_f =
-\chi_z/\chi_{zb}$. Evaluating  $\chi_z$ and $\chi_{zb}$ for $N=64$ at
$\beta_f=0.53$ and $\beta_b=0.81$, we find a slope value of 0.62. The
line-segment of slope 0.62  included in Fig.~\ref{N64_bs0phase} is 
a good approximation to the stretching transition line in the
globule-range of $\beta_b$ values. 
A linear relationship between $\beta_f$ and $\beta_b$ at the
stretching transition is consistent with the experimental observation
by Li {\it et al.}\cite{li10} that the force plateau depends linearly
on the interfacial energy between polymer and solvent.

\subsection{Effect of surface attraction}\label{adsorption}  

The presence of an attractive surface affects the mechanical response
of a tethered chain to a force perpendicular to the
surface. To study the effects of surface attraction on 
force-extension relations and to investigate the effect of tension on
the adsorption transition, we focus on athermal solvent conditions and
keep the bead-contact field at zero, $\beta_b=0$.
An analysis of the density of states 
shows that the adsorption transition, identified from a maximum in
$\chi_s$ (Eq.~\ref{chis}), occurs 
near $\beta_{sa} = 1.18$ for $N=64$ and $\beta_{sa} = 1.14$ for
$N=128$ (see Fig.~\ref{N128_bb0_bs0} (b)); as $N\rightarrow\infty$,
$\beta_{sa}\rightarrow 0.98\pm 0.03$\cite{de04,de08}.
For $\beta_s<0.8$, we find the force extension relations 
to be very similar to the hard surface case (see
Fig.~\ref{allNforce}). We thus refer to 
surface contact fields with $\beta_s<0.8$ as slightly attractive and
investigate larger $\beta_s$ values in detail.

\begin{figure}[!htbp]
\includegraphics[width=3.4in]{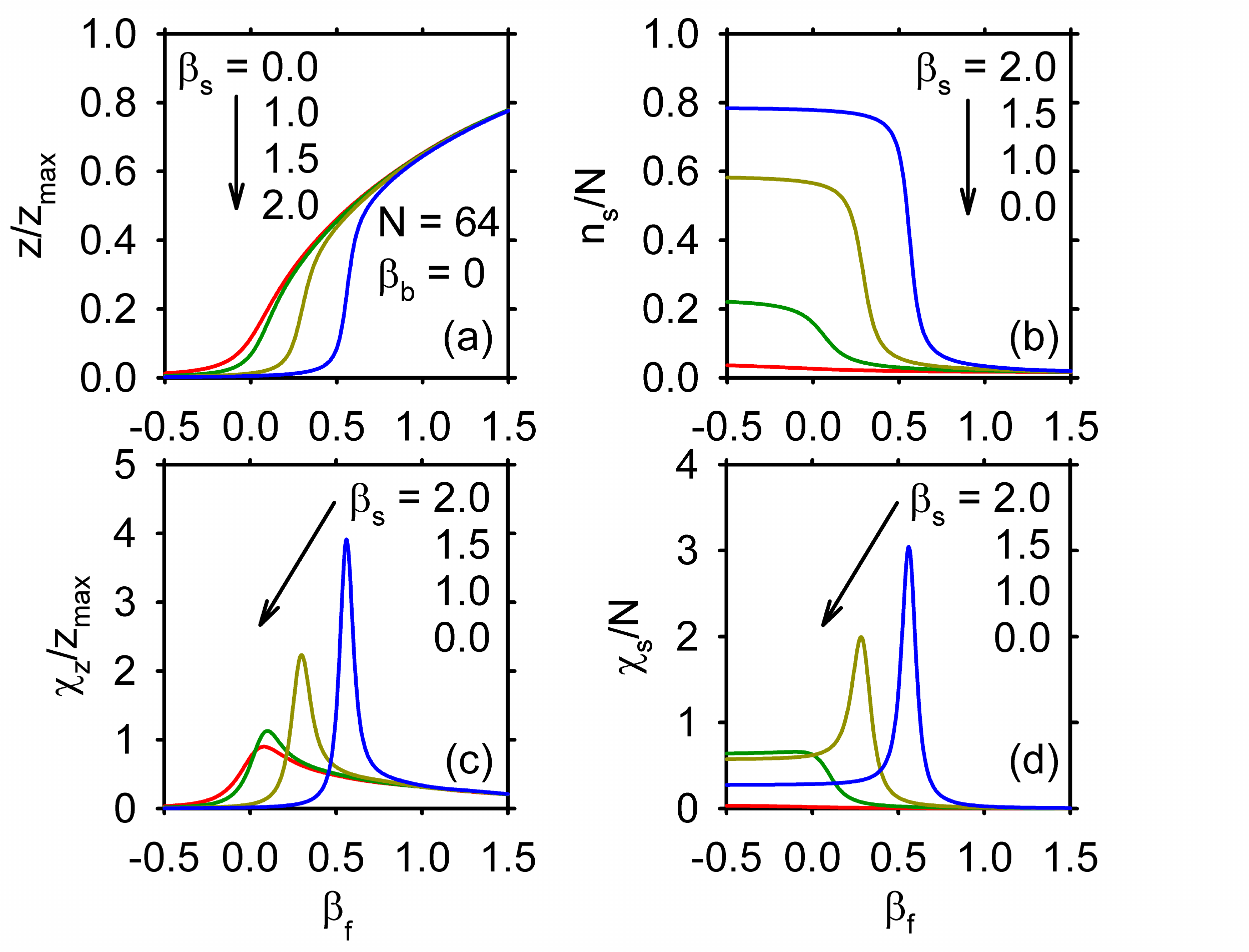}
\caption{Effect of surface attraction on tethered chains under tension
  for chain length $N=64$ in athermal solvent conditions,
  $\beta_b=0$. (a) Normalized extension $z/z_\mathrm{max}$ as a
  function of tension field, $\beta_f$, for surfaces with increasingly
  attractive interactions, $\beta_s = 0$, 1.0, 1.5, and 2.0. 
  (b) Normalized surface contacts $n_s/N$, (c) normalized 
  extension fluctuations $\chi_z/z_\mathrm{max}$, and (d) 
  normalized surface-contact fluctuations  $\chi_s/N$ 
  as a function of $\beta_f$ for 
  the same surface-contact fields $\beta_s$. 
\label{N64bb0_manybs}}
\end{figure}

In Fig.~\ref{N64bb0_manybs} (a) we present force-extension curves for
chains of length $N=64$ tethered to surfaces with several surface
contact fields from $\beta_s = 0$ (hard surface see
Sec.~\ref{athermal}) to $\beta_s = 2.0$; 
Fig.~\ref{N64bb0_manybs} (b) shows the normalized number of surface
contacts, $n_s/N$, as the chains are being pulled. 
At high tension fields, the chains are stretched away from the
surface; $n_s/N$ approaches $1/N$ and the extensions become
independent of the surface interactions 
as $\beta_f$ becomes very large.
For low forces, that is for $\beta_f$ values that are too small to
change the number of surface contacts significantly, the adsorbed
chains ($\beta_s=1.5$ and $\beta_s=2.0$ in Fig.~\ref{N64bb0_manybs}) 
have much lower extensions and much higher (differential)
spring constants than the desorbed chains. This is expected because, 
for adsorbed chains, a small applied force
changes only the conformation of the unadsorbed tail of the chain. 
Since the length of the tail decreases with increasing $\beta_s$ and 
vanishes as $\beta_s$ becomes very large (strong coupling limit), both 
zero-force extension and fluctuations decrease with increasing
$\beta_s$. 
Fig.~\ref{z0chiz0} includes data for the zero-force extension,
$z_0$, and inverse spring constant, $\chi_{z0}$, at $\beta_s=2$ for a
range of chain lengths from $N=16$ to $N=128$. 
The results show that $z_0$ and $\chi_{z0}$ are independent of $N$, 
consistent with the length of the desorbed tail being
independent of the chain length.\cite{ei82,de08}

For adsorbed chains, an increase in the tension field is expected to
lead to force-induced desorption. The graphs for $\beta_s=1.5$ and
$\beta_s=2.0$ in Fig.~\ref{N64bb0_manybs} (a) and (b) show a sharp
increase in extension and a simultaneous loss of surface contacts
within a narrow force range. 
Increase in chain extension and loss of surface contacts are more
gradual for $\beta_s=1.0$, where the chains are desorbed but close to
the adsorption transition. For chains tethered to hard surfaces,
$\beta_s=0.0$, the number of surface contacts is already small at zero
force.
In Fig.~\ref{N64bb0_manybs} (c) and (d), we present extension
fluctuations, $\chi_z$, and surface contact fluctuations, $\chi_s$,
respectively, corresponding to the force-extension curves in panel (a).
For all surface conditions, the extension fluctuation graphs in panel
(c) show a maximum. These peaks move to higher tension fields and 
become sharper as $\beta_s$ increases. 
The surface-contact fluctuations shown in panel (d), on the other
hand, have well defined peaks only for adsorbing surfaces. 
For hard surfaces, $\chi_s$ decreases monotonically as $\beta_f$
increases. As $\beta_s$ approaches the adsorption value, the graph of
$\chi_s$ develops first a plateau at low $\beta_f$ and then a
maximum. 
The $\chi_s$ graph for $\beta_s=1.0$ in Fig.~\ref{N64bb0_manybs} (d) is
right on the verge of having a maximum, consistent with
$\beta_{sa}\simeq 1.0$.  
A comparison of the $\chi_z$ and $\chi_s$ graphs for $\beta_s=1.5$ and
$\beta_s=2.0$ shows similar behavior for both types of fluctuations in
the transition region. With increasing surface attraction, the peak
heights grow and the peaks occur at higher fields. The 
peaks in the extension fluctuation and the surface-contact
fluctuations occur at somewhat different tension fields; 
the difference between the peak locations decreases rapidly as
the surface attraction increases and it also decreases with
increasing chain length. 
The gap between the transition fields determined from peaks in
$\chi_s$ and $\chi_z$ is another example of the finite-size effects 
typical for small systems near phase transitions \cite{fe91c}.

\subsubsection{Microcanonical evaluation}\label{microcan_adsorption}

\begin{figure}[!htbp]
\includegraphics[width=3.4in]{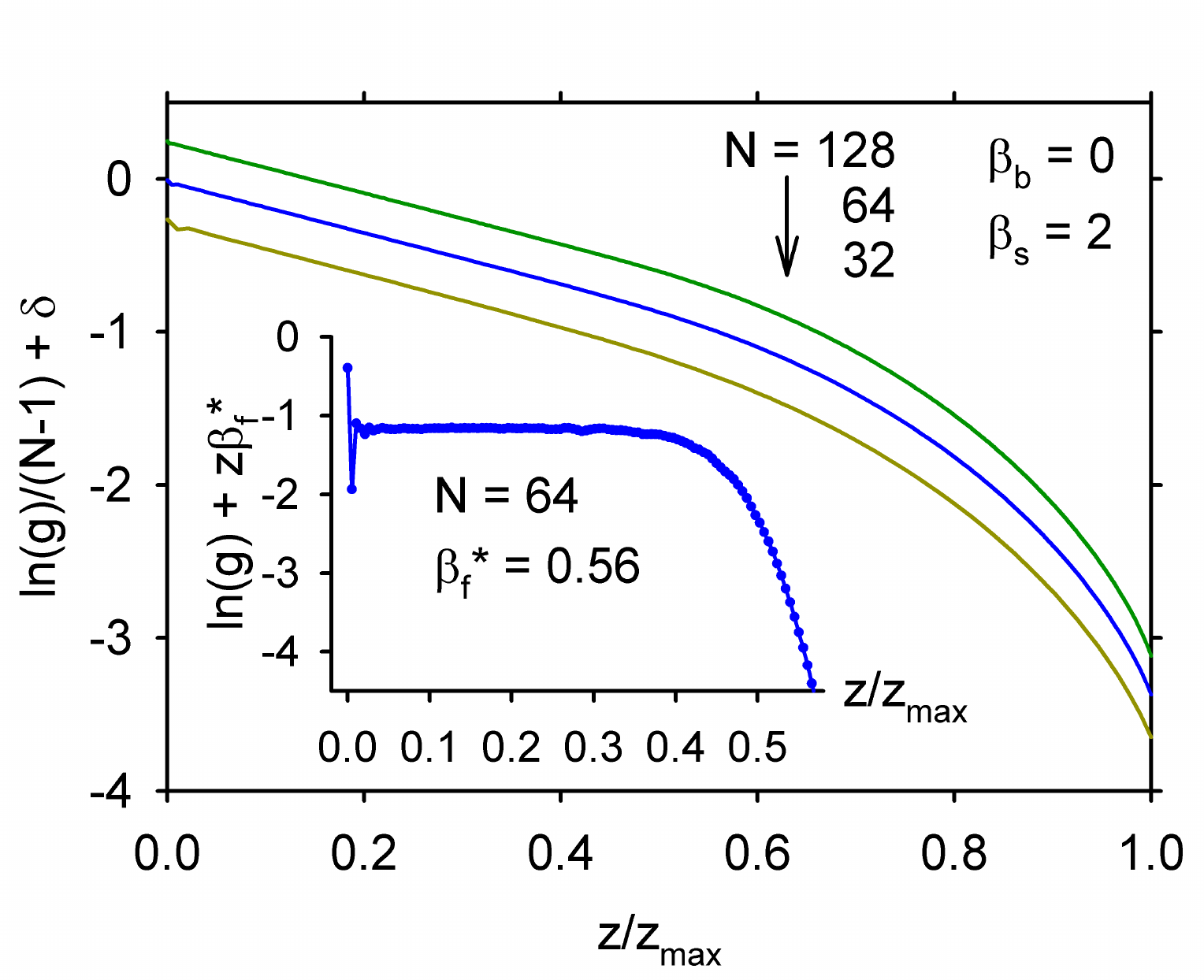}
\caption{Densities of state for an adsorbing surface. The solid lines
  represent the normalized logarithm of the density of states
  $g(z;\beta_s=2,\beta_b=0)$ as a function of the normalized
  extension $z/z_\mathrm{max}$. To separate the graphs for the
  different chain lengths, constant values $\delta=0.25$, 0, -0.25 
  have been added to $\ln(g)$ for $N=128$, 64, and 32, respectively.
  The steps in $\ln(g)$ near $z=0$, evident for $N=32$ and clearly
  visible in the inset, are due to the discrete positions the chain
  end can assume on the lattice. 
  The inset shows, for $N=64$ and the transition field
  $\beta_f^*=0.56$, the logarithm of the reweighted density of states,
  $g\exp(z\beta_f^*)$, which is proportional to the canonical
  probability distribution. 
\label{bs2bb0_lngs}}
\end{figure}

To investigate
the nature of the transition between adsorbed and
stretched chain conformations, we present in
Fig.~\ref{bs2bb0_lngs} results for the density of states 
$g(z;\beta_s,\beta_b)$, for $\beta_s=2.0$, $\beta_b=0$ and
chain lengths $N=32$, 64, and 128. 
We find discrete steps, up and down, in  $\ln(g)$ for the first few  
extension values, just visible for $N=32$ in Fig.~\ref{zavgWL}, before
$\ln(g)$ decreases monotonically.  
These steps reflect the discrete nature of
our lattice model; for high $\beta_s$, the beads of the chain,
including the chain end and its bonded neighbor, have a large
probability to be at or near the surface.  Due to bond-length
restrictions and excluded-volume interactions, the probability to find
the last bead at $z=1$, for example, is smaller than the probability
for either $z=0$ or $z=2$ leading to non-monotonous behavior of 
$\ln(g)$ for very low extensions. The range of relative extensions 
where steps in $g(z)$ occur decreases with increasing chain
length. 

For intermediate
extensions, the graphs of $\ln(g)$ for $\beta_s=2$ are linear 
within the statistical uncertainties. 
The slope of the lines is nearly identical
for all three chain lengths and the extension range where linear
behavior is observed increases with increasing chain length. 
According to Eq.~(\ref{betafmicro}), the slope of a $\ln(g(z))$
graph represents the negative tension field as a function of
extension.
This implies that the extensions in the straight-line
portions of the graphs all belong to the same tension field, which we
call $\beta_f^*$.  
For $N=64$, we show in the inset of Fig.~\ref{bs2bb0_lngs} the
logarithm of the reweighted density of states, $g\exp(z\beta_f^*)$,
which, up to a normalization factor, represents the canonical
probability distribution defined in Eq.~(\ref{prob1d}).
The discrete steps in the probability distribution at low extensions
are now clearly visible.
For intermediate extensions, we find a flat probability distribution,
implying that conformations with a wide range of extensions have the
same probability for being realized. 

\begin{figure}[!htbp]
\includegraphics[width=3.4in]{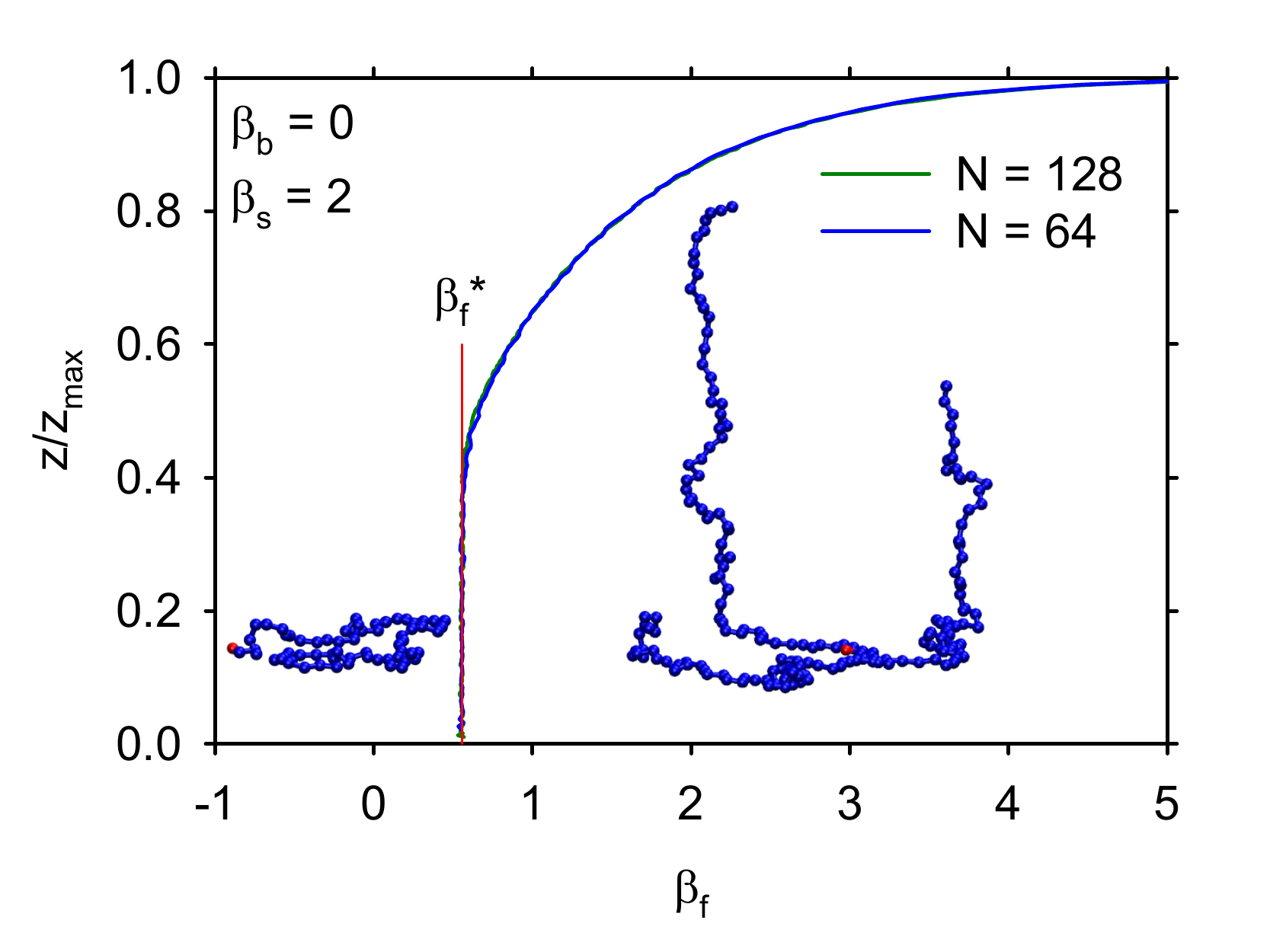}
\caption{Force-induced desorption for surface-contact field
  $\beta_s=2$ and athermal solvent conditions, $\beta_b=0$. 
  The solid lines represent results from a microcanonical evaluation
  of the densities of state for chain lengths $N=128$ and $N=64$.
  The vertical line segment indicates the tension field at the
  transition, $\beta_f^*=0.56$, and highlights the 
  vertical part of the extension curve, which signals phase
  coexistence. The results for the two chain 
  lengths are just barely distinguishable at the end of the
  coexistence region. The simulation snapshots show conformations for
  chains 
  of length $N=64$; an adsorbed chain (left) and a set of
  conformations for the coexistence region (right).
\label{bs2bb0_microcan}}
\end{figure}

The extension results in Figs.~\ref{N64bb0_manybs} and
\ref{z0chiz0} were obtained with a canonical evaluation of the
density of states, 
according to Eqs.~(\ref{zavgWL}) and (\ref{chizWL}). 
In Fig.~\ref{bs2bb0_microcan}, we present results from a micro-canonical
evaluation of the density of states for chains of length $N=64$ and
$N=128$. The tension field $\beta_f$ is calculated from a numerical
derivative of the logarithmic density of states according to
Eq.~(\ref{betafmicro}), where the smallest extensions were omitted due
to the discrete jumps in $g(z)$ discussed above. 
In Fig.~\ref{bs2bb0_microcan}, we show the graph with $\beta_f$ on the
horizontal axis to facilitate comparison with the force
extension-curves in Fig.~\ref{N64bb0_manybs} (a). 
In this representation, the region of constant slope of $\ln(g)$
becomes a vertical line at the transition field, 
$\beta_f^*\simeq 0.56$, and shows clearly the discontinuous nature of
the transition. 
According to Eq.~(\ref{chizmicro}), the second derivative of $\ln(g)$
with respect to $z$ yields the inverse of 
the extension fluctuations. 
Since the first derivative of $\ln(g)$ is constant, we find that
$\chi_z^{-1}$ vanishes in the coexistence region, corresponding to a 
$\delta$-function peak in $\chi_z$ at the transition field. This
singularity is analogous to the $\delta$-peak in the isobaric
heat-capacity at vapor-liquid coexistence and typical for
discontinuous transitions. 

Comparing the force-extension results in Fig.~\ref{bs2bb0_microcan}
with those obtained by a canonical evaluation of the density of states
in Fig.~\ref{N64bb0_manybs} (a), we note that the results are in excellent
agreement at high tensions and start to deviate only near the
transition. In the canonical evaluation, where $\beta_f$ is
controlled, the average extension is calculated as a weighted sum over
all possible extensions. For finite systems, this leads to a
broadening of the transition and a transition region, whose size
decreases with increasing chain length. In the thermodynamic limit,
where $N\rightarrow\infty$, both evaluation methods
are expected to give identical results. 

In contrast to other first-order transitions in
finite-size systems (see, for example, the results for stretching
chains from globular and crystalline phases in Sec.~\ref{poor}) the
probability distribution at the transition field is 
flat rather than bimodal, suggesting that a large number of states
coexist at the transition field $\beta_f^*$ \cite{sk12}.  
As we discuss further in Sec.~\ref{conclusion}, 
this is due to the absence of an interfacial barrier between
coexisting states. At the transition between adsorbed and stretched
states, the coexistence is between adsorbed and stretched parts of the
chain with a negligible interface between the domains.
In the inset of Fig.~\ref{bs2bb0_microcan} we show a set of such
conformations for chain length $N=64$.

\subsubsection{Phase portrait for finite chains near attractive
  surfaces}\label{phase_adsorption} 

\begin{figure}[!htbp]
\includegraphics[width=3.4in]{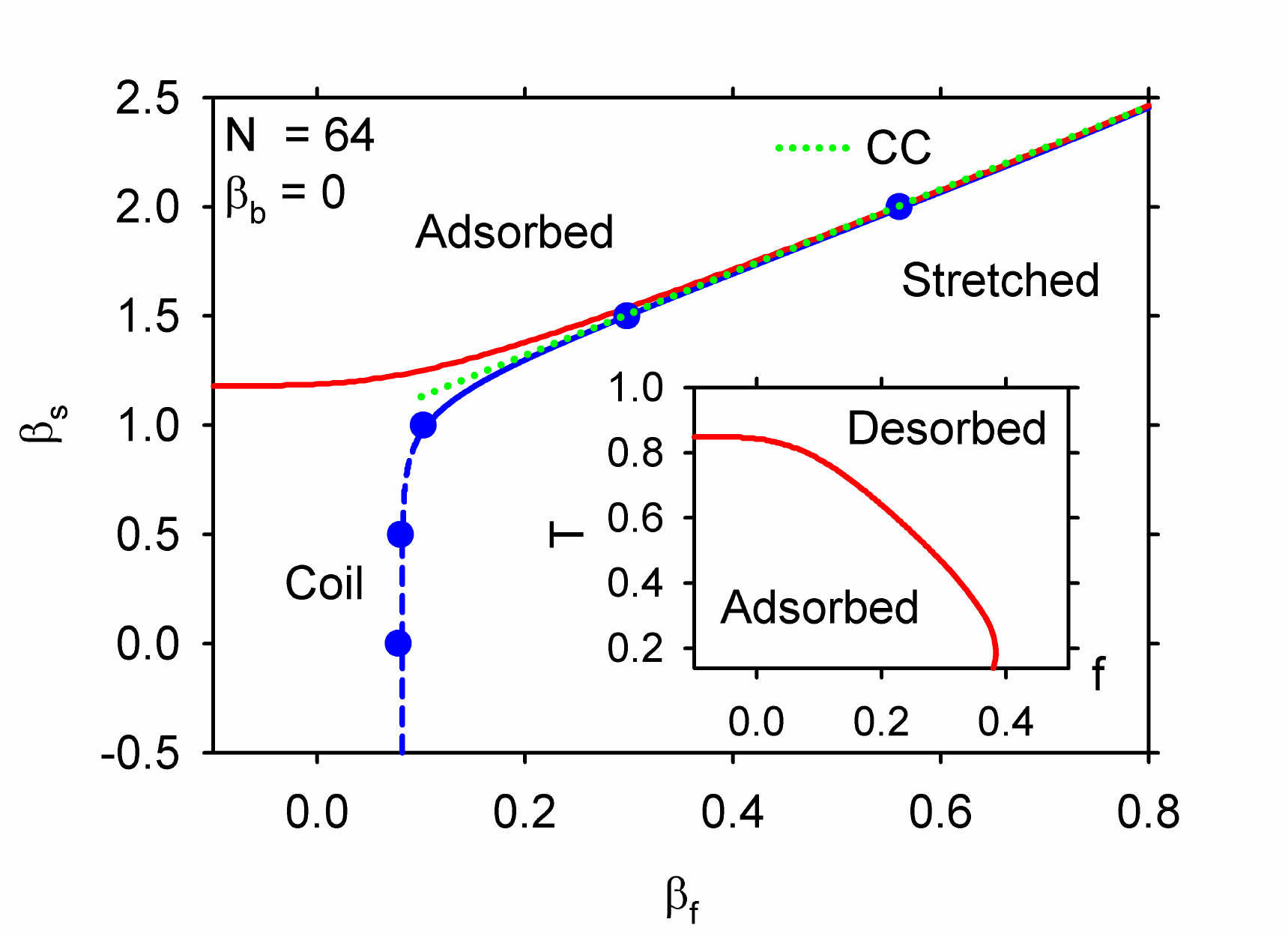}
\caption{Phase portrait for tethered chains of length $N=64$ in the
  space of tension field, $\beta_f$, and surface-contact field,
  $\beta_s$. The solid and dashed lines show the location of
  fluctuation maxima, determined from an evaluation of the 
  3-d density of states $g(n_s,n_b,z)$. The red solid line delineating
  the region of adsorbed conformations represents maxima of
  surface-contact fluctuations, $\chi_s$. 
  Desorbed chains have expanded coil and stretched
  conformations for low and high tension fields, respectively. 
  The solid and dashed blue lines show maxima in extension
  fluctuations, $\chi_z$, which indicate the force-induced
  desorption transition at high $\beta_s$ (solid lines) but 
  do not represent phase transitions for low $\beta_s$ (dashed
  lines). 
  The filled symbols show inflection point results from Wang-Landau simulations
  at fixed $\beta_s$ and  $\beta_b=0$. 
  The dotted line labeled CC is a straight line with slope 1.9 
  estimated from a Clausius-Clapeyron relation. 
  The inset shows the phase portrait in the force--temperature plane,
  where $T = 1/\beta_s$ and $f = \beta_f/\beta_s$ for surface
  interaction energy $\epsilon_s=-1$. 
\label{N64_bb0phase}}
\end{figure}

To investigate the relationship between tension and adsorption over a
range of conditions, we present a phase portrait for chains of
length $N=64$ in the $\beta_f - \beta_s$ plane of tension and surface
contact fields in Fig.~\ref{N64_bb0phase}.
We distinguish between two phases, adsorbed and desorbed, where the
desorbed phase includes both expanded coil and stretched
chain conformations. 

The blue lines in Fig.~\ref{N64_bb0phase} indicate maxima in the
extension fluctuations, $\chi_z$. At low    
surface contact fields $\beta_s$, these maxima are not related to 
phase transitions and therefore a dashed line is used to show their
location in the $\beta_f$-$\beta_s$ plane.
The location of $\chi_z$
maxima is independent of $\beta_s$ for repulsive and slightly
attractive surfaces. As $\beta_s$ approaches the adsorption value,
$\beta_{sa}\simeq 1.18$ for $N=64$, chain stretching starts to interfere
with adsorption. In the discussion of Fig.~\ref{N64bb0_manybs} (d) we
noted that the surface-contact fluctuations as a function of $\beta_f$
first develop a maximum for $\beta_s\gtrsim 1.0$. Around this value,
we see in Fig.~\ref{N64_bb0phase} the line of $\chi_z$ maxima starting to
move to higher tension fields and approach the line of $\chi_s$
maxima. 
The results presented in Figs.~\ref{bs2bb0_lngs} and 
\ref{bs2bb0_microcan} show that the transition between adsorbed and
stretched chains is discontinuous. For chains of length $N=64$, we
find evidence for phase coexistence down to about $\beta_s=1.3$. The
lines of fluctuation maxima at high fields  
therefore represent coexistence lines, whose slope may be estimated
from a Clausius Clapeyron (CC) relation. For coexistence between
adsorbed and stretched chains, the CC relation reads 
$d\beta_s/d\beta_f = -\Delta z/\Delta n_s$, where $\Delta z$ and
$\Delta n_s$ are the differences in extension and number of surface
contacts of the coexisting states, respectively. 
For $\beta_s=2.0$ we estimate a value of 1.9 for the slope,
which is indicated by the dotted line in the figure and seen to give a
good representation of the coexistence curve in the range shown
here. As $\beta_s$ increases further, the slope increases and reaches
a value of about 2.7 for the largest fields where we evaluate our
data. In the limit $\beta_s\rightarrow\infty$, the slope is expected
to approach a value of 3, since the largest bond length in the model
is 3 and its value will be added to the extension each time a surface
contact is broken.

The red solid line separating adsorbed conformations at high $\beta_s$
values from desorbed conformations at low $\beta_s$ values represents
maxima in the surface contact fluctuations, $\chi_s$, which we use to
identify the adsorption transition. 
At low tension fields, the adsorption transition is continuous and the
line of $\chi_s$ maxima almost independent of $\beta_f$. For tension
fields $\beta_f\gtrsim 0.1$, the adsorption transition moves to higher
$\beta_s$ values and for fields $\beta_f\gtrsim 0.2$, we find evidence
for phase coexistence. 
The phase diagram has a region where finite-size effects are
particularly evident; for $N=64$ it is the area 
$1.0 \lesssim \beta_s \lesssim 1.6$ and 
$0.1 \lesssim \beta_f \lesssim 0.4$. In this region, the transition
field values obtained from maxima in surface-contact and extension
fluctuations are not the same (see Fig.~\ref{N64bb0_manybs}) leading
to a gap between the transition lines. A comparison with results from
other chain lengths shows that size of the region where the lines
approach each other decreases with increasing chain length. 

In the inset of Fig.~\ref{N64_bb0phase}, we
present the line of adsorption transitions in the more familiar
force-temperature plane. For the conversion, we set $\epsilon_s = -1$
for the interaction energy, which yields $T=1/\beta_s$ and
$f=\beta_f/\beta_s$. In this representation, the chain is adsorbed at
low temperatures and forces and desorbs by increasing temperature or
tension force. The adsorption transition is continuous at low forces
and high temperatures and becomes discontinuous as the tension
increases. Near the lowest temperatures accessible to us, the 
phase diagram shows reentrant behavior, in agreement with theoretical
predictions and simulation results in the literature, (see, for
example, Refs.~[\onlinecite{sk12,bh09}]). 

\subsubsection{Effect of solvent condition on force-induced
  desorption}\label{desorbpoor} 

To investigate the effect of solvent condition on force-induced
desorption, we performed 1-d Wang-Landau simulations for chains of
length $N=64$ tethered to an adsorbing surface, $\beta_s = 2$ in
near-$\theta$ ($\beta_b = 0.5$) and poor ($\beta_b = 0.7$)
conditions. 
In the absence of tension, these surface and solvent conditions
yield adsorbed extended chain conformations.\cite{lu08} This means
that even though a desorbed chain at  $\beta_b = 0.7$ is collapsed,
the adsorbed chain is in an extended quasi-two dimensional
conformation. Compared to athermal solvent conditions ($\beta_b = 0$)
the lateral extension of an adsorbed chain in poor solvent is smaller
and the 
perpendicular extension is slightly larger as a result of competition
between bead-bead and surface contacts. 

\begin{figure}[!htbp]
\includegraphics[width=3.4in]{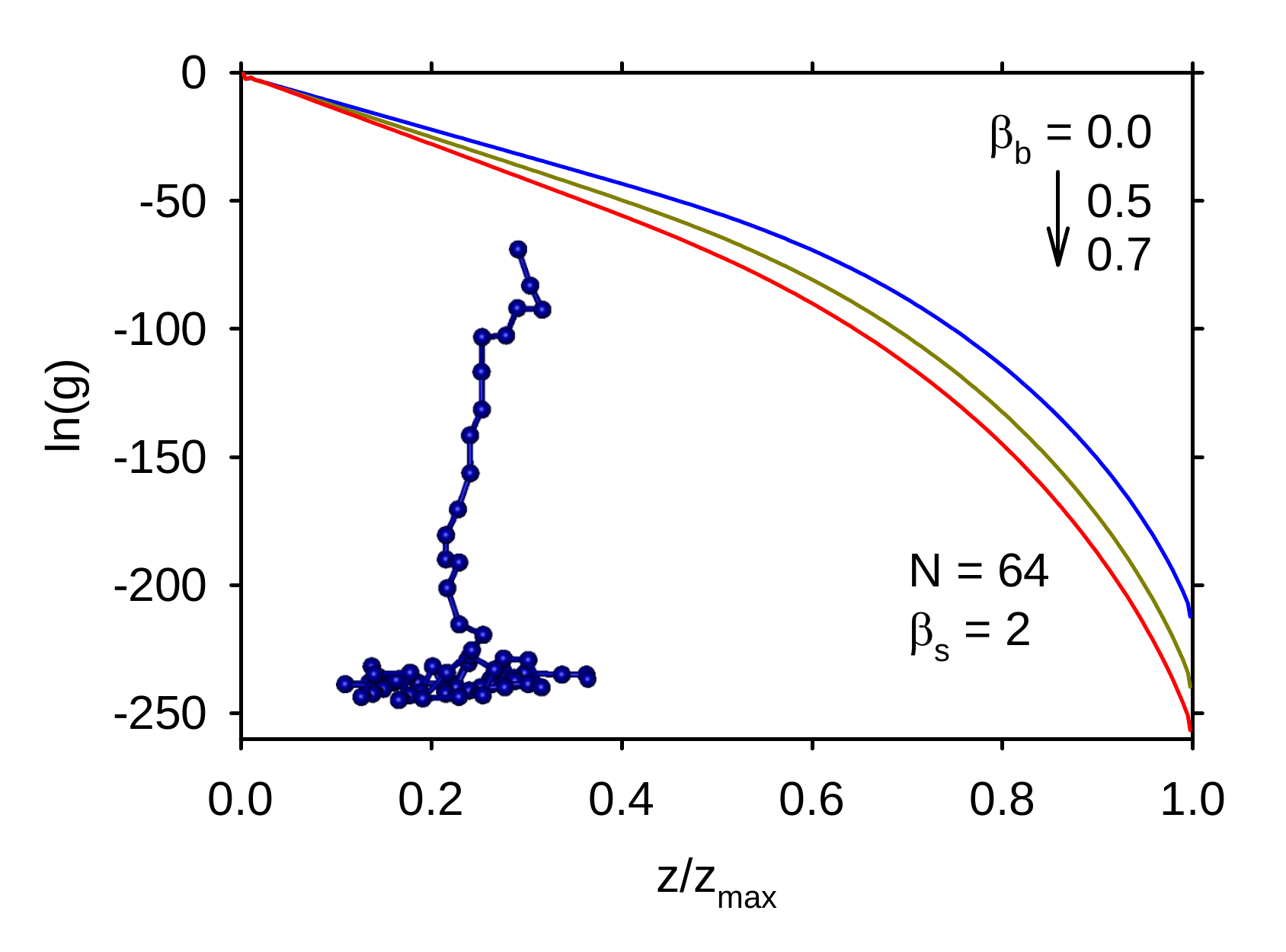}
\caption{Densities of state for an adsorbing surface, $\beta_s = 2$,
  and three different solvent conditions for chains of length $N=64$. 
  The solid lines represent the logarithm of the normalized density of
  states $g(z;\beta_s,\beta_b)$ as a function of the normalized
  extension $z/z_\mathrm{max}$ for athermal solvent, $\beta_b = 0$,
  near $\theta$ solvent, $\beta_b = 0.5$, 
  and poor solvent $\beta_b = 0.7$. The simulation snapshot shows a
  chain 
  conformation in the transition region (linear part of the graph) for
  $\beta_b = 0.7$. 
\label{lngadsorbpoor}
}
\end{figure}

In Fig.~\ref{lngadsorbpoor} we present 
data for the densities of state for $\beta_b = 0$, $\beta_b = 0.5$,
and $\beta_b = 0.7$.
The inset of Fig.~\ref{lngadsorbpoor} shows a chain conformation in
the transition region for $\beta_b = 0.7$, which may be compared with
athermal-solvent conformations in Fig.~\ref{bs2bb0_microcan}.
For all solvent conditions, 
the results show an extended range of extensions, where $\ln(g)$
decreases linearly. 
As discussed in Sec.~\ref{microcan_adsorption}
(see Figs.~\ref{bs2bb0_lngs} and \ref{bs2bb0_microcan}) this
implies coexistence of states at a transition  
field $\beta_f*$, which is given by the negative slope of the
graphs. We find values of $\beta_f* = 0.56$, 0.64, and 0.73 for the
solvent conditions $\beta_b = 0$, 0.5, and 0.7, respectively.

\begin{figure}[!htbp]
\includegraphics[width=3.4in]{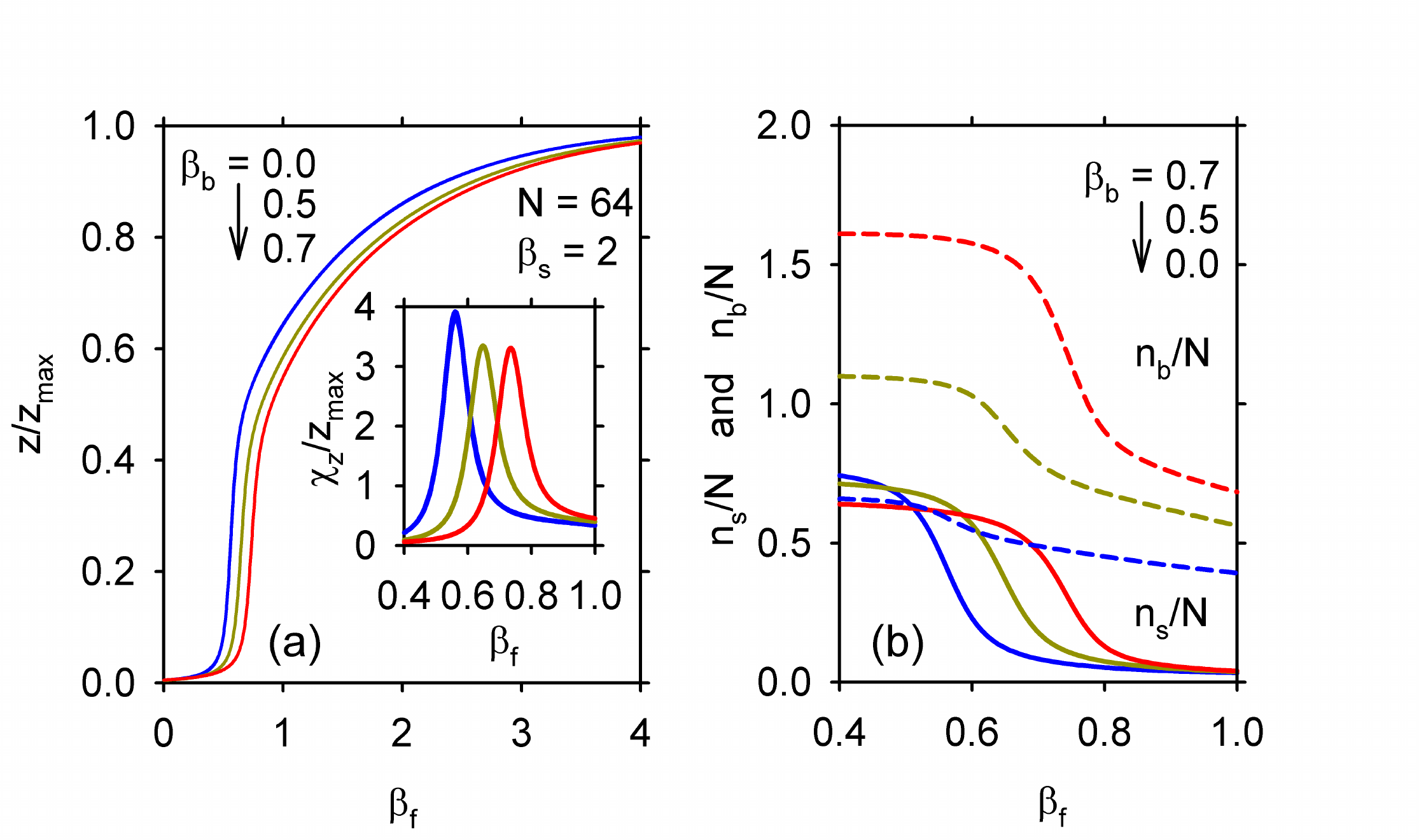}
\caption{Force-induced desorption in good and poor solvent. 
(a) Normalized extension $z/z_\mathrm{max}$ as a
  function of tension field, $\beta_f$, for adsorbing surface,
  $\beta_s = 2.0$, and increasingly poorer solvent, $\beta_b = 0$, 0.5,
  and 0.7 for $N=64$; the inset shows the normalized 
  extension fluctuations, $\chi_z/z_\mathrm{max}$, in the transition
  region. 
(b) Normalized surface contacts, $n_s/N$, (solid lines) and bead
  contacts, $n_b/N$, (dashed lines) in the transition region.
\label{adsorbpoor}
}
\end{figure}

In Fig.~\ref{adsorbpoor} (a) we present force-extension results from
the canonical evaluation of the densities of state; the inset shows
extension fluctuations in the transition region.
The graphs confirm that the transition field increases with increasing
$\beta_b$ and that the transition is sharpest for the athermal
solvent. 
Panel (b) shows the bead-bead and surface contact numbers in the
transition region. The initial number of surface contacts is largest
for the chain in athermal solvent and smallest for poor-solvent
conditions. Conversely, bead contact numbers are smallest in athermal
solvent and largest in poor solvent. 
As the tension increases, both types of contact numbers decrease first
gradually and then rapidly as the chain undergoes the transition to
the stretched state. 

To summarize, force-induced desorption in poor solvent 
occurs at higher tension than in good solvent conditions. The nature
of the transition is discontinuous with a broad range of coexisting
states for chains that are stretched from adsorbed extended initial
states.

\section{Summary and conclusion}\label{conclusion}

In this work, we developed a simulation approach to investigate
tethered chain molecules subject to an applied tension force. To
describe systems with  
a broad range of surface interactions and solvent conditions, 
we employed a bond-fluctuation lattice model with bead-surface and 
bead-bead interactions, where the effect of solvent interactions is
included implicitly in the interaction parameters of the model.
The thermodynamic properties of the chains are completely determined by
the density of states $g(n_s,n_b,z)$ over the three-dimensional state
space of surface contacts ($n_s$), bead-bead contacts ($n_b$), and
extensions ($z$); the fields conjugate to these quantities are the 
bead-contact field $\beta_b=-\epsilon_b/T$,   
the surface contact field $\beta_s=-\epsilon_s/T$, and the tension
field $\beta_f=f/T$, where $f$ is the applied force. 
For fixed surface and bead contact fields, the 1-d density of states
$g(z;\beta_s,\beta_b)$ determines the force-extension relations of the
chains. 

With the aid of the Monte Carlo simulation techniques described in
Appendix~\ref{g3}, we constructed $g(n_s,n_b,z)$ for chains of length
$N=32$, 64, and 128. Force-extension curves calculated from
$g(n_s,n_b,z)$ are most reliable for low to intermediate tension
fields. They allow us to investigate tethered chains for continuously
varying solvent and surface conditions and thereby identify states of
interest.   
For selected solvent and surface conditions, we performed Wang-Landau
simulations at fixed $\beta_s$ and $\beta_b$ 
to determine the 1-d density of states $g(z;\beta_s,\beta_b)$ as
described in Appendix ~\ref{g1}.
These simulations provide access to the high tension regime and allow us
to study details of phase transitions under tension.
To validate the method, we compared with scaling laws in the low
and intermediate force regime and exact results in independent-bond
(IB) approximation for high tensions. We find that our simulation
results are consistent with scaling and IB predictions.

Single-chain pulling experiments give insight into the effect of
solvent conditions on the mechanical response of chain molecules. 
Since simple polymer models, such as the freely-jointed chain
\cite{gr94c} and the wormlike chain \cite{bu94,ma95c,bo99}, have 
fixed solvent conditions, we wanted to explore if our coarse-grained
model is able to describe force-extension relations 
over the wide range of conditions, where experiments have been
performed. 
In the intermediate force regime, the force-extension curve has 
universal properties that allow us to compare our model with 
comparatively short chains to experiments on long biological chain
molecules.    
The contour length and the characteristic segment length connect the
model to the biomolecule and are determined from simulation results
and experimental data, respectively. The remaining
model parameter describes the solvent condition and is estimated from a 
comparison with experimental data.   
We obtain a good qualitative description of the experimental data 
except at the highest tensions (see Fig.~\ref{Danil_comp}), where the
bonds of the biological molecules become extensible \cite{sm96c}
requiring more detailed molecular models.
Our experience suggests that the computational method presented in 
Appendix~\ref{g1} can be applied without prohibitive computational
cost to obtain force-extension relations for more realistic models of
biomolecules in good to moderate solvents and thus become a useful
tool to interpret rupture-type single-molecule experiments.

In poor solvent conditions, a sufficiently large pulling force
stretches the chain out of the globular state. 
In agreement with theoretical predictions and experimental
observation, our model yields first-order transitions from the globule
to the stretched state for long chains (see
Fig.~\ref{N128goodpoor}).
We also find that the tension field at the
stretching transition increases linearly with increasing bead-contact
field (see Fig.~\ref{N64_bs0phase}), which is consistent with
experimental results by Li  {\it et al.}\cite{li10} 
For force-induced transitions from states with crystalline order 
we discovered that there is no
intermediate globular state; instead, we find coexistence between
crystalline and stretched conformations and transition states, where 
crystallites and stretched chain sections coexist on the same chain
(see Fig.~\ref{N32crystal}). 
Our results for the effect of solvent condition on the stretching
transition and the effect of tension on chain collapse and
crystallization are summarized in the phase
portrait presented in Fig.~\ref{N64_bs0phase}. 

To study the effects of surface attractions, we start with athermal 
solvent conditions ($\beta_b=0$) and vary the surface contact field
$\beta_s$. 
Slightly attractive surface interactions lead to a distortion of the
expanded coil conformations found near hard surfaces and to
force-extension relations that are very similar to those of hard
surfaces. 
For adsorbing surfaces, a small applied force
changes only the conformation of the unadsorbed tail of the chain,
whose size is independent of the chain length $N$ and 
decreases with increasing surface attraction. This leads to 
zero-force extensions, $z_0$, and
inverse spring constants, $\chi_{z0}$, that are 
independent of the chain length and decrease with increasing
$\beta_s$ (see Figs.~\ref{z0chiz0} and 
\ref{N64bb0_manybs}).  

The transition between adsorbed and stretched states is discontinuous
with a flat (rather than bimodal) probability distribution at the
transition field $\beta_f^*$. 
The negative logarithm of the probability distribution is the
variational free energy. Normally, first order transitions in finite
systems lead to bimodal distribution functions, which yield free
energy functions with two minima corresponding to the two coexisting
phases. However, the ``hump'' in between the minima is due to
interfacial contributions in the mixed phase region of the finite
system. In our case, however, the ``interface'' is not an extended
object, but corresponds to a single monomer only, and hence there is
no free energy cost due to the interface for phase coexistence here.  
Our findings for the transition are consistent with those of 
Skvortsov {\it et al.} \cite{sk12}, who discuss the nature of the
desorption transition in detail.
We have also performed simulations in $\theta$ and poor solvent
conditions, where the chains are in adsorbed extended states in the
absence of tension, and find that, while force-induced desorption
occurs at higher tension than in good-solvent conditions, the nature of
the transition is unchanged. 

The phase portrait in Fig.~\ref{N64_bb0phase} illustrates the relation
between adsorption and chain stretching. 
For weakly adsorbing surfaces, we find expanded coil conformations
that become stretched (without passing through a phase transition)
with increasing tension field. 
At high $\beta_s$, the chains are adsorbed and undergo a discontinuous
transition to stretched states with increasing $\beta_f$. The
adsorption transition at low tension is continuous and little
affected by the tension field $\beta_f$, until $\beta_f$ is
sufficiently large to desorb the chains. While 
finite-size effects are evident in the region where adsorption
and chain stretching first interfere with each other, we expect the
general conclusions drawn from this work to be applicable in the
long-chain limit.

\section*{Acknowledgments}

The authors would like to thank Mark Taylor for many helpful
discussions and the Buchtel College of
Arts and Sciences at the University of Akron for providing 
computational facilities.  
Financial support through the Deutsche Forschungsgemeinschaft (grant
No. SFB 625/A3) is gratefully acknowledged. 

\appendix
\section{Construction of the 3-d density of states $g(n_s,n_b,z)$}\label{g3}

In the absence of tension, a state of the system is described by the
pair of contact numbers $(n_s,n_b)$, where the accessible contact
numbers form a two-dimensional state space. \cite{lu08}
The density of states, $g(n_s,n_b)$, defined as the 
number of chain conformations (micro states) for state $(n_s,n_b)$, 
contains the complete information about the equilibrium thermodynamics
of the system.
In earlier work, we employed Wang-Landau algorithms
\cite{wa01b,la04b,zh05} to construct the density of states $g(n_s,n_b)$
over the two-dimensional state space of monomer-monomer and
monomer-surface contacts for tethered chains in the absence of tension
for chain lengths up to $N=64$ \cite{lu08} and $N=128$.
The elementary moves for the simulations consisted of displacement
of individual beads to nearest-neighbor sites, pivot moves about the
$z$ axis, and so-called cut-and-permute moves,
where the chain is cut at a random bead, top and
bottom are interchanged, and the chain is reassembled and tethered to
the surface if the move is accepted.\cite{ca02c}
For our simulations of tethered chains over a two-dimensional state
space, we employed several different strategies 
to find the 2-d density of states $g(n_s,n_b)$,
within reasonable computation times \cite{lu08}. 
For each chain length, at least two independent densities of states
were generated and extensively umbrella sampled to avoid systematic
errors and achieve a quality that allows evaluation by numerical
differentiation. 

To construct $g(n_s,n_b,z)$ we take advantage of the fact that,
up to an overall factor that cancels in the evaluations, the density
of states  $g(n_s,n_b,z)$ may be written as the product
\beq\label{g3d}
g(n_s,n_b,z) = p(z;n_s,n_b)g(n_s,n_b) ,
\enq
where $p(z;n_s,n_b)$ is the probability that the extension equals 
$z$ for given contact numbers $(n_s,n_b)$. 
To determine the probability distributions $p(z;n_s,n_b)$ we 
use the previously calculated 2-d density of states $g(n_s,n_b)$
\cite{lu08} in ``production'' simulations where we build histograms
of the height $z$ of the last bead without updating $g(n_s,n_b)$.
In practice, the probabilities $p(z;n_s,n_b)$ become very small for
large values of $z$ and have large relative uncertainties, which 
restricts the evaluation of the 3-d density of states to states with
very low tension forces.  
In order to improve the statistics for chain conformations with larger
extensions, we performed production simulations under fixed field(s) 
and evaluated the results with histogram reweighting techniques.
When only the tension field $\beta_f$ is fixed, 
the acceptance criterion for the simulations is
\ber
p\left( (n_s,n_b,z;\beta_f) \rightarrow (n'_s,n'_b,z';\beta_f) \right)
\mbox{\hspace{3em}} & & \nonumber \\
=
\min\left(\frac{g(n_s,n_b)}{g(n'_s,n'_b)}e^{-\beta_f(z-z')},1\right),
&&   
\label{tens_accpt}
\enr
where $z'$ and $z$ are the last bead's $z$-coordinate of the trial and
original chain conformations, respectively.  
To improve sampling of partially desorbed or extended chains under
tension, we also performed production simulations,  
where a contact field, for example the surface field $\beta_s$, is also
held constant. In this case, the acceptance criterion becomes
\ber
p\left( (n_s,n_b,z;\beta_s,\beta_f) \rightarrow%
(n'_s,n'_b,z';\beta_s,\beta_f) \right) 
\mbox{\hspace{3em}} & & \nonumber \\
=
\min\left(\frac{g(n_s,n_b)}{g(n'_s,n'_b)}%
e^{-\beta_s(n_s-n'_s)-\beta_f(z-z')},1\right), &&
\label{tens_betas_accpt}
\enr
and similarly for fixed bead-contact field $\beta_b$.
For chain lengths $N = 16$, 32, and 128, production
simulations with the acceptance criterion (\ref{tens_accpt}) were
carried out for tension fields $\beta_f = 0.0, 0.1, \ldots, 0.6$ and a
few additional fields.    
To explore the effect of tension on adsorption and
chain collapse, more extensive simulations were performed for chains of length
$N=64$. Tension fields $\beta_f = 0.0, 0.1, \ldots, 1.0$ 
were employed with Eq.~(\ref{tens_accpt}) and 
simulations with acceptance criterion 
(\ref{tens_betas_accpt}) were performed for a few $\beta_f$ values and
surface fields up to  $\beta_s = 2.84$ as well as for fixed
bead-contact fields up to $\beta_b = 0.91$.
Production simulations for finite tension typically have $10^9$ Monte
Carlo steps, divided into ten blocks, 
where the first block is used for equilibration  and results from the 
remaining nine blocks are evaluated to find the average profiles and
uncertainties. For zero tension, 
the total production time is typically 10$^{10}$ MC steps and  
all ten blocks are evaluated.

Histograms collected during the simulations yield values for the 
probability $p(z;n_s,n_b,\beta_f)$ to find the last bead at height
$z$ for given contact numbers $(n_s,n_b)$ and tension field
$\beta_f$. To compensate for the applied field, the profiles
$p(z;n_s,n_b,\beta_f)$ are multiplied with $z$-dependent weights
$e^{-\beta_f z}$. 
For each pair $(n_s,n_b)$, the reweighted profiles from different
tensions are combined by applying constant shift factors to assure a
good match in the overlap regions.
In the process, data at the edges of the profiles are examined and
data points with very large uncertainty are discarded.
Finally, data derived from different tensions are averaged in their
regions of overlap to arrive at the combined probability profiles
$p(z;n_s,n_b)$, which are entered into Eq.~(\ref{g3d}) 
to determine the density of states $g(n_s,n_b,z)$ over the
three-dimensional state space. 
In Fig.~S2 of the supplementary material \cite{supp} we
present an 
example for a combined profile. These results illustrate how the
technique of  production under tension with reweighting extends the profile 
to much larger heights than are accessible in zero-tension
simulations.

\section{1-d densities of state}\label{g1}

While the combination of profiles described above increases the range
of accessible heights of the last bead, it is not sufficient to reach
the most highly extended chain conformations. In order to explore the
full range of tension forces and chain extensions we perform Wang-Landau
simulations that sample all possible heights of the last bead at fixed
surface and bead contact fields. The acceptance criterion for these
simulations is 
\ber
p\left( (n_s,n_b,z;\beta_s,\beta_b) \rightarrow%
(n'_s,n'_b,z';\beta_s,\beta_b) \right) 
\mbox{\hspace{4em}} & & \nonumber \\
=
\min\left(\frac{g(z;\beta_s,\beta_b)}{g(z';\beta_s,\beta_b)}%
e^{-\beta_s(n_s-n'_s)-\beta_b(n_b-n'_b)},1\right),   &&
\label{betas_betab_accpt}
\enr
where $g(z;\beta_s,\beta_b)$ is the current estimate for the density
of state for the one-dimensional state space of $z$ coordinates of the
last bead for the given fields $\beta_s$ and $\beta_b$. 
For these simulations, the density of states is updated with the
original Wang-Landau prescription\cite{wa01b,la04b} with refinement
levels $f_k = \exp(2^{-(k-1)})$ for $k=4,5,\ldots,30$. We experimented
with different flatness criteria and found that the traditional 80\%
rule works well for these simulations.
We performed simulations for chains of length $N=16$, 32, 64, 128, and
256 for several contact fields. 
Typically, the density of states converged in simulations over the
whole range of $z$. However, in cases 
where the range of probabilities is very large, for example for
$N=256$,  simulations were carried out in overlapping windows and
their results combined to give the final density of states. 
For contact fields that lead to discontinuous stretching transitions,
for example for $N=32$ and $\beta_b=2$, care must be taken that the
entire coexistence region is covered by a window. 
Uncertainty estimates for the densities of states are obtained
either from multiple independent simulations or by comparing results
from the last few iteration levels.  

In Fig.~S3 of the supplementary material \cite{supp} we
present the 1-d density of states $g(z;\beta_s,\beta_b)$ for chain
length $N=64$ and fields $\beta_s = \beta_b = 0$ and compare it with
the corresponding probability distribution, $p(z;\beta_s,\beta_b)$, 
calculated by partial summation of the 3-d density of states
$g(n_s,n_b,z)$. The figure illustrates two typical observations.
First, the 1-d density of states $g(z;\beta_s,\beta_b)$ extends smoothly
to the highest possible extension, $z_\mathrm{max} = 3(N-1)$, while 
$p(z;\beta_s,\beta_b)$ has a limited range and shows
large statistical errors at higher extension. Secondly, 
in the range of $z$ values where both probabilities are
available, $p(z;\beta_s,\beta_b)$ and
$g(z;\beta_s,\beta_b)$ are in excellent agreement.
In this work, we use 3-d and 1-d densities of states in a
complimentary way: 
With results from $g(n_s,n_b,z)$, we investigate tethered chains for
continuous ranges of the thermodynamic fields $\beta_s$, $\beta_b$,
and $\beta_f$, we construct phase diagrams, and identify states of
interest. For these, we perform 1-d Wang-Landau simulations to
investigate their force-extension relations in detail, including the 
highest possible extensions.  


\begin{thebibliography}{65}
\expandafter\ifx\csname natexlab\endcsname\relax\def\natexlab#1{#1}\fi
\expandafter\ifx\csname bibnamefont\endcsname\relax
  \def\bibnamefont#1{#1}\fi
\expandafter\ifx\csname bibfnamefont\endcsname\relax
  \def\bibfnamefont#1{#1}\fi
\expandafter\ifx\csname citenamefont\endcsname\relax
  \def\citenamefont#1{#1}\fi
\expandafter\ifx\csname url\endcsname\relax
  \def\url#1{\texttt{#1}}\fi
\expandafter\ifx\csname urlprefix\endcsname\relax\def\urlprefix{URL }\fi
\providecommand{\bibinfo}[2]{#2}
\providecommand{\eprint}[2][]{\url{#2}}

\bibitem[{\citenamefont{Ritort}(2006)}]{ri06}
\bibinfo{author}{\bibfnamefont{F.}~\bibnamefont{Ritort}}, \bibinfo{journal}{J.
  Phys.: Cond. Mat.} \textbf{\bibinfo{volume}{18}}, \bibinfo{pages}{R531}
  (\bibinfo{year}{2006}).

\bibitem[{\citenamefont{Kumar and Li}(2010)}]{ku10}
\bibinfo{author}{\bibfnamefont{S.}~\bibnamefont{Kumar}} \bibnamefont{and}
  \bibinfo{author}{\bibfnamefont{M.~S.} \bibnamefont{Li}},
  \bibinfo{journal}{Physics Reports} \textbf{\bibinfo{volume}{486}},
  \bibinfo{pages}{1} (\bibinfo{year}{2010}).

\bibitem[{\citenamefont{Evans and Ritchie}(1997)}]{ev97}
\bibinfo{author}{\bibfnamefont{E.}~\bibnamefont{Evans}} \bibnamefont{and}
  \bibinfo{author}{\bibfnamefont{K.}~\bibnamefont{Ritchie}},
  \bibinfo{journal}{Biophys. J.} \textbf{\bibinfo{volume}{72}},
  \bibinfo{pages}{1541} (\bibinfo{year}{1997}).

\bibitem[{\citenamefont{Ray et~al.}(2007)\citenamefont{Ray, Brown, and
  Akhremitchev}}]{ra07}
\bibinfo{author}{\bibfnamefont{C.}~\bibnamefont{Ray}},
  \bibinfo{author}{\bibfnamefont{J.~R.} \bibnamefont{Brown}}, \bibnamefont{and}
  \bibinfo{author}{\bibfnamefont{B.~B.} \bibnamefont{Akhremitchev}},
  \bibinfo{journal}{J. Phys. Chem. B} \textbf{\bibinfo{volume}{111}},
  \bibinfo{pages}{1963} (\bibinfo{year}{2007}).

\bibitem[{\citenamefont{Dougan et~al.}(2009)\citenamefont{Dougan, Li, Badilla,
  Berne, and Fernandez}}]{do09}
\bibinfo{author}{\bibfnamefont{L.}~\bibnamefont{Dougan}},
  \bibinfo{author}{\bibfnamefont{J.}~\bibnamefont{Li}},
  \bibinfo{author}{\bibfnamefont{C.~J.} \bibnamefont{Badilla}},
  \bibinfo{author}{\bibfnamefont{B.~J.} \bibnamefont{Berne}}, \bibnamefont{and}
  \bibinfo{author}{\bibfnamefont{J.~M.} \bibnamefont{Fernandez}},
  \bibinfo{journal}{Proc. Natl. Acad. Sci. USA} \textbf{\bibinfo{volume}{106}},
  \bibinfo{pages}{12605} (\bibinfo{year}{2009}).

\bibitem[{\citenamefont{Vrbov{\'{a}} and Whittington}(1996)}]{vr96}
\bibinfo{author}{\bibfnamefont{T.}~\bibnamefont{Vrbov{\'{a}}}}
  \bibnamefont{and} \bibinfo{author}{\bibfnamefont{S.~G.}
  \bibnamefont{Whittington}}, \bibinfo{journal}{J. Phys. A.: Math. Gen.}
  \textbf{\bibinfo{volume}{29}}, \bibinfo{pages}{6253} (\bibinfo{year}{1996}).

\bibitem[{\citenamefont{Rajesh et~al.}(2002)\citenamefont{Rajesh, Dhar, Giri,
  Kumar, and Singh}}]{ra02b}
\bibinfo{author}{\bibfnamefont{R.}~\bibnamefont{Rajesh}},
  \bibinfo{author}{\bibfnamefont{D.}~\bibnamefont{Dhar}},
  \bibinfo{author}{\bibfnamefont{D.}~\bibnamefont{Giri}},
  \bibinfo{author}{\bibfnamefont{S.}~\bibnamefont{Kumar}}, \bibnamefont{and}
  \bibinfo{author}{\bibfnamefont{Y.}~\bibnamefont{Singh}},
  \bibinfo{journal}{Phys. Rev. E} \textbf{\bibinfo{volume}{65}},
  \bibinfo{pages}{056124} (\bibinfo{year}{2002}).

\bibitem[{\citenamefont{Krawczyk et~al.}(2005)\citenamefont{Krawczyk, Owczarek,
  Prellberg, and Rechnitzer}}]{kr05}
\bibinfo{author}{\bibfnamefont{J.}~\bibnamefont{Krawczyk}},
  \bibinfo{author}{\bibfnamefont{A.~L.} \bibnamefont{Owczarek}},
  \bibinfo{author}{\bibfnamefont{T.}~\bibnamefont{Prellberg}},
  \bibnamefont{and}
  \bibinfo{author}{\bibfnamefont{A.}~\bibnamefont{Rechnitzer}},
  \bibinfo{journal}{Europhys. Lett.} \textbf{\bibinfo{volume}{70}},
  \bibinfo{pages}{726} (\bibinfo{year}{2005}).

\bibitem[{\citenamefont{Bachmann and Janke}(2006)}]{ba06}
\bibinfo{author}{\bibfnamefont{M.}~\bibnamefont{Bachmann}} \bibnamefont{and}
  \bibinfo{author}{\bibfnamefont{W.}~\bibnamefont{Janke}},
  \bibinfo{journal}{Phys. Rev. E} \textbf{\bibinfo{volume}{73}},
  \bibinfo{pages}{041802} (\bibinfo{year}{2006}).

\bibitem[{\citenamefont{Luettmer-Strathmann
  et~al.}(2008)\citenamefont{Luettmer-Strathmann, Rampf, Paul, and
  Binder}}]{lu08}
\bibinfo{author}{\bibfnamefont{J.}~\bibnamefont{Luettmer-Strathmann}},
  \bibinfo{author}{\bibfnamefont{F.}~\bibnamefont{Rampf}},
  \bibinfo{author}{\bibfnamefont{W.}~\bibnamefont{Paul}}, \bibnamefont{and}
  \bibinfo{author}{\bibfnamefont{K.}~\bibnamefont{Binder}},
  \bibinfo{journal}{J. Chem. Phys.} \textbf{\bibinfo{volume}{128}},
  \bibinfo{pages}{064903} (\bibinfo{year}{2008}).

\bibitem[{\citenamefont{Binder et~al.}(2008)\citenamefont{Binder, Paul,
  Strauch, Rampf, Ivanov, and Luettmer-Strathmann}}]{bi08}
\bibinfo{author}{\bibfnamefont{K.}~\bibnamefont{Binder}},
  \bibinfo{author}{\bibfnamefont{W.}~\bibnamefont{Paul}},
  \bibinfo{author}{\bibfnamefont{T.}~\bibnamefont{Strauch}},
  \bibinfo{author}{\bibfnamefont{F.}~\bibnamefont{Rampf}},
  \bibinfo{author}{\bibfnamefont{V.}~\bibnamefont{Ivanov}}, \bibnamefont{and}
  \bibinfo{author}{\bibfnamefont{J.}~\bibnamefont{Luettmer-Strathmann}},
  \bibinfo{journal}{J. Phys.: Condens. Matter} \textbf{\bibinfo{volume}{20}},
  \bibinfo{pages}{494215} (\bibinfo{year}{2008}).

\bibitem[{\citenamefont{M{\"{o}}ddel et~al.}(2011)\citenamefont{M{\"{o}}ddel,
  Janke, and Bachmann}}]{mo11}
\bibinfo{author}{\bibfnamefont{M.}~\bibnamefont{M{\"{o}}ddel}},
  \bibinfo{author}{\bibfnamefont{W.}~\bibnamefont{Janke}}, \bibnamefont{and}
  \bibinfo{author}{\bibfnamefont{M.}~\bibnamefont{Bachmann}},
  \bibinfo{journal}{Macromolecules} \textbf{\bibinfo{volume}{44}},
  \bibinfo{pages}{9013} (\bibinfo{year}{2011}).

\bibitem[{\citenamefont{Knotts~IV et~al.}(2008)\citenamefont{Knotts~IV,
  Rathore, and de~Pablo}}]{kn08}
\bibinfo{author}{\bibfnamefont{T.~A.} \bibnamefont{Knotts~IV}},
  \bibinfo{author}{\bibfnamefont{N.}~\bibnamefont{Rathore}}, \bibnamefont{and}
  \bibinfo{author}{\bibfnamefont{J.}~\bibnamefont{de~Pablo}},
  \bibinfo{journal}{Biophys. J.} \textbf{\bibinfo{volume}{94}},
  \bibinfo{pages}{4473} (\bibinfo{year}{2008}).

\bibitem[{\citenamefont{Heinz et~al.}(2009)\citenamefont{Heinz, Farmer, Pandey,
  Slocik, Patnaik, Pachter, and Naik}}]{he09}
\bibinfo{author}{\bibfnamefont{H.}~\bibnamefont{Heinz}},
  \bibinfo{author}{\bibfnamefont{B.~L.} \bibnamefont{Farmer}},
  \bibinfo{author}{\bibfnamefont{R.~B.} \bibnamefont{Pandey}},
  \bibinfo{author}{\bibfnamefont{J.~M.} \bibnamefont{Slocik}},
  \bibinfo{author}{\bibfnamefont{S.~S.} \bibnamefont{Patnaik}},
  \bibinfo{author}{\bibfnamefont{R.}~\bibnamefont{Pachter}}, \bibnamefont{and}
  \bibinfo{author}{\bibfnamefont{R.~R.} \bibnamefont{Naik}},
  \bibinfo{journal}{J. Am. Chem. Soc.} \textbf{\bibinfo{volume}{131}},
  \bibinfo{pages}{9704} (\bibinfo{year}{2009}).

\bibitem[{\citenamefont{Swetnam and Allen}(2012)}]{sw12}
\bibinfo{author}{\bibfnamefont{A.}~\bibnamefont{Swetnam}} \bibnamefont{and}
  \bibinfo{author}{\bibfnamefont{M.~P.} \bibnamefont{Allen}},
  \bibinfo{journal}{Phys. Rev. E} \textbf{\bibinfo{volume}{85}},
  \bibinfo{pages}{062901} (\bibinfo{year}{2012}).

\bibitem[{\citenamefont{Radhakrishna et~al.}(2012)\citenamefont{Radhakrishna,
  Sharma, and Kumar}}]{ra12}
\bibinfo{author}{\bibfnamefont{M.}~\bibnamefont{Radhakrishna}},
  \bibinfo{author}{\bibfnamefont{S.}~\bibnamefont{Sharma}}, \bibnamefont{and}
  \bibinfo{author}{\bibfnamefont{S.~K.} \bibnamefont{Kumar}},
  \bibinfo{journal}{J. Chem. Phys.} \textbf{\bibinfo{volume}{136}},
  \bibinfo{pages}{114114} (\bibinfo{year}{2012}).

\bibitem[{\citenamefont{Bustamante et~al.}(1994)\citenamefont{Bustamante,
  Marko, Siggia, and Smith}}]{bu94}
\bibinfo{author}{\bibfnamefont{C.}~\bibnamefont{Bustamante}},
  \bibinfo{author}{\bibfnamefont{J.~F.} \bibnamefont{Marko}},
  \bibinfo{author}{\bibfnamefont{E.~D.} \bibnamefont{Siggia}},
  \bibnamefont{and} \bibinfo{author}{\bibfnamefont{S.}~\bibnamefont{Smith}},
  \bibinfo{journal}{Science} \textbf{\bibinfo{volume}{265}},
  \bibinfo{pages}{1599} (\bibinfo{year}{1994}).

\bibitem[{\citenamefont{Smith et~al.}(1996)\citenamefont{Smith, Cui, and
  Bustamante}}]{sm96c}
\bibinfo{author}{\bibfnamefont{S.~B.} \bibnamefont{Smith}},
  \bibinfo{author}{\bibfnamefont{Y.}~\bibnamefont{Cui}}, \bibnamefont{and}
  \bibinfo{author}{\bibfnamefont{C.}~\bibnamefont{Bustamante}},
  \bibinfo{journal}{Science} \textbf{\bibinfo{volume}{271}},
  \bibinfo{pages}{795} (\bibinfo{year}{1996}).

\bibitem[{\citenamefont{Haupt et~al.}(2002)\citenamefont{Haupt, Senden, and
  Sevick}}]{ha02c}
\bibinfo{author}{\bibfnamefont{B.~J.} \bibnamefont{Haupt}},
  \bibinfo{author}{\bibfnamefont{T.~J.} \bibnamefont{Senden}},
  \bibnamefont{and} \bibinfo{author}{\bibfnamefont{E.~M.}
  \bibnamefont{Sevick}}, \bibinfo{journal}{Langmuir}
  \textbf{\bibinfo{volume}{18}}, \bibinfo{pages}{2174} (\bibinfo{year}{2002}).

\bibitem[{\citenamefont{Dessinges et~al.}(2002)\citenamefont{Dessinges, Maier,
  Zhang, Peliti, Bensimon, and Croquette}}]{de02d}
\bibinfo{author}{\bibfnamefont{M.-N.} \bibnamefont{Dessinges}},
  \bibinfo{author}{\bibfnamefont{B.}~\bibnamefont{Maier}},
  \bibinfo{author}{\bibfnamefont{Y.}~\bibnamefont{Zhang}},
  \bibinfo{author}{\bibfnamefont{M.}~\bibnamefont{Peliti}},
  \bibinfo{author}{\bibfnamefont{D.}~\bibnamefont{Bensimon}}, \bibnamefont{and}
  \bibinfo{author}{\bibfnamefont{V.}~\bibnamefont{Croquette}},
  \bibinfo{journal}{Phys. Rev. Lett.} \textbf{\bibinfo{volume}{89}},
  \bibinfo{pages}{248102} (\bibinfo{year}{2002}).

\bibitem[{\citenamefont{Cocco et~al.}(2003)\citenamefont{Cocco, Marko,
  Monasson, Sarkar, and Yan}}]{co03}
\bibinfo{author}{\bibfnamefont{S.}~\bibnamefont{Cocco}},
  \bibinfo{author}{\bibfnamefont{J.~F.} \bibnamefont{Marko}},
  \bibinfo{author}{\bibfnamefont{R.}~\bibnamefont{Monasson}},
  \bibinfo{author}{\bibfnamefont{A.}~\bibnamefont{Sarkar}}, \bibnamefont{and}
  \bibinfo{author}{\bibfnamefont{J.}~\bibnamefont{Yan}}, \bibinfo{journal}{Eur.
  Phys. J. E} \textbf{\bibinfo{volume}{10}}, \bibinfo{pages}{249}
  (\bibinfo{year}{2003}).

\bibitem[{\citenamefont{Danilowicz et~al.}(2007)\citenamefont{Danilowicz, Lee,
  Coljee, and Prentiss}}]{da07}
\bibinfo{author}{\bibfnamefont{C.}~\bibnamefont{Danilowicz}},
  \bibinfo{author}{\bibfnamefont{C.~H.} \bibnamefont{Lee}},
  \bibinfo{author}{\bibfnamefont{V.~W.} \bibnamefont{Coljee}},
  \bibnamefont{and} \bibinfo{author}{\bibfnamefont{M.}~\bibnamefont{Prentiss}},
  \bibinfo{journal}{Phys. Rev. E} \textbf{\bibinfo{volume}{75}},
  \bibinfo{pages}{030902} (\bibinfo{year}{2007}).

\bibitem[{\citenamefont{Gunari et~al.}(2007)\citenamefont{Gunari, Balazs, and
  Walker}}]{gu07b}
\bibinfo{author}{\bibfnamefont{N.}~\bibnamefont{Gunari}},
  \bibinfo{author}{\bibfnamefont{A.~C.} \bibnamefont{Balazs}},
  \bibnamefont{and} \bibinfo{author}{\bibfnamefont{G.~C.}
  \bibnamefont{Walker}}, \bibinfo{journal}{J. Am. Chem. Soc.}
  \textbf{\bibinfo{volume}{129}}, \bibinfo{pages}{10046}
  (\bibinfo{year}{2007}).

\bibitem[{\citenamefont{Saleh et~al.}(2009)\citenamefont{Saleh, McIntosh,
  Pincus, and Ribeck}}]{sa09}
\bibinfo{author}{\bibfnamefont{O.~A.} \bibnamefont{Saleh}},
  \bibinfo{author}{\bibfnamefont{D.~B.} \bibnamefont{McIntosh}},
  \bibinfo{author}{\bibfnamefont{P.}~\bibnamefont{Pincus}}, \bibnamefont{and}
  \bibinfo{author}{\bibfnamefont{N.}~\bibnamefont{Ribeck}},
  \bibinfo{journal}{Phys. Rev. Lett.} \textbf{\bibinfo{volume}{102}},
  \bibinfo{pages}{068301} (\bibinfo{year}{2009}).

\bibitem[{\citenamefont{Li and Walker}(2010)}]{li10}
\bibinfo{author}{\bibfnamefont{I.~T.~S.} \bibnamefont{Li}} \bibnamefont{and}
  \bibinfo{author}{\bibfnamefont{G.~C.} \bibnamefont{Walker}},
  \bibinfo{journal}{J. Am. Chem. Soc.} \textbf{\bibinfo{volume}{132}},
  \bibinfo{pages}{6530} (\bibinfo{year}{2010}).

\bibitem[{\citenamefont{Dittmore et~al.}(2011)\citenamefont{Dittmore, McIntosh,
  Halliday, and Saleh}}]{di11}
\bibinfo{author}{\bibfnamefont{A.}~\bibnamefont{Dittmore}},
  \bibinfo{author}{\bibfnamefont{D.~B.} \bibnamefont{McIntosh}},
  \bibinfo{author}{\bibfnamefont{S.}~\bibnamefont{Halliday}}, \bibnamefont{and}
  \bibinfo{author}{\bibfnamefont{O.~A.} \bibnamefont{Saleh}},
  \bibinfo{journal}{Phys. Rev. Lett.} \textbf{\bibinfo{volume}{107}},
  \bibinfo{pages}{148301} (\bibinfo{year}{2011}).

\bibitem[{\citenamefont{Pincus}(1976)}]{pi76}
\bibinfo{author}{\bibfnamefont{P.}~\bibnamefont{Pincus}},
  \bibinfo{journal}{Macromolecules} \textbf{\bibinfo{volume}{9}},
  \bibinfo{pages}{386} (\bibinfo{year}{1976}).

\bibitem[{\citenamefont{Halperin and Zhulina}(1991{\natexlab{a}})}]{ha91c}
\bibinfo{author}{\bibfnamefont{A.}~\bibnamefont{Halperin}} \bibnamefont{and}
  \bibinfo{author}{\bibfnamefont{E.~B.} \bibnamefont{Zhulina}},
  \bibinfo{journal}{Europhys. Lett.} \textbf{\bibinfo{volume}{15}},
  \bibinfo{pages}{417} (\bibinfo{year}{1991}{\natexlab{a}}).

\bibitem[{\citenamefont{Halperin and Zhulina}(1991{\natexlab{b}})}]{ha91b}
\bibinfo{author}{\bibfnamefont{A.}~\bibnamefont{Halperin}} \bibnamefont{and}
  \bibinfo{author}{\bibfnamefont{E.~B.} \bibnamefont{Zhulina}},
  \bibinfo{journal}{Macromolecules} \textbf{\bibinfo{volume}{24}},
  \bibinfo{pages}{5393} (\bibinfo{year}{1991}{\natexlab{b}}).

\bibitem[{\citenamefont{Marko and Siggia}(1995)}]{ma95c}
\bibinfo{author}{\bibfnamefont{J.~F.} \bibnamefont{Marko}} \bibnamefont{and}
  \bibinfo{author}{\bibfnamefont{E.~D.} \bibnamefont{Siggia}},
  \bibinfo{journal}{Macromolecules} \textbf{\bibinfo{volume}{28}},
  \bibinfo{pages}{8759} (\bibinfo{year}{1995}).

\bibitem[{\citenamefont{Ha and Thirumalai}(1997)}]{ha97f}
\bibinfo{author}{\bibfnamefont{B.-Y.} \bibnamefont{Ha}} \bibnamefont{and}
  \bibinfo{author}{\bibfnamefont{D.}~\bibnamefont{Thirumalai}},
  \bibinfo{journal}{J. Chem. Phys.} \textbf{\bibinfo{volume}{106}},
  \bibinfo{pages}{4243} (\bibinfo{year}{1997}).

\bibitem[{\citenamefont{Bouchiat et~al.}(1999)\citenamefont{Bouchiat, Wang,
  Allemand, Strick, Block, and Croquette}}]{bo99}
\bibinfo{author}{\bibfnamefont{C.}~\bibnamefont{Bouchiat}},
  \bibinfo{author}{\bibfnamefont{M.~D.} \bibnamefont{Wang}},
  \bibinfo{author}{\bibfnamefont{J.-F.} \bibnamefont{Allemand}},
  \bibinfo{author}{\bibfnamefont{T.}~\bibnamefont{Strick}},
  \bibinfo{author}{\bibfnamefont{M.}~\bibnamefont{Block}}, \bibnamefont{and}
  \bibinfo{author}{\bibfnamefont{V.}~\bibnamefont{Croquette}},
  \bibinfo{journal}{Biophysical Journal} \textbf{\bibinfo{volume}{76}},
  \bibinfo{pages}{409} (\bibinfo{year}{1999}).

\bibitem[{\citenamefont{Livadru et~al.}(2003)\citenamefont{Livadru, Netz, and
  Kreuzer}}]{li03}
\bibinfo{author}{\bibfnamefont{L.}~\bibnamefont{Livadru}},
  \bibinfo{author}{\bibfnamefont{R.~R.} \bibnamefont{Netz}}, \bibnamefont{and}
  \bibinfo{author}{\bibfnamefont{H.~J.} \bibnamefont{Kreuzer}},
  \bibinfo{journal}{J. Chem. Phys.} \textbf{\bibinfo{volume}{118}},
  \bibinfo{pages}{1404} (\bibinfo{year}{2003}).

\bibitem[{\citenamefont{Bhattacharya
  et~al.}(2009{\natexlab{a}})\citenamefont{Bhattacharya, Rostiashvili, Milchev,
  and Vilgis}}]{bh09}
\bibinfo{author}{\bibfnamefont{S.}~\bibnamefont{Bhattacharya}},
  \bibinfo{author}{\bibfnamefont{V.~G.} \bibnamefont{Rostiashvili}},
  \bibinfo{author}{\bibfnamefont{A.}~\bibnamefont{Milchev}}, \bibnamefont{and}
  \bibinfo{author}{\bibfnamefont{T.~A.} \bibnamefont{Vilgis}},
  \bibinfo{journal}{Macromolecules} \textbf{\bibinfo{volume}{42}},
  \bibinfo{pages}{2236} (\bibinfo{year}{2009}{\natexlab{a}}).

\bibitem[{\citenamefont{Klushin and Skvortsov}(2011)}]{kl11}
\bibinfo{author}{\bibfnamefont{L.~I.} \bibnamefont{Klushin}} \bibnamefont{and}
  \bibinfo{author}{\bibfnamefont{A.~M.} \bibnamefont{Skvortsov}},
  \bibinfo{journal}{J. Phys. A: Math. Theor.} \textbf{\bibinfo{volume}{44}},
  \bibinfo{pages}{473001} (\bibinfo{year}{2011}).

\bibitem[{\citenamefont{Skvortsov et~al.}(2012)\citenamefont{Skvortsov,
  Klushin, Polotsky, and Binder}}]{sk12}
\bibinfo{author}{\bibfnamefont{A.~M.} \bibnamefont{Skvortsov}},
  \bibinfo{author}{\bibfnamefont{L.~I.} \bibnamefont{Klushin}},
  \bibinfo{author}{\bibfnamefont{A.~A.} \bibnamefont{Polotsky}},
  \bibnamefont{and} \bibinfo{author}{\bibfnamefont{K.}~\bibnamefont{Binder}},
  \bibinfo{journal}{Phys. Rev. E} \textbf{\bibinfo{volume}{85}},
  \bibinfo{pages}{031803} (\bibinfo{year}{2012}).

\bibitem[{\citenamefont{Wittkop et~al.}(1994)\citenamefont{Wittkop, Sommer,
  Kreitmeier, and G{\"{o}}ritz}}]{wi94c}
\bibinfo{author}{\bibfnamefont{M.}~\bibnamefont{Wittkop}},
  \bibinfo{author}{\bibfnamefont{J.-U.} \bibnamefont{Sommer}},
  \bibinfo{author}{\bibfnamefont{S.}~\bibnamefont{Kreitmeier}},
  \bibnamefont{and}
  \bibinfo{author}{\bibfnamefont{D.}~\bibnamefont{G{\"{o}}ritz}},
  \bibinfo{journal}{Phys. Rev. E} \textbf{\bibinfo{volume}{49}},
  \bibinfo{pages}{5472} (\bibinfo{year}{1994}).

\bibitem[{\citenamefont{Wittkop et~al.}(1996)\citenamefont{Wittkop, Kreitmeier,
  and G{\"{o}}ritz}}]{wi96c}
\bibinfo{author}{\bibfnamefont{M.}~\bibnamefont{Wittkop}},
  \bibinfo{author}{\bibfnamefont{S.}~\bibnamefont{Kreitmeier}},
  \bibnamefont{and}
  \bibinfo{author}{\bibfnamefont{D.}~\bibnamefont{G{\"{o}}ritz}},
  \bibinfo{journal}{Phys. Rev. E} \textbf{\bibinfo{volume}{53}},
  \bibinfo{pages}{838} (\bibinfo{year}{1996}).

\bibitem[{\citenamefont{Grassberger and Hsu}(2002)}]{gr02}
\bibinfo{author}{\bibfnamefont{P.}~\bibnamefont{Grassberger}} \bibnamefont{and}
  \bibinfo{author}{\bibfnamefont{H.-P.} \bibnamefont{Hsu}},
  \bibinfo{journal}{Phys. Rev. E} \textbf{\bibinfo{volume}{65}},
  \bibinfo{pages}{031807} (\bibinfo{year}{2002}).

\bibitem[{\citenamefont{Frisch and Verga}(2002)}]{fr02b}
\bibinfo{author}{\bibfnamefont{T.}~\bibnamefont{Frisch}} \bibnamefont{and}
  \bibinfo{author}{\bibfnamefont{A.}~\bibnamefont{Verga}},
  \bibinfo{journal}{Phys. Rev. E} \textbf{\bibinfo{volume}{65}},
  \bibinfo{pages}{041801} (\bibinfo{year}{2002}).

\bibitem[{\citenamefont{Celestini et~al.}(2004)\citenamefont{Celestini, Frisch,
  and Oyharcabal}}]{ce04}
\bibinfo{author}{\bibfnamefont{F.}~\bibnamefont{Celestini}},
  \bibinfo{author}{\bibfnamefont{T.}~\bibnamefont{Frisch}}, \bibnamefont{and}
  \bibinfo{author}{\bibfnamefont{X.}~\bibnamefont{Oyharcabal}},
  \bibinfo{journal}{Phys. Rev. E} \textbf{\bibinfo{volume}{70}},
  \bibinfo{pages}{012801} (\bibinfo{year}{2004}).

\bibitem[{\citenamefont{Morrison et~al.}(2007)\citenamefont{Morrison, Hyeon,
  Toan, Ha, and Thirumalai}}]{mo07b}
\bibinfo{author}{\bibfnamefont{G.}~\bibnamefont{Morrison}},
  \bibinfo{author}{\bibfnamefont{C.}~\bibnamefont{Hyeon}},
  \bibinfo{author}{\bibfnamefont{N.~M.} \bibnamefont{Toan}},
  \bibinfo{author}{\bibfnamefont{B.-H.} \bibnamefont{Ha}}, \bibnamefont{and}
  \bibinfo{author}{\bibfnamefont{D.}~\bibnamefont{Thirumalai}},
  \bibinfo{journal}{Macromolecules} \textbf{\bibinfo{volume}{40}},
  \bibinfo{pages}{7343} (\bibinfo{year}{2007}).

\bibitem[{\citenamefont{Bhattacharya
  et~al.}(2009{\natexlab{b}})\citenamefont{Bhattacharya, Rostiashvili, Milchev,
  and Vilgis}}]{bh09b}
\bibinfo{author}{\bibfnamefont{S.}~\bibnamefont{Bhattacharya}},
  \bibinfo{author}{\bibfnamefont{V.~G.} \bibnamefont{Rostiashvili}},
  \bibinfo{author}{\bibfnamefont{A.}~\bibnamefont{Milchev}}, \bibnamefont{and}
  \bibinfo{author}{\bibfnamefont{T.~A.} \bibnamefont{Vilgis}},
  \bibinfo{journal}{Phys. Rev. E} \textbf{\bibinfo{volume}{79}},
  \bibinfo{pages}{030802} (\bibinfo{year}{2009}{\natexlab{b}}).

\bibitem[{\citenamefont{Toan and Thirumalai}(2010)}]{to10}
\bibinfo{author}{\bibfnamefont{N.~M.} \bibnamefont{Toan}} \bibnamefont{and}
  \bibinfo{author}{\bibfnamefont{D.}~\bibnamefont{Thirumalai}},
  \bibinfo{journal}{Macromolecules} \textbf{\bibinfo{volume}{43}},
  \bibinfo{pages}{4394} (\bibinfo{year}{2010}).

\bibitem[{\citenamefont{Hsu and Binder}(2012)}]{hs12}
\bibinfo{author}{\bibfnamefont{H.-P.} \bibnamefont{Hsu}} \bibnamefont{and}
  \bibinfo{author}{\bibfnamefont{K.}~\bibnamefont{Binder}},
  \bibinfo{journal}{J. Chem. Phys.} \textbf{\bibinfo{volume}{136}},
  \bibinfo{pages}{024901} (\bibinfo{year}{2012}).

\bibitem[{\citenamefont{Wang and Landau}(2001{\natexlab{a}})}]{wa01}
\bibinfo{author}{\bibfnamefont{F.}~\bibnamefont{Wang}} \bibnamefont{and}
  \bibinfo{author}{\bibfnamefont{D.~P.} \bibnamefont{Landau}},
  \bibinfo{journal}{Phys. Rev. Lett.} \textbf{\bibinfo{volume}{86}},
  \bibinfo{pages}{2050} (\bibinfo{year}{2001}{\natexlab{a}}).

\bibitem[{\citenamefont{Wang and Landau}(2001{\natexlab{b}})}]{wa01b}
\bibinfo{author}{\bibfnamefont{F.}~\bibnamefont{Wang}} \bibnamefont{and}
  \bibinfo{author}{\bibfnamefont{D.~P.} \bibnamefont{Landau}},
  \bibinfo{journal}{Phys. Rev. E} \textbf{\bibinfo{volume}{64}},
  \bibinfo{pages}{056101} (\bibinfo{year}{2001}{\natexlab{b}}).

\bibitem[{\citenamefont{Paul et~al.}(2007)\citenamefont{Paul, Strauch, Rampf,
  and Binder}}]{pa07}
\bibinfo{author}{\bibfnamefont{W.}~\bibnamefont{Paul}},
  \bibinfo{author}{\bibfnamefont{T.}~\bibnamefont{Strauch}},
  \bibinfo{author}{\bibfnamefont{F.}~\bibnamefont{Rampf}}, \bibnamefont{and}
  \bibinfo{author}{\bibfnamefont{K.}~\bibnamefont{Binder}},
  \bibinfo{journal}{Phys. Rev. E} \textbf{\bibinfo{volume}{75}},
  \bibinfo{pages}{060801} (\bibinfo{year}{2007}).

\bibitem[{\citenamefont{W{\"{u}}st et~al.}(2011)\citenamefont{W{\"{u}}st, Li,
  and Landau}}]{wu11}
\bibinfo{author}{\bibfnamefont{T.}~\bibnamefont{W{\"{u}}st}},
  \bibinfo{author}{\bibfnamefont{Y.~W.} \bibnamefont{Li}}, \bibnamefont{and}
  \bibinfo{author}{\bibfnamefont{D.~P.} \bibnamefont{Landau}},
  \bibinfo{journal}{J. Stat. Phys.} \textbf{\bibinfo{volume}{144}},
  \bibinfo{pages}{638} (\bibinfo{year}{2011}).

\bibitem[{\citenamefont{Taylor et~al.}(2009)\citenamefont{Taylor, Paul, and
  Binder}}]{ta09}
\bibinfo{author}{\bibfnamefont{M.~P.} \bibnamefont{Taylor}},
  \bibinfo{author}{\bibfnamefont{W.}~\bibnamefont{Paul}}, \bibnamefont{and}
  \bibinfo{author}{\bibfnamefont{K.}~\bibnamefont{Binder}},
  \bibinfo{journal}{J. Chem. Phys.} \textbf{\bibinfo{volume}{131}},
  \bibinfo{pages}{114907} (\bibinfo{year}{2009}).

\bibitem[{\citenamefont{Decas et~al.}(2008)\citenamefont{Decas, Sommer, and
  Blumen}}]{de08}
\bibinfo{author}{\bibfnamefont{R.}~\bibnamefont{Decas}},
  \bibinfo{author}{\bibfnamefont{J.-U.} \bibnamefont{Sommer}},
  \bibnamefont{and} \bibinfo{author}{\bibfnamefont{A.}~\bibnamefont{Blumen}},
  \bibinfo{journal}{Macromol. Theory Simul.} \textbf{\bibinfo{volume}{17}},
  \bibinfo{pages}{429} (\bibinfo{year}{2008}).

\bibitem[{\citenamefont{Grosberg and Khokhlov}(1994)}]{gr94c}
\bibinfo{author}{\bibfnamefont{A.~Y.} \bibnamefont{Grosberg}} \bibnamefont{and}
  \bibinfo{author}{\bibfnamefont{A.~R.} \bibnamefont{Khokhlov}},
  \emph{\bibinfo{title}{Statistical Physics of Macromolecules}}, AIP Series in
  Polymers and Complex Materials (\bibinfo{publisher}{American Institute of
  Physics}, \bibinfo{address}{Woodbury, NY}, \bibinfo{year}{1994}).

\bibitem[{\citenamefont{Cifra and Bleha}(1995)}]{ci95}
\bibinfo{author}{\bibfnamefont{P.}~\bibnamefont{Cifra}} \bibnamefont{and}
  \bibinfo{author}{\bibfnamefont{T.}~\bibnamefont{Bleha}},
  \bibinfo{journal}{Macromol. Theory Simul.} \textbf{\bibinfo{volume}{4}},
  \bibinfo{pages}{233} (\bibinfo{year}{1995}).

\bibitem[{\citenamefont{Brak et~al.}(2009)\citenamefont{Brak, Dyke, Lee,
  Owczarek, Prellberg, Rechnitzer, and Whittington}}]{br09}
\bibinfo{author}{\bibfnamefont{R.}~\bibnamefont{Brak}},
  \bibinfo{author}{\bibfnamefont{P.}~\bibnamefont{Dyke}},
  \bibinfo{author}{\bibfnamefont{J.}~\bibnamefont{Lee}},
  \bibinfo{author}{\bibfnamefont{A.~L.} \bibnamefont{Owczarek}},
  \bibinfo{author}{\bibfnamefont{T.}~\bibnamefont{Prellberg}},
  \bibinfo{author}{\bibfnamefont{A.}~\bibnamefont{Rechnitzer}},
  \bibnamefont{and} \bibinfo{author}{\bibfnamefont{S.~G.}
  \bibnamefont{Whittington}}, \bibinfo{journal}{J. Phys. A}
  \textbf{\bibinfo{volume}{42}}, \bibinfo{pages}{085001}
  (\bibinfo{year}{2009}).

\bibitem[{\citenamefont{Eisenriegler et~al.}(1982)\citenamefont{Eisenriegler,
  Kremer, and Binder}}]{ei82}
\bibinfo{author}{\bibfnamefont{E.}~\bibnamefont{Eisenriegler}},
  \bibinfo{author}{\bibfnamefont{K.}~\bibnamefont{Kremer}}, \bibnamefont{and}
  \bibinfo{author}{\bibfnamefont{K.}~\bibnamefont{Binder}},
  \bibinfo{journal}{J. Chem. Phys.} \textbf{\bibinfo{volume}{77}},
  \bibinfo{pages}{6296} (\bibinfo{year}{1982}).

\bibitem[{\citenamefont{Carmesin and Kremer}(1988)}]{ca88b}
\bibinfo{author}{\bibfnamefont{I.}~\bibnamefont{Carmesin}} \bibnamefont{and}
  \bibinfo{author}{\bibfnamefont{K.}~\bibnamefont{Kremer}},
  \bibinfo{journal}{Macromolecules} \textbf{\bibinfo{volume}{21}},
  \bibinfo{pages}{2819} (\bibinfo{year}{1988}).

\bibitem[{\citenamefont{Binder}(1995)}]{bi95}
\bibinfo{author}{\bibfnamefont{K.}~\bibnamefont{Binder}},
  \emph{\bibinfo{title}{Monte Carlo and Molecular Dynamics Simulations in
  Polymer Science}} (\bibinfo{publisher}{Oxford University Press},
  \bibinfo{address}{Oxford}, \bibinfo{year}{1995}).

\bibitem[{\citenamefont{Landau and Binder}(2000)}]{la00b}
\bibinfo{author}{\bibfnamefont{D.~P.} \bibnamefont{Landau}} \bibnamefont{and}
  \bibinfo{author}{\bibfnamefont{K.}~\bibnamefont{Binder}},
  \emph{\bibinfo{title}{A Guide to Monte Carlo Simulations in Statistical
  Physics}} (\bibinfo{publisher}{Cambridge University},
  \bibinfo{address}{Cambridge, UK}, \bibinfo{year}{2000}).

\bibitem[{sup()}]{supp}
\bibinfo{note}{See Supplemental Material at the end of this manuscript
  for details on the scaling of
  the zero-force extension fluctuations and for figures illustrating the
  construction and validation of the density-of-states obtained by the
  algorithms described in the appendices.}

\bibitem[{\citenamefont{Rampf et~al.}(2005)\citenamefont{Rampf, Paul, and
  Binder}}]{ra05b}
\bibinfo{author}{\bibfnamefont{F.}~\bibnamefont{Rampf}},
  \bibinfo{author}{\bibfnamefont{W.}~\bibnamefont{Paul}}, \bibnamefont{and}
  \bibinfo{author}{\bibfnamefont{K.}~\bibnamefont{Binder}},
  \bibinfo{journal}{Europhys. Lett.} \textbf{\bibinfo{volume}{70}},
  \bibinfo{pages}{628} (\bibinfo{year}{2005}).

\bibitem[{\citenamefont{Ferrenberg and Landau}(1991)}]{fe91c}
\bibinfo{author}{\bibfnamefont{A.~M.} \bibnamefont{Ferrenberg}}
  \bibnamefont{and} \bibinfo{author}{\bibfnamefont{D.~P.}
  \bibnamefont{Landau}}, \bibinfo{journal}{Phys. Rev. B}
  \textbf{\bibinfo{volume}{44}}, \bibinfo{pages}{5081} (\bibinfo{year}{1991}).

\bibitem[{\citenamefont{Descas et~al.}(2004)\citenamefont{Descas, Sommer, and
  Blumen}}]{de04}
\bibinfo{author}{\bibfnamefont{R.}~\bibnamefont{Descas}},
  \bibinfo{author}{\bibfnamefont{J.-U.} \bibnamefont{Sommer}},
  \bibnamefont{and} \bibinfo{author}{\bibfnamefont{A.}~\bibnamefont{Blumen}},
  \bibinfo{journal}{J. Chem. Phys.} \textbf{\bibinfo{volume}{120}},
  \bibinfo{pages}{8831} (\bibinfo{year}{2004}).

\bibitem[{\citenamefont{Landau et~al.}(2004)\citenamefont{Landau, Tsai, and
  Exler}}]{la04b}
\bibinfo{author}{\bibfnamefont{D.~P.} \bibnamefont{Landau}},
  \bibinfo{author}{\bibfnamefont{S.-H.} \bibnamefont{Tsai}}, \bibnamefont{and}
  \bibinfo{author}{\bibfnamefont{M.}~\bibnamefont{Exler}},
  \bibinfo{journal}{Am. J. Phys.} \textbf{\bibinfo{volume}{72}},
  \bibinfo{pages}{1294} (\bibinfo{year}{2004}).

\bibitem[{\citenamefont{Zhou et~al.}(2006)\citenamefont{Zhou, Schulthess,
  Torbr{\"{u}}gge, and Landau}}]{zh05}
\bibinfo{author}{\bibfnamefont{C.}~\bibnamefont{Zhou}},
  \bibinfo{author}{\bibfnamefont{T.~C.} \bibnamefont{Schulthess}},
  \bibinfo{author}{\bibfnamefont{S.}~\bibnamefont{Torbr{\"{u}}gge}},
  \bibnamefont{and} \bibinfo{author}{\bibfnamefont{D.~P.}
  \bibnamefont{Landau}}, \bibinfo{journal}{Phys. Rev. Lett.}
  \textbf{\bibinfo{volume}{96}}, \bibinfo{pages}{120201}
  (\bibinfo{year}{2006}).

\bibitem[{\citenamefont{Causo}(2002)}]{ca02c}
\bibinfo{author}{\bibfnamefont{M.~S.} \bibnamefont{Causo}},
  \bibinfo{journal}{J. Stat. Phys.} \textbf{\bibinfo{volume}{108}},
  \bibinfo{pages}{247} (\bibinfo{year}{2002}).

\end{thebibliography}

\newpage

\renewcommand{\thefigure}{S\arabic{figure}}
\renewcommand{\theequation}{S\arabic{equation}}
\setcounter{figure}{0}

\section*{Supplemental Material for: Transitions of tethered chain molecules under tension}

\subsubsection*{\bf Scaling of extension-fluctuation maxima with chain length}

\begin{figure}[!hbt]
\includegraphics[width=3.4in]{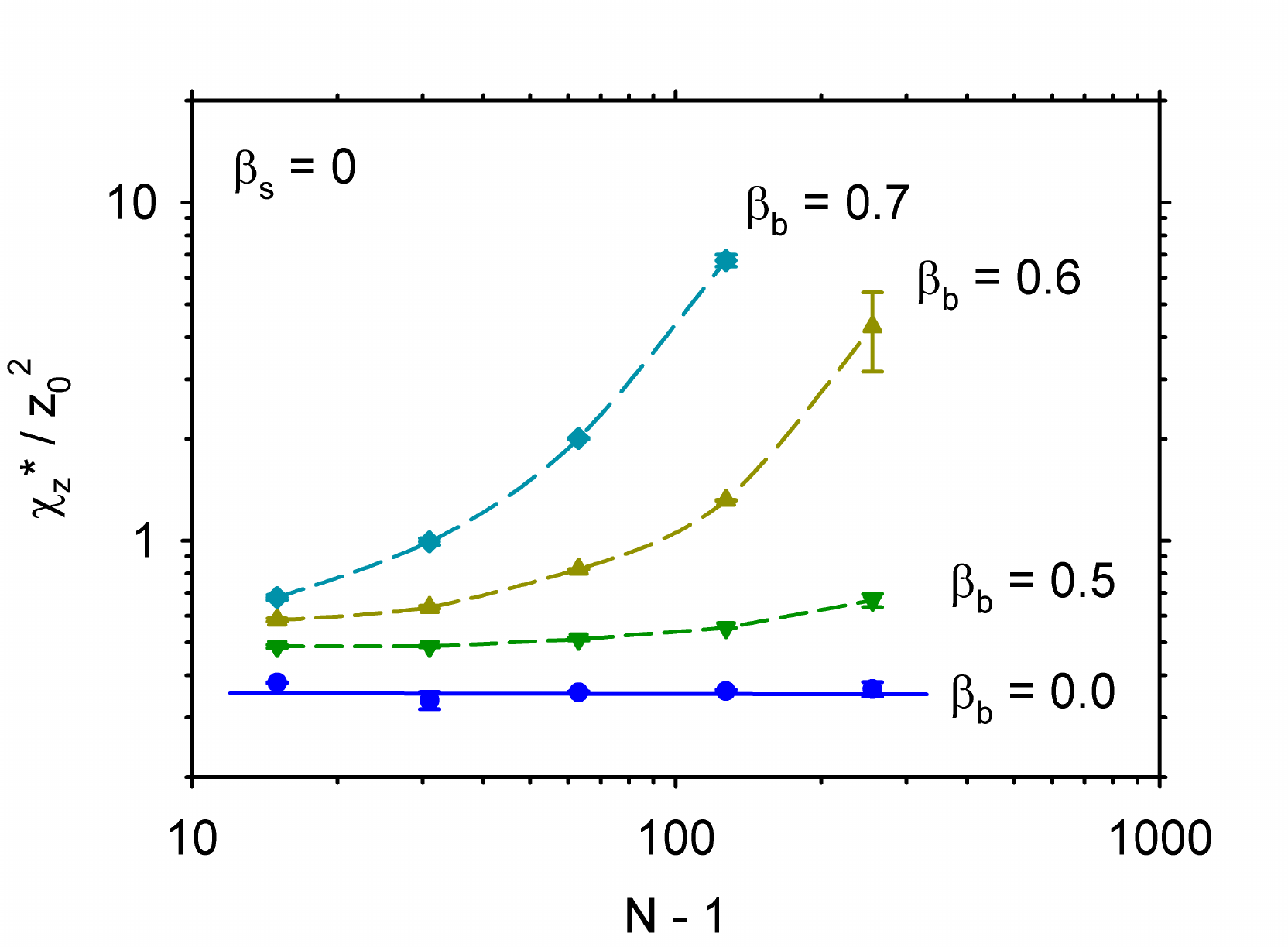}
\caption{Scaled maxima of extension fluctuations $\chi_z^*/z_0^2$ 
as a function of chain length, $N-1$, for four solvent conditions,
$\beta_b = 0$, 0.5, 0.6, and 0.7. The symbols represent simulation
results, the dashed lines are a guide to the eye. 
The solid line represents 
the average value of the simulation results for $\beta_b=0$. 
\label{chizi}}
\end{figure}

To discuss the significance of maxima in the extension
fluctuations, 
we present in Fig.~\ref{chizi} chain-length dependent results for
scaled fluctuation maxima, $\chi_z^*/z_0^2$, where $z_0$ is the
zero-force extension.  
For athermal solvent, $\beta_b=0$, the 
fluctuation maxima $\chi_z^*$ occur in the low-force scaling region 
and are expected to scale with chain length in the same
way as the zero-force fluctuations, i.e.\ 
$\chi_{z}^*\sim z_0^2\sim N^{2\nu}$ so that 
$\chi_{z}^*/z_0^2$ is expected to be independent of the chain length.  
This behavior is indeed observed for the $\beta_b = 0$ data in 
Fig.~\ref{chizi}. 
For $\beta_b=0.5$, the scaled fluctuation maxima 
in Fig.~\ref{chizi} are just starting to show an
increase with increasing chain length. 
For poorer solvents, $\beta_b = 0.6$ and $\beta_b = 0.7$,
$\chi_z^*/z_0^2$ grows markedly with $N$ implying that 
the fluctuation maxima grow much faster than the chain length. 
Since this is the expected behavior near phase transitions, these
results suggest that chain stretching acquires the character of a phase
transition as the solvent quality decreases.

\clearpage

\subsubsection*{\bf Probability distributions from 1-d 
and 3-d densities of states}

\begin{figure}[!htbp]
\includegraphics[width=3.4in]{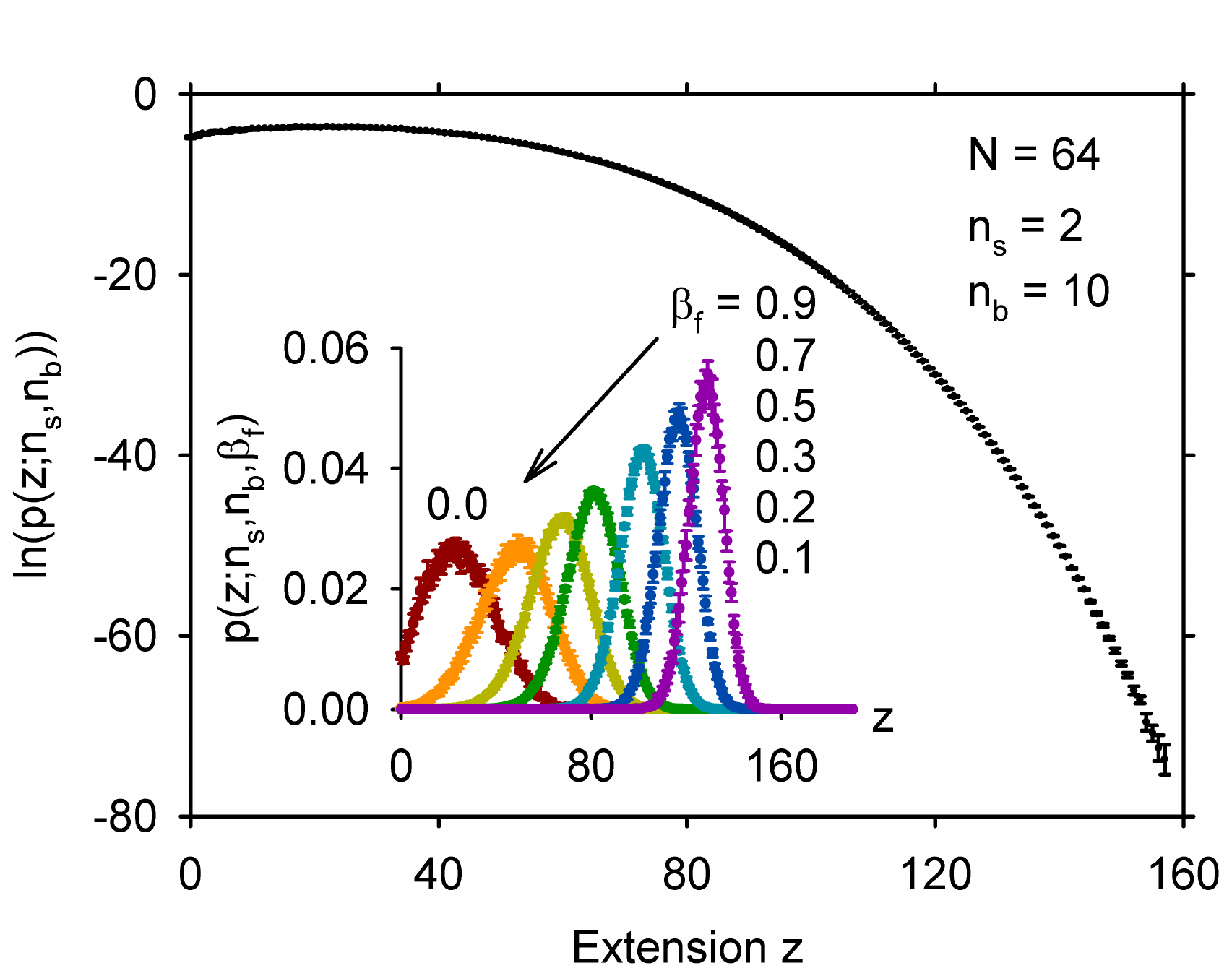}
\caption{
Logarithm of the combined probability profile $p(z;n_s,n_b)$ as a
function of extension, $z$, for chain length $N=64$, 
$n_s=2$ surface contacts, and $n_b=10$ bead contacts. The symbols with
error bars show the probability distribution obtained by reweighting
profiles from production simulations under different fields.
The combined profile shows that there is a good match between
the reweighted probabilities in the overlap regions. 
The inset shows probability profiles $p(z;n_s,n_b,\beta_f)$ (in linear
scale) from simulations for seven different tension fields 
between $\beta_f=0.0$ and $\beta_f=0.9$ that
contributed to the combined profile. 
As one would expect, with increasing tension force in the
upward direction, the maximum of the probability 
shifts to larger extensions and the width of the profiles decreases. 
These results illustrate how the technique of 
production under tension with reweighting extends the profile 
to much larger heights than are accessible in zero-tension
simulations; for the contact numbers in this figure, $n_s = 2$, $n_b = 10$, 
from a maximum height of about
$z=60$ ($\beta_f=0.0$ profile in inset) to one of about $z=155$. 
\label{profiles}}
\end{figure}

Fig.~\ref{profiles} illustrates the construction of the 3-d density of states.

\begin{figure}[!hbp]
\includegraphics[width=3.4in]{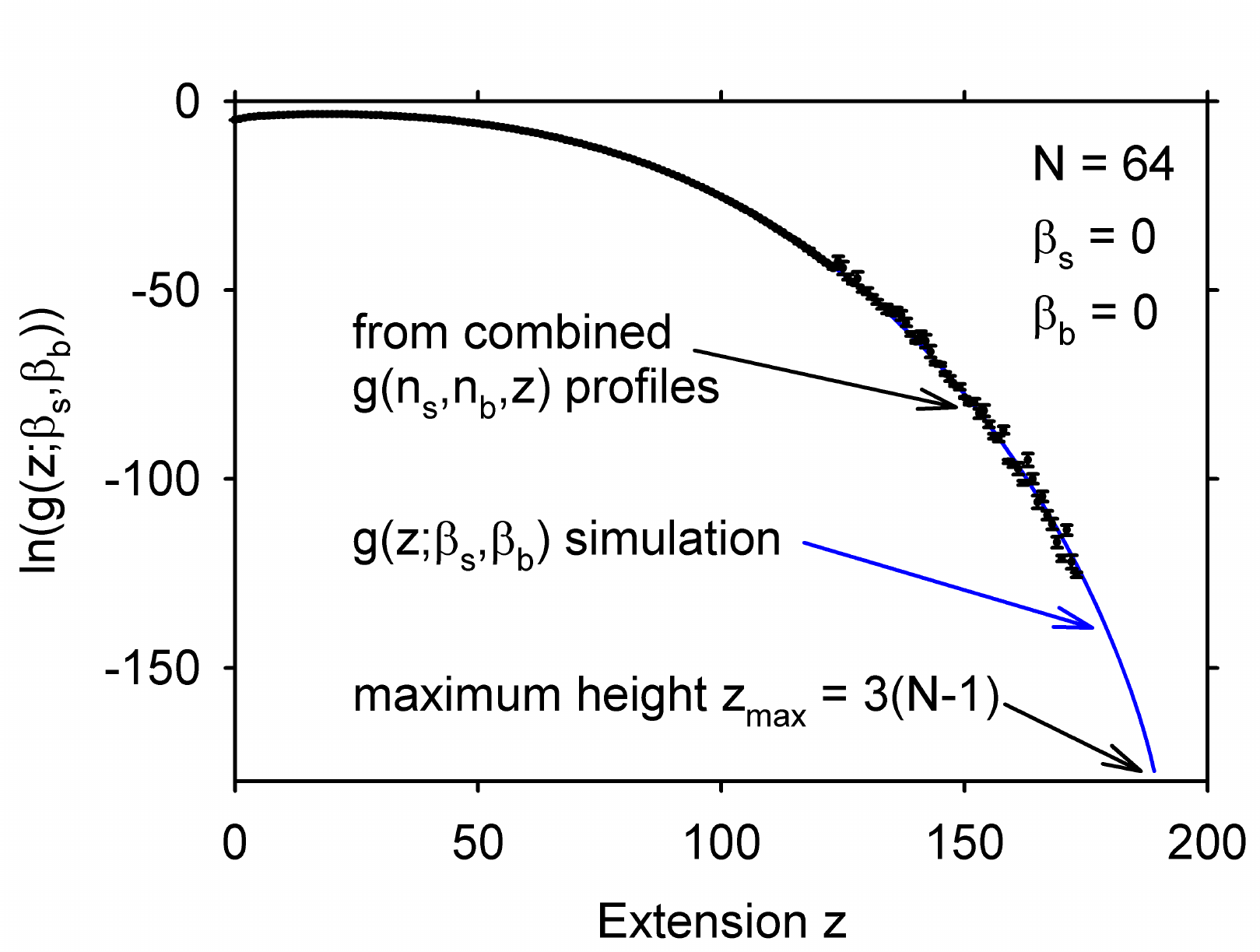}
\caption{Logarithm of the normalized density of states
  $g(z;\beta_s,\beta_b)$ for chain length $N=64$ and contact fields
  $\beta_s = \beta_b = 0$. The largest possible
  extension, $z_\mathrm{max} = 3(N-1) = 189$, for this chain length is
  indicated by an arrow. The
  solid line represents results from one 
  Wang-Landau simulation over the one-dimensional space of heights
  $z$; results from two other simulations are indistinguishable on
  this scale. The symbols with 
  error bars represent the probability distribution $p(z;\beta_s,\beta_b)$
  calculated with Eq.~(\ref{zprofile}) from the density of states
  $g(n_s,n_b,z)$ over the three-dimensional space of states
  $(n_s,n_b,z)$. 
  The range of extensions for $p(z;\beta_s,\beta_b)$ is
  limited and the uncertainties at higher $z$ are large due to
  insufficient sampling of highly stretched chain conformations during
  the production simulations. 
In the range where both are available, $p(z;\beta_s,\beta_b)$ and
$g(z;\beta_s,\beta_b)$ are in excellent agreement.
\label{lngz}}
\end{figure}

\newpage
When the 1-d density of states is normalized to unity so that 
$\sum_z g(z;\beta_s,\beta_b) = 1$, it represents the
probability to find the last bead at height $z$ for contact fields 
$\beta_s$ and $\beta_b$. 
The probability distribution $p(z;\beta_s,\beta_b)$ for the extension
at constant contact fields may also be calculated by partial summation
of the 3-d density of states $g(n_s,n_b,z)$ 
\begin{equation}\label{zprofile}
p(z;\beta_s,\beta_b) = \frac{\sum_{n_s,n_b}g(n_s,n_b,z)
\mathrm{e}^{\beta_s n_s}\mathrm{e}^{\beta_b n_b}}
{\sum_{n_s,n_b,z}g(n_s,n_b,z)
\mathrm{e}^{\beta_s n_s}\mathrm{e}^{\beta_b n_b}}. 
\end{equation}
In Fig.~\ref{lngz} we show a comparison of $p(z;\beta_s,\beta_b)$ and
normalized $g(z;\beta_s,\beta_b)$ results.

\end{document}